\documentclass[twocolumns]{emulateapj-rtx4}

\usepackage{float}
\usepackage{graphicx}
\usepackage{latexsym}
\usepackage{amsfonts,amsmath,amssymb}
\usepackage{url}
\usepackage[utf8]{inputenc}
\usepackage{textcomp}
\usepackage{comment}
\usepackage{longtable}
\usepackage{multirow,booktabs}
\usepackage[english]{babel}
\usepackage[usenames]{color}
\usepackage{epstopdf}

\usepackage{txfonts} 
\usepackage{natbib} 
\usepackage{xspace}
\usepackage{color}
\usepackage[colorlinks,urlcolor=blue,citecolor=blue,linkcolor=blue]{hyperref}

\def\lsim{\mathrel{\raise.3ex\hbox{$<$\kern-.75em\lower1ex\hbox{$\sim$}}}}
\def\gsim{\mathrel{\raise.3ex\hbox{$>$\kern-.75em\lower1ex\hbox{$\sim$}}}}
\def\gtwid{\mathrel{\raise.3ex\hbox{$>$\kern-.75em\lower1ex\hbox{$\sim$}}}}
\def\proptwid{\mathrel{\raise.3ex\hbox{$\propto$\kern-.75em\lower1ex\hbox{$\sim$}}}}

\shorttitle{SMA observations of the CH$_3$OH maser outflow in DR21(OH)}
\shortauthors{Orozco-Aguilera Ma. T. et al.}

\begin{document}


\title{SMA line observations of the CH$_3$OH-maser outflow in DR21(OH)}


\author{Ma. T. Orozco-Aguilera\altaffilmark{1}}

\author{A. Hern\'andez-G\'omez\altaffilmark{2,3}  }

\author{Luis A. Zapata\altaffilmark{2}}


\altaffiltext{1}{Instituto Nacional de Astrof\'{i}sica, \'{O}ptica y Electr\'{o}nica, 72840, Tonantzintla, Puebla, M\'exico}

\altaffiltext{2}{Instituto de Radioastronom\'ia y Astrof\'isica, Universidad
Nacional Aut\'onoma de M\'exico, Morelia 58089, Mexico}

\altaffiltext{3}{IRAP, Universit\'e de Toulouse, CNRS, UPS, CNES, Toulouse, France}

\begin{abstract}
We present a (sub)millimeter line survey of the methanol maser outflow located in the massive star-forming 
region DR21(OH) carried out with the Submillimeter Array (SMA) at 217/227 GHz and 337/347 GHz. 
We find transitions from several molecules towards the maser outflow such as CH$_3$OH, H$_2$CS, 
C$^{17}$O, H$^{13}$CO$^+$ and C$^{34}$S. However, with the present observations, we cannot discard 
the possibility that some of the observed species such as C$^{17}$O, C$^{34}$S, and H$_2$CS,
might be instead associated with the compact and dusty continuum sources located in the MM2 region.
Given that most of transitions correspond 
to methanol lines,  we have computed a rotational diagram with CASSIS and a LTE synthetic spectra with XCLASS for 
the detected methanol lines in order to estimate the rotational temperature and column density in small solid angle of the outflow 
where enough lines are present. We obtain a rotational temperature of $28\pm 2.5$K and a column 
density of $6.0\pm 0.9 \times 10^{15}$ cm$^{-2}$. These values are comparable to those column densities/rotational temperatures reported 
in outflows emanating from low-mass stars. Extreme and moderate physical conditions 
to excite the maser and thermal emission coexist within the CH$_3$OH flow. 
Finally, we do not detect any complex molecules associated with the flow,  e.g., CH3OCHO, (CH3)2CO, and CH$_3$CH$_2$CN.
\end{abstract}


\section{Introduction}

DR21(OH) (also known as W75S) is a prominent high-mass star forming region with a bolometric luminosity 
of about 1.5 $\times$10$^4$ L$_\odot$ and with a total mass of about 1$\times$10$^4$ M$_\odot$ \citep{har1977,chan1993}  
located about $3\arcmin$ north of  DR21 in the Cygnus X molecular cloud complex 
\citep{downes1966, motte2007, jakob2007, reipurth2008,zapata2013}. From DR21 emanates 
an energetic molecular outflow, but its nature seems to be different from the typical flows excited mediate disks \citep{zapata2013,zapata2017}. 
DR21(OH) and DR21 is embedded in a 4 pc long and massive filamentary ridge that extends in a north-south orientation 
\citep{harvey1986, vallee2006, csengeri2011, schneider2010, hennemann2012}.  
The distance to DR21(OH) has been accurately determined recently by trigonometric 
parallax of its associated methanol masers as 1.50 $\pm$ 0.08 kpc \citep{rygl2012}. 
This region has been extensively studied at infrared, millimeter and submillimeter wavelengths
\citep{lai2003, davis2007, kum2007, araya2009, minh2011, zapata2012, minh2012, girart2013}.\\

DR21(OH) contains two main dusty condensations MM1 and MM2 that are warm $\sim$ 50 and 30 K and 
massive 350 and 570 M$_\odot$, respectively, see {\it e.g.} \citet{man1991}. Recently, nine compact millimeter 
sources were revealed within both condensations, and with masses in a range of $4-25$ M$_\odot$ see  {\it e.g.}  
\citet{zapata2012, girart2013}. 
Five of the compact sources are associated with the extended millimeter source MM1 (SMA $5-9$) and 
four with MM2 (SMA $1-4$). These sources are likely to be large dusty disks/envelopes around high mass protostars. 
Two of the compact sources associated with MM1 (SMA6 and SMA7) have spectral features consistent with 
hot molecular cores, see  {\it e.g.}  \citet{minh2011}, \citet{minh2012}, and \citet{zapata2012}. 

Several dense molecular outflows have been reported to emanate from within the MM1 and MM2 cores, traced by 
CO, SiO, H$_2$CO, H$_2$CS and CH$_3$OH emission \citep{lai2003, minh2011, zapata2012, girart2013}. 
In particular, a well-collimated east--west bipolar maser and thermal CH$_3$OH  flow driven from within the 
SMA4 has been reported and discussed by \citet{plambeck1990}, \citet{kogan1998}, \citet{kurtz2004}, 
\citet{araya2009}, \citet{fish2011}, and \citet{zapata2012}. The LSR radial velocity of CH$_3$OH flow 
is nearly ambient (10 to $-$5 km s$^{-1}$) and is elongated in an east-west direction. 
This might suggest that this outflow is close to the plane of the sky. 
The outflow in DR21(OH) has also been detected in formaldehyde transitions at 
millimeter wavelengths \citep{zapata2012}.  

It is thought that the molecular outflows provide a chemical enrichment to the surroundings of the young stellar objects
through shocks, which can then destroy and/or release complex molecules trapped in the grains into the gas phase. 
This process is done by  compressing and then heating the interstellar medium \citep{garay1998, bachiller2001, jorgensen2007, arce2008}. 
The molecular outflows that present an overabundance of species such as SiO, CH$_3$OH, H$_2$CO, HCO$^+$, HCN, and H$_2$O, 
or even complex molecules (with six or more atoms) are classified as ``chemically active outflows" \citep{bachiller2001}. 
Therefore, the E-W outflow in DR21(OH) could be considered as chemically active. However, as we will see 
in the next sections no complex molecules are detected in the DR21(OH) outflow (e.g., CH3OCHO, (CH3)2CO, and CH$_3$CH$_2$CN.

\begin{figure*}[ht!]
\begin{center}
\includegraphics[width=1.05\textwidth]{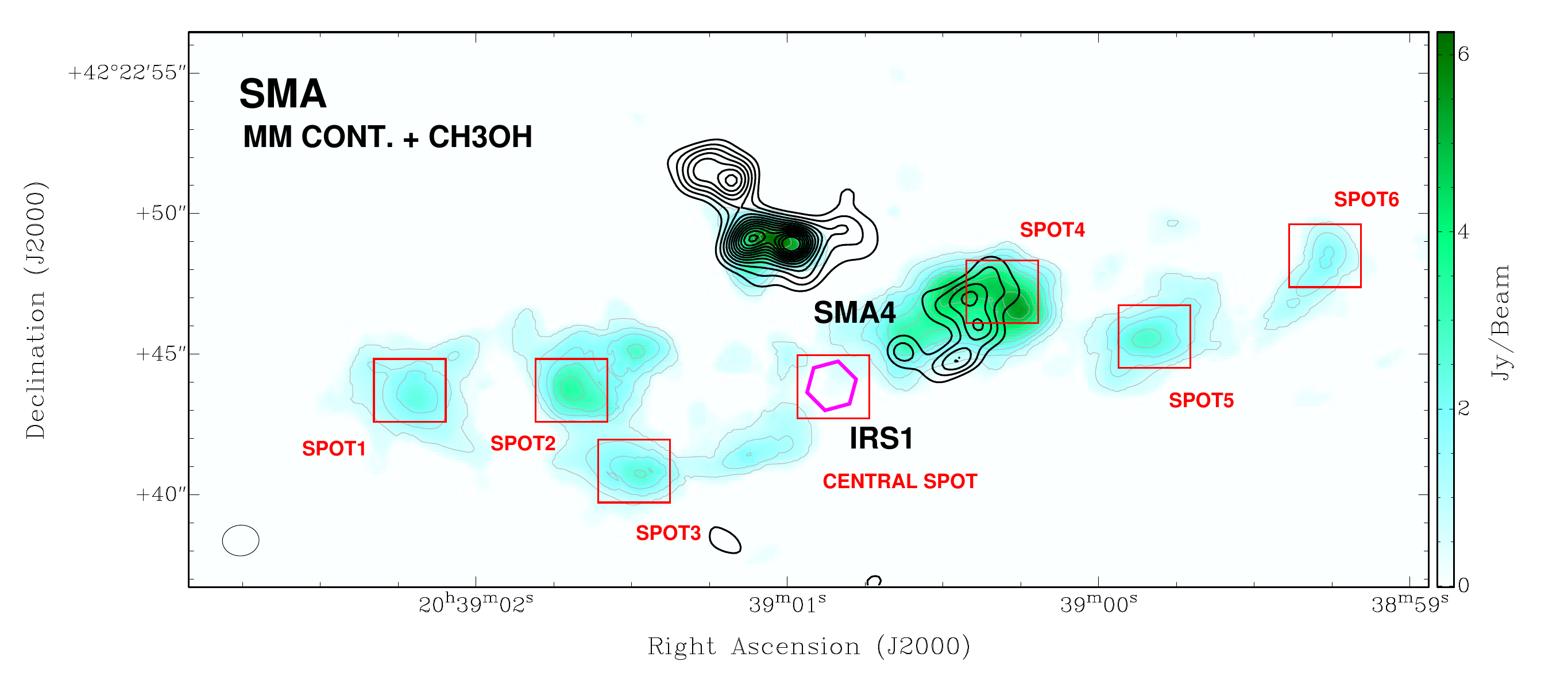}  
\caption{SMA moment zero map of the CH$_3$OH [4(2,2)-3(1,2)-E] emission (green scale and grey contours) overlaid on an 1.4 mm
continuum emission map (black contours). The SMA data were obtained from \citet{zapata2012}. 
The red squares mark the position where the spectra were obtained (see the appendix). 
The magenta hexagon marks the position of the infrared source IRS1 observed with Spitzer at 8 $\mu$m \citep{araya2009}. 
The synthesized beam of the image is shown in the bottom left corner of the image. 
The black contours are from 30\% to 95\% with steps of 5\% of the 1.4 mm continuum emission peak; the peak is at 150 mJy beam$^{-1}$. 
The grey contours are from 5\% to 85\% with steps of 10\% of the line emission peak. 
The color-scale bar on the right indicates the emission peak of the line emission. }
\end{center}
\label{figSpots}
\end{figure*}
 
There are only a few cases where ``chemically active outflows" have been reported in the literature.  
The list includes the L1157, B1-b, SMM4-W, BHR71, HH 114 MMS and IRAS 20126+4104 outflows 
\citep{arce2008,ober2011,taf2013,palau2017}. These flows are ejected from low- and high-mass young stars.  
In particular, in these outflows complex molecules have been detected, for example the HCOOCH$_3$ or the C$_2$H$_5$OH. 
Chemical models indicate that complex molecules  can form on grain surfaces \citep{gar2006,her2009}, 
present probably in the outflows.

Here, we have carried out a molecular line survey toward the CH$_3$OH outflow in DR21(OH), 
aiming to determinate its physical conditions, and to find evidence of complex organic molecules present in the flow. 

\section{Observations}

The observations of DR21(OH) were made with the SMA \citep{ho2004} in May 2006  at $\nu \sim$ 217/227 GHz 
and in August 2006  at $\nu \sim$ 337/347 GHz in its extended and compact configuration, respectively. 
The total observing time on-source with the SMA at 337/347 GHz was about 1.38 hrs., while for the 217/227 GHz 
band was around 2.22 hrs.  The pointing center for both datasets was $\alpha(J2000.0) = 20^h 39^m 01^s$, 
and $\delta(J2000) = 42 \arcdeg 22' 48''$. The Full Width Half Maximum (FWHM) is $55 \arcsec$ 
and $36 \arcsec$ at 230 and 345 GHz, respectively. The molecular emission arising from the outflow, 
and the emission from the continuum condensations fall well within both FWHMs. 

The SMA digital correlator was configured to cover 4 GHz of bandwidth, 2 GHz in the Upper Sideband (USB) and 2 GHz 
in the Lower Sideband (LSB). Each 2 GHz sidebands was covered by 24 chunks with a width of 104 MHz each and divided by 
128 resolution channels.  This correlator configuration provided a spectral resolution of about 0.8125 MHz 
({\it i.e.}, $\sim 0.7$ km s$^{-1}$ at 345 GHz and $\sim1$ km s$^{-1}$ at 230 GHz). 
Heterodyne SIS receivers were used and system temperatures ranged between 120 to 240 K 
for the different antennas at both frequencies.  

The zenith opacity ($\tau_{230 GHz}$), measured with the NRAO tipping radiometer located at the Caltech 
Submillimeter Observatory, was between 0.1 and 0.25 at both frequencies, indicating  reasonable weather 
conditions during the observations. Uranus and Callisto provided the absolute scale for the flux density calibration. 
The phase-gain calibrators were the quasars BL Lac and MCW349. The bandpass calibration was done using 
the quasar 3C454.3. The uncertainty in the flux scale is estimated to be about $15\%-20\%$, 
based on SMA monitoring of quasars\footnote{http://sma1.sma.hawaii.edu/callist/callist.html}.

The initial data calibration and reduction were made using the IDL superset MIR\footnote{http://cfa-www.harvard.edu/$\sim$cqi/mircook.html}. 
The spectra, molecular lines maps and the image analysis were made using the MIRIAD and KARMA softwares \citep{sau1995,goo96}. 
The robust weighting parameter was set to 2 giving a synthesized beam for 230 GHz of $1\arcsec .3 \times 1 \arcsec .0$ (PA = $-74 \arcdeg$) 
and for the 345 GHz $2 \arcsec .75 \times 1 \arcsec .96$ (PA = $-66.65 \arcdeg$). 
The resulting image {\it r.m.s.} noise for the line images were $\sim$20 mJy beam$^{-1}$ 
for each velocity channel at 230 GHz, and $\sim$50 mJy beam$^{-1}$ at 345 GHz.

\begin{deluxetable*}{llllclcc}
\tablecolumns{8}
\tablewidth{0pc}
\tablecaption{Species identified in the molecular outflow}
\tablehead{
\colhead{Frequency}&\colhead{Species}&\colhead{Transition} &\colhead{E$_u$} &\colhead{$A_{ij}$} &\colhead{Spot$^{(*)}$}&Range of Velocities$^{(**)}$ & Gaussian Fit\\
\colhead {(MHz)} & \colhead{}  &\colhead{} & \colhead{(K)} &\colhead{($s^{-1}$)} &\colhead{} &\colhead{km s$^{-1}$} & Center and FWHM [km $s^{-1}$]}
\startdata 
217104.919					&	SiO$^{(***)}$	       	&	$5_0 - 4_0$						& 31.26			&	5.20$\times 10^{-4}$		&	 C		&	-22, \, +8           &  -6.1$\pm$0.5; ~20.0$\pm$4.0  \\          
218222.192					&	H$_2$CO$^{(***)}$	&	$3_{ 0, 3} - 2_{ 0, 2}$				&	20.96 		&	2.82$\times 10^{-4}$		&	1,2,3,4,5		& 	-10, \, +6   &  -2.5$\pm$0.6 ; ~14.5$\pm$2.0 \\
218440.050					&	E-CH$_3$OH$^{(***)}$ &	 4$_{ +2, \, 2, \, 0} -3_{+1, \, 2, \, 0}$		&	45.46		&	4.69$\times 10^{-5}$		&	1,2,3,4,5,6 	& 	-9, \, +4      & -2.7$\pm$0.5 ; ~10.5$\pm$1.0  \\
218475.632					&	H$_2$CO$^{(***)}$	&	$3_{ 2, 2} - 2_{ 2, 1}$				&	68.09		&	1.57$\times 10^{-4}$		&	1,2,3,4,5		&	-10, \, +3    &  -3.3$\pm$0.3 ; ~13.5$\pm$1.0 \\
336865.110					&	A-CH$_3$OH			&	$12_{ 1, \, 11, \, -0} - 12_{ 0, \, 12, \, +0}$ 	&	197.07		&	4.07$\times 10^{-4}$		&	1,4,C 		& 	-6,\, +6  & -0.5$\pm$0.2 ; ~11.3$\pm$0.8  \\
337061.471				& 	C$^{17}$O 			&	 $3-2$							&	 32.35 		& 	1.11$\times 10^{-6}$ 	&	 4,C 			& 	 -4,\, +5  &    -0.3$\pm$0.2 ; ~4.6$\pm$0.5 \\
337396.459				 	& 	C$^{34}$S 			&	 $7-6$							&	 64.77 		& 	8.0$\times 10^{-4}$ 		&	 C			&	-8,\, +2  & -1.4$\pm$0.5 ; ~9.6$\pm$0.7 \\ 
338083.195 				 	& 	H$_{2}$CS 			& 	$10_{ 1, \, 10} - 9_{ 1, \, 9}$			& 	102.43 		& 	5.77$\times 10^{-4}$ 	& 	C			&	-8,\, +2   &-1.5$\pm$0.5 ; ~9.2$\pm$0.7 \\ 		
338124.502					&	E-CH$_3$OH			&	$7_{+0,\, 7, \,0} - 6_{ +0,\, 6, \, 0}$		&	78.08		&	1.70$\times 10^{-4}$		&	1,2,3,4,C		&	-8,\, +4    & -1.8$\pm$0.5 ; ~8.4$\pm$0.5  \\ 
338344.628					&	E-CH$_3$OH			&	$7_{ -1,\, 7, \, 0} - 6_{ -1,\, 6, \,0}$		&	70.55		&	1.67$\times 10^{-4}$		&	1,2,3,4,5,6,C 	&	-8,\, +4     & -2.8$\pm$0.5 ; ~8.5$\pm$0.4  \\ 
338404.580$^{(a)}$				&	E-CH$_3$OH			&	$7_{ +6,\, 2, \, 0} - 6_{ +6,\, 1,\,  0}$		&	243.79		&	4.48$\times 10^{-5}$		&	1,2,3,4,5,6,C	&	-8,\, +6     & -1.9$\pm$0.2 ; ~10.5$\pm$0.5  \\ 
338512.627$^{(b)}$				&	A-CH$_3$OH			&	$7_{ 4, \, 4,\, -0} - 6_{ 4, \, 3,\, -0}$		&	145.33		&	1.15$\times 10^{-4}$		&	2,3,4,C		&	-8,\, +5  &  -1.4$\pm$0.2 ; ~9.5$\pm$0.6   \\ 
338540.795$^{(c)}$			&	A-CH$_3$OH			&	$7_{ 3, \,5, \, +0} - 6_{ 3,\, 4,\,+0}$		&	114.79		&	1.39$\times 10^{-4}$		&	4,C			&	-9,\, +4  & -1.9$\pm$0.2 ; ~9.0$\pm$0.7 \\  
338559.928					&	E-CH$_3$OH			&	$7_{ -3, \, 5, \, 0} - 6_{ -3, \, 4, \, 0}$		&	 127.71		&	1.40$\times 10^{-4}$		& 	 4,C			&	-8, \, +2  & -2.1$\pm$0.4 ; ~9.4$\pm$0.7 \\ 
338583.195$^{(d)}$ 				&	E-CH$_3$OH			&	$7_{ +3, \, 4,  \, 0} - 6_{ +3,\,  3, \, 0}$	&	112.71		&	1.39$\times 10^{-4}$		&	4,C			&	-7,\, +5  &   -0.8$\pm$0.6 ; ~9.0$\pm$0.5   \\  
338614.999$^{(e)}$ 				&	E-CH$_3$OH			&	$7_{ +1, \, 6 \, 0} - 6_{ +1, \, 5, \, 0}$		&	86.05		&	1.71$\times 10^{-4}$		&	1,2,3,4,C		&	-9,\, +4  & -2.2$\pm$0.5 ; ~10.5$\pm$0.6 \\    
338639.939$^{(f)}$				&	A-CH$_3$OH			&	$7_{ 2,\, 5, \, +0} - 6_{ 2, \, 4, \, +0}$		&	102.72		&	1.57$\times 10^{-4}$		&	4,C			&	-8,\, +4  &  -1.6$\pm$0.3 ; ~9.8$\pm$0.3  \\  
338721.693$^{(g)}$				&	E-CH$_3$OH			&	$7_{ +2, \, 5, \, 0} - 6_{ +2, \, 4, \, 0}$		&	87.26		&	1.55$\times 10^{-4}$		&	1,2,3,4,5,6,C	&	-8,\, +6  &  -3.1$\pm$0.3 ; ~8.7$\pm$0.4 \\  
346998.344					&	H$^{13}$CO$^+$		&	$4 - 3$							&	41.63		&	3.29$\times 10^{-3}$		&	4,C			&	-8,\,  +2 &  -1.5$\pm$0.2 ; ~4.7$\pm$0.3 \\
347330.578					&	SiO					&	$8 - 7$							&	75.01		&	2.20$\times 10^{-3}$		&	C			&	 -10, \, +6  &  -3.1$\pm$0.5 ; ~12.5$\pm$4.0  \\
348534.364 					& 	H$_{2}$CS 			&	$10_{1, \, 9} - 9_{1, \, 8}$				&	105.19		&	6.32$\times 10^{-4}$		&	C			&	-6, \, +2    & -1.0$\pm$0.4 ; ~10.7$\pm$0.7  \\
\enddata
\tablecomments{\scriptsize Columns are frequency, quantum numbers, upper level energy, A$_{ij}$ Einstein coefficient, spot where the transition was identified and the range of velocities for the molecule. These values are based on CDMS database. $^{(a)}$ Line blended with CH$_3$OH at 338408.698 MHz. $^{(b)}$ Line blended with CH$_3$OH at 338512.856 MHz. $^{(c)}$ Line blended with $^{33}$SO at 338541.265 MHz and CH$_3$OH at 338543.204 MHz. $^{(d)}$ Line blended with $^{33}$SO at 338591.071 MHz. $^{(e)}$ Line blended with SO$_2$ at 338611.807 MHz.$^{(f)}$ Line blended with  $^{33}$SO at 338639.671 MHz. $^{(g)}$ Line blended with CH$_3$OH at 338722.898 MHz. $^{(*)}$ The numbers correspond to the spots described in Figure \ref{figSpots}. $^{(**)}$ The velocity range is approximate since the exact values are complicated to determine. These values were obtained from gaussian fittings to the spectra data. $^{(***)}$ These lines were already reported in \citet{zapata2012}.}
\label{tablaMol}
\end{deluxetable*}

\section{Results and Discussion}

In Figure \ref{figSpots}, we present the millimeter CH$_3$OH thermal emission from the maser outflow in DR21(OH).  
This image was adapted from \citet{zapata2012}, and shows the integrated intensity map (or the zero moment) of the E-CH$_3$OH [4(2,2)-3(1,2)] 
molecule at 218.44005 GHz, and the 1.4 mm continuum emission from objects present in DR21(OH). 
The millimeter methanol emission is tracing the bipolar low-velocity east--west maser outflow revealed for the first time by \citet{plambeck1990} 
at centimeter wavelengths. The emission appears to be concentrated into compact bow-shock structures along the bipolar outflow. 
Overall, the methanol millimeter emission follows a morphology very similar to that seen in the 36 and 44 GHz methanol 
maser lines \citep{araya2009, kogan1998, plambeck1990}, with the millimeter source (SMA4) being well in the middle of the flow. 
\citet{zapata2012} proposed SMA4 as its excitation source (see Figure \ref{figSpots}). However, an infrared source (IRAC1) that is prominent at 8 $\mu$m 
was also proposed by  \citet{araya2009} as a possible candidate for its exciting source.

In Figure \ref{figSpots}, we have also included the regions from the methanol outflow where we have obtained the spectra 
(SPOT1-6 and THE CENTRAL SPOT) to compute their physical conditions.  The thermal emission from the detected 
molecules, see Table \ref{tablaMol}, is very sparse and complicated to make a single averaged spectra. However, the technique applied here allow us 
to find different exciting conditions along the flow. We extracted the spectra from the red boxes defined in Figure \ref{figSpots} at both 
analyzed frequencies (217 \& 337 GHz)  and identified a total of 21 molecular lines, 12 of which correspond to methanol. 
In Table \ref{tablaMol}, we show the identified lines and give the parameters for each line, such as the transition, 
the energy level $E_{u}$, the Einstein coefficient $A_{ij}$, and the spot number in which it was found. 
These parameters are based on CDMS \footnote{https://www.astro.uni-koeln.de/cdms} database \citep{muller2001,muller2005}.

\begin{figure*}[ht!]
\includegraphics[width=0.92\textwidth, angle=0, clip=True, trim= 0 0 0 0]{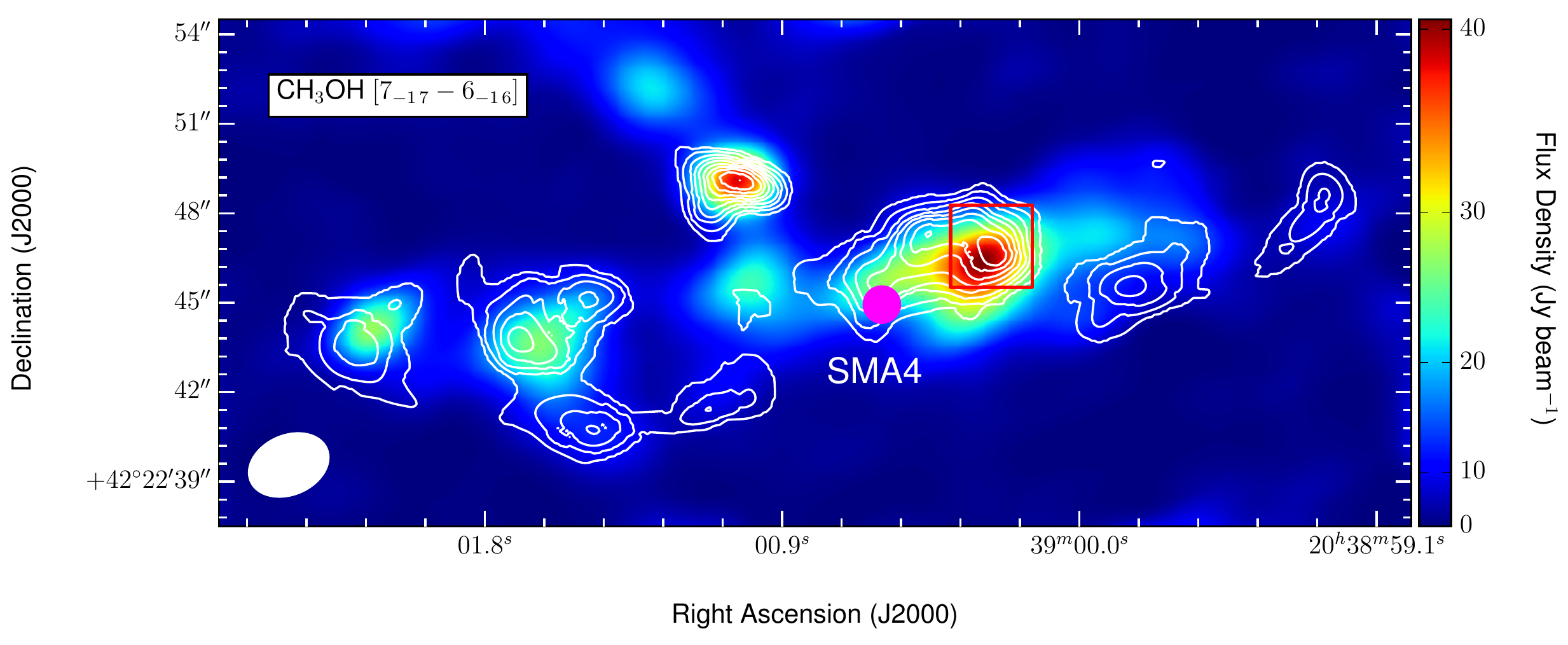}\\
\includegraphics[width=0.93\textwidth, angle=0, clip=True, trim= 0 0 0 0]{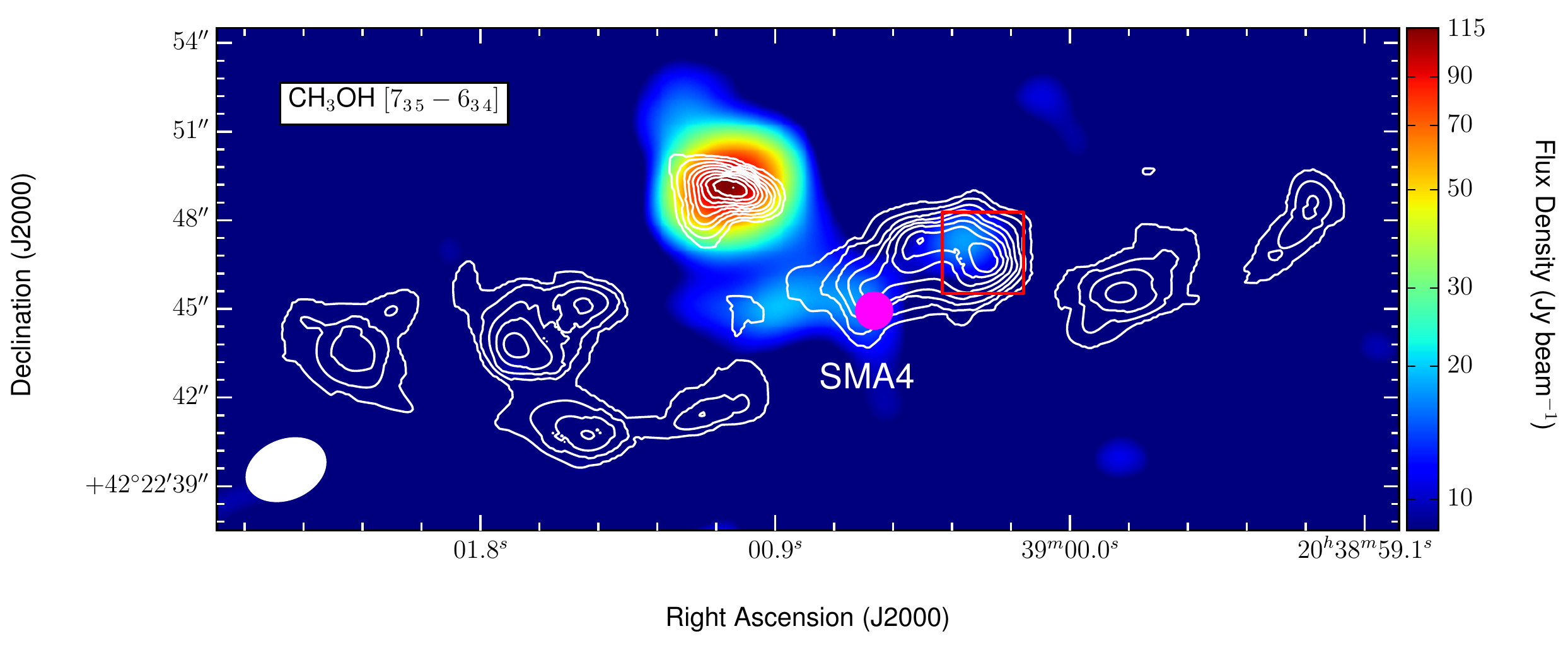}
\caption{SMA moment zero maps of the CH$_3$OH [$7_{-1,7,0}-6_{-1,6,0}$] (top) and CH$_3$OH [$7_{+3,5,0}-6_{+3,4,0}$] (bottom). 
The synthesized beams of the observations are shown in the lower right corner of each image. 
The black contours are the same as in Figure 1. The color-scale bar on the right 
indicates the emission peak of the line emission. The integrated range of velocities are shown in Table 1.}
\label{figMom0}
\end{figure*}
 
In the Appendix section, we include all the spectra obtained from the red boxes shown in Figure \ref{figSpots}.  These spectra include
the four bands analized in this study (217/227 and 337/347 GHz).  As it can be seen from the Appendix section and
Table \ref{tablaMol}, the line emission from the different species is not found in all the box regions.  This difficulties our study of the CH$_3$OH-outflow.   
Even taken only the methanol line, it is still the same problem. We have included two images in Figure \ref{figMom0} showing the morphology 
of methanol lines to observe its sparsity.  From this image, it is clear that only in the SPOT4 can make our analysis where
most of the methanol lines are present. In the other positions we only found some methanol lines, but not enough to make a
precise estimation for the column density and rotational temperatures. Some lines are even contaminated or blended with other transitions or species.      

From the spectra of SPOT4, we have made a search in the SPITZER images for a compact infrared source located in this 
position,  since we might have contamination from the sub-millimeter condensations associated with young protostars 
reported by \citet{zapata2012, girart2013} or other sources in MM2 cluster. 

 With a magenta hexagon in Figure \ref{figSpots} we have included the position of the closest infrared SPITZER source to MM2, which is 
the infrared source IRS1 detected with Spitzer at 8 $\mu$m by \citet{araya2009}. We did not find any infrared source within MM2 cluster 
(especially associated with the millimeter compact source SMA2), so at this point the possibility of protostellar contamination seems to be difficult.
In addition, as the molecular emission reported in this paper is very sparse towards
the MM2 cluster, especially the high density tracer CH$_3$OH (see Figures 12-15), the presence of a protostellar hot corino could also be discarded.  
 However,  the species such as C$^{17}$O, C$^{34}$S, and H$_2$CS could be tracing cold and compact dusty condensations within MM2.

\begin{figure}[ht!]
\centering
\includegraphics[width=0.51\textwidth, trim= 0 0 5 5, clip]{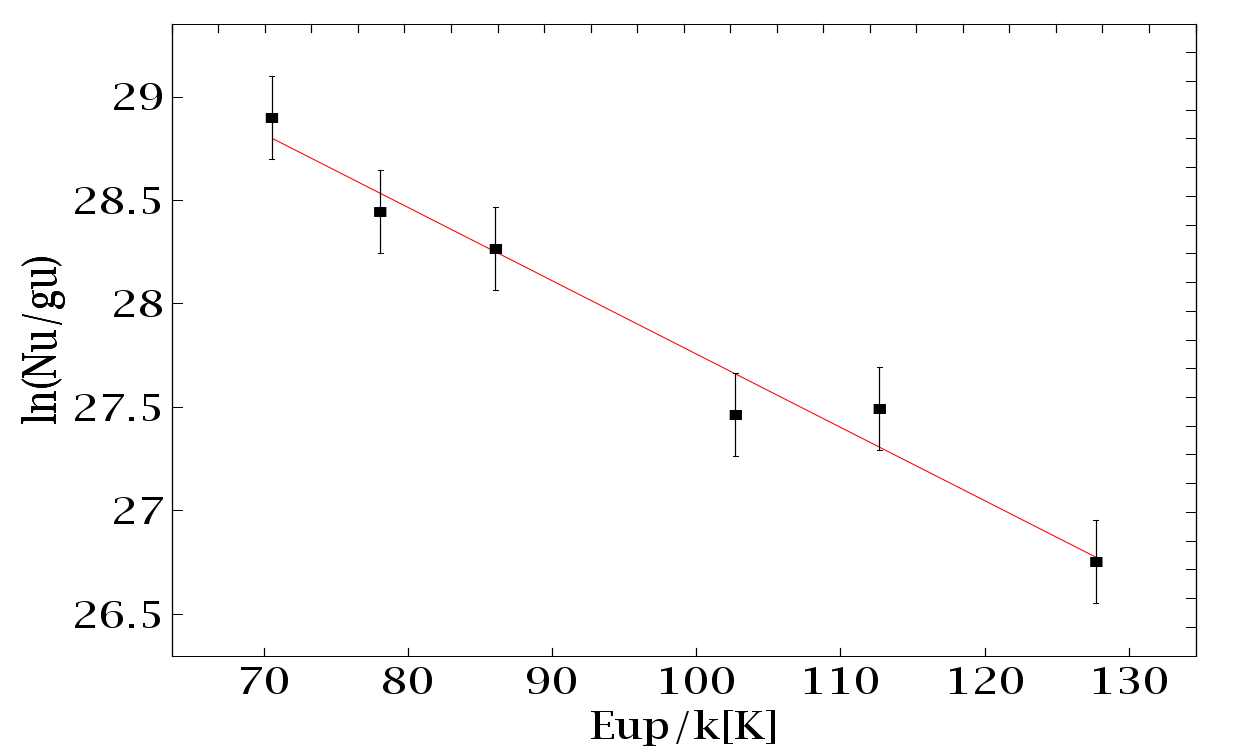} 
\caption{CH$_3$OH methanol rotational diagram for the SPOT4 in the CH$_3$OH outflow. 
We obtained a rotational temperature $28\pm 2.5$K and a column density of $6\pm 1.7 \times 10^{15}$ cm$^{-2}$. 
The error bars are associated with the error from the gaussian fitting to compute the integrated intensity. 
These bars includes 15\% of calibration error from the observations. We have not included blended lines in the fitting.}
\label{diagRot}
\end{figure}

Within the methanol outflow we only found simple molecules such as CH$_3$OH, H$_2$CS, 
H$_2$CO, SiO and H$^{13}$CO$^+$ (see Table 1). More complex molecules 
were not found at 4-$\sigma$ level of 80 mJy at 230 GHz and 200 mJy at 345 GHz. These complex 
molecules include Methyl Formate (CH3OCHO), Acetone ((CH3)2CO), and Ethyl Formate (CH$_3$CH$_2$CN). 
This is in agreement to that proposed in \citet{gar2006,her2009}, given that these simple molecules considered 
as first generation, or parent molecules, are probably released by shocks directly into the 
gas phase from the icy and dusty mantles.

\begin{figure*}[ht!]
\begin{center}
\includegraphics[width=0.9\textwidth, trim= 0 0 0 0, clip]{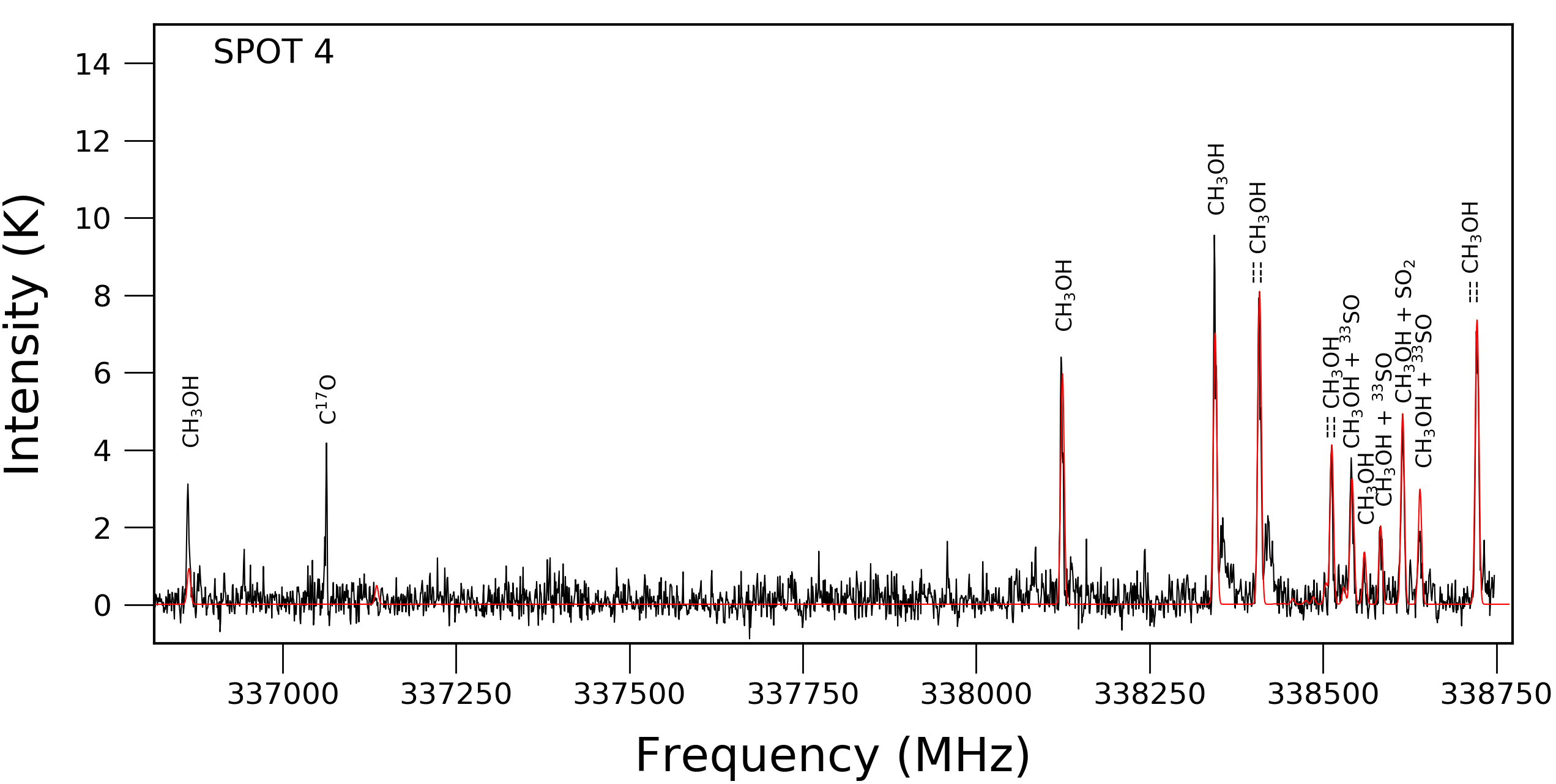} \\ 
\caption{SMA spectra from the outflow at 337 GHz from the SPOT4 (black line). 
We show in red the model made with XCLASS taken into account 
the values from the rotational diagram. All spectra obtained from 
the spots are shown in the appendix. }
\label{especs}
\end{center}
\vspace{0.5cm}
\end{figure*}

state of art tools (as cassis and xclass for (sub)millimeter observations) to make a powerful analysis of the physical conditions of the outflow.     

 \subsection{Physical parameters of the outflow in the SPOT4}
 
We have used the methanol lines identified in the outflow (SPOT4) to determine its physical parameters. 
We used the rotational diagram method to calculate the column density and rotational 
temperature associated with the lines. This physical method 
takes into account that the measured integrated intensity of the lines, $\int I_\nu dv$ (Jy beam$^{-1}$ km s$^{-1}$) 
is related to the column densities of the molecules in the upper level N$_u$ through the next equation: 

$$
\frac{N_u}{g_u}=\frac{N_{tot}}{Q(T_{tot})}e^{{-E_u/T_{tot}}}=\frac{1.7 \times 10^{14}}{\nu \mu^2 S} \int I_\nu dv, 
$$

where g$_u$ is the statistical weight of level $u$; N$_{tot}$ is the total column density in cm$^{-2}$; Q(T$_{tot}$) is 
the partition function for the rotational temperature T$_{rot}$; E$_u$ is the energy of the upper level in Kelvin; $\mu$ 
is the permanent dipole moment in Debye; and $S$ is the strength value. Therefore, a logarithmic plot of the quantity 
on the right-hand side of equation of above as a function of E$_u$ provides a straight line with slope (1/T$_{rot}$) 
and intercepts in N$_{tot}$/Q(T$_{rot}$). This gives the rotational temperature and column density. 
This method assumes that all level populations can be characterized by a single rotational temperature T$_{rot}$, 
and the lines are optically thin. 

In Figure \ref{diagRot} we show the rotational diagram obtained with CASSIS\footnote{http://cassis.irap.omp.eu} 
(Centre d'Analyse Scientifique de Spectres Instrumentaux et Synth\'etiques, \citealt{caux2011}).  We have avoided blended methanol 
lines to not over-estimate the integrated flux for these lines. From the fit we obtain a rotational temperature 
of $28\pm 2.5$K and a column density of $6 \pm 0.9 \times 10^{15}$ cm$^{-2}$ for the SPOT4 at 337GHz.
 We did the rotational diagram calculation for all the spots in the two bands, but because in the other spots there 
 were a few methanol transitions or they were blended, the calculation could not be done with good certainty.
 
Using these values obtained from the rotational diagram for the column density and rotational temperature, 
we have computed a synthetic model with XCLASS \citep{muller2017} to compare it with the observations. In Figure \ref{especs}
 we show the results. The synthetic model reproduces very well the spectra for the SPOT4 and in the window
 of 337 GHz where many methanol are present. The CH$_3$OH lines that are not well fitted is because
 of line blending or contamination of other spectral lines. We also tried to make LTE XCLASS synthetic 
 spectra for the rest of the spots, but again the blending, and the sparsity of the molecular emission along
 the outflow did not allow it.
 
The physical values for the rotational temperature obtained here are similar to those reported in the peak 
B1 in the blue lobe of the chemically rich active outflow L1157 ejected by a low-mass young 
star \citep{arce2008}. However, it is not clear if the physical conditions are the same since 
the column densities differ by a factor of 100. These values for the column density and rotational 
temperature are very low (more obvious for the rotational temperatures) compared with those values reported for hot molecular cores 
\citep[40 to 485K, and $10^{13}$ to $10^{17}$ cm$^{-2}$;][]{hernandez2014}.  Moreover, \citet{palau2017} for the
the case of the massive outflow in IRAS 20126+4104 reported column densities between $10^{15}$ to $10^{17}$ cm$^{-2}$,  
and rotational temperatures of 100 to 235 K. These values are also quite high to these reported here.   

Compared with other molecular outflows reported in the literature (L1157, HH 114 MMS and IRAS 20126+4104), 
no complex molecules were found in the methanol outflow. A possibility for this non-detection could be 
related with the age of the molecular outflow. 
The dynamical age of the outflow is about of 15,000 years. This time is too short for the formation 
of complex molecules in the gas phase after the shock-induced sputtering of the grain mantles 
\citep[see][]{gar2006,her2009}. However, \citet{arce2008} noted that is it most likely that the 
complex species are formed in the surface of grains and then are ejected from the grain 
mantles by the shocks. 

There could be the possibility that the emission from the C$^{34}$S and H$_2$CS may be contaminated from the  
sulfur-dominated region discovery by \citet{plambeck1990}, likely originated from the northern part of the flow.
However, our maps with better angular resolution from the H$_2$CS line reveal some emission very close to SMA4, which could be tracing
the outflow material ejected from this object. This possibility favours our interpretation that this emission (C$^{34}$S and H$_2$CS) is
excited by the outflow. However, we think that more sensitive observations from sulfur molecules could help to reveal its origin.  

\section{Conclusions} 

The main results of our work can be summarized as follows:

\begin{itemize}

\item 
We report the detection of 21 molecular lines, 12 of which correspond to methanol toward the methanol outflow in DR21(OH). 
These lines include CH$_3$OH, H$_2$CS, C$^{17}$O, H$^{13}$CO$^+$, and C$^{34}$S. 
This is the first time that these lines are detected within the outflow.  However,  we cannot discard the possibility that some 
of the observed species such as C$^{17}$O, C$^{34}$S, and H$_2$CS, might be instead associated with compact dust continuum 
emission from cores in MM2 region. We suggest that this outflow is chemically active.

\item
Given that most of the detected transitions correspond to methanol lines,  we have computed a rotational diagram with CASSIS 
and a LTE synthetic spectra with XCLASS for 
the detected methanol lines in order to estimate the rotational temperature and column density in small solid angle of the outflow 
where enough lines are present. We obtain a rotational temperature of $28\pm 2.5$K and a column 
density of $6\pm 0.9 \times 10^{15}$ cm$^{-2}$. These values are comparable to those column densities/ rotational temperatures reported 
in outflows emanating from low-mass stars.

\item
No complex molecules were found in the methanol outflow, e.g., CH3OCHO, (CH3)2CO, and CH$_3$CH$_2$CN.

\end{itemize}

\acknowledgments
M. T. O., A. H-G., and L.A.Z acknowledge the financial support from DGAPA, UNAM, and CONACyT, M\'exico. We would like to thank Aina Palau for discussing
multiple times the LTE models used in the study and for the infrared images that she made in older versions of the paper.   
We are very thankful for the thoughtful suggestions of the anonymous referees that helped to improve our manuscript.

\section{Appendix}

\subsection{Molecular spectra}

In this appendix we show the obtained spectra from the spots (see figure \ref{figSpots}). 

\begin{figure*}[ht!]
  \centering
	\begin{tabular}{ll}
	\includegraphics[width=0.5\textwidth]{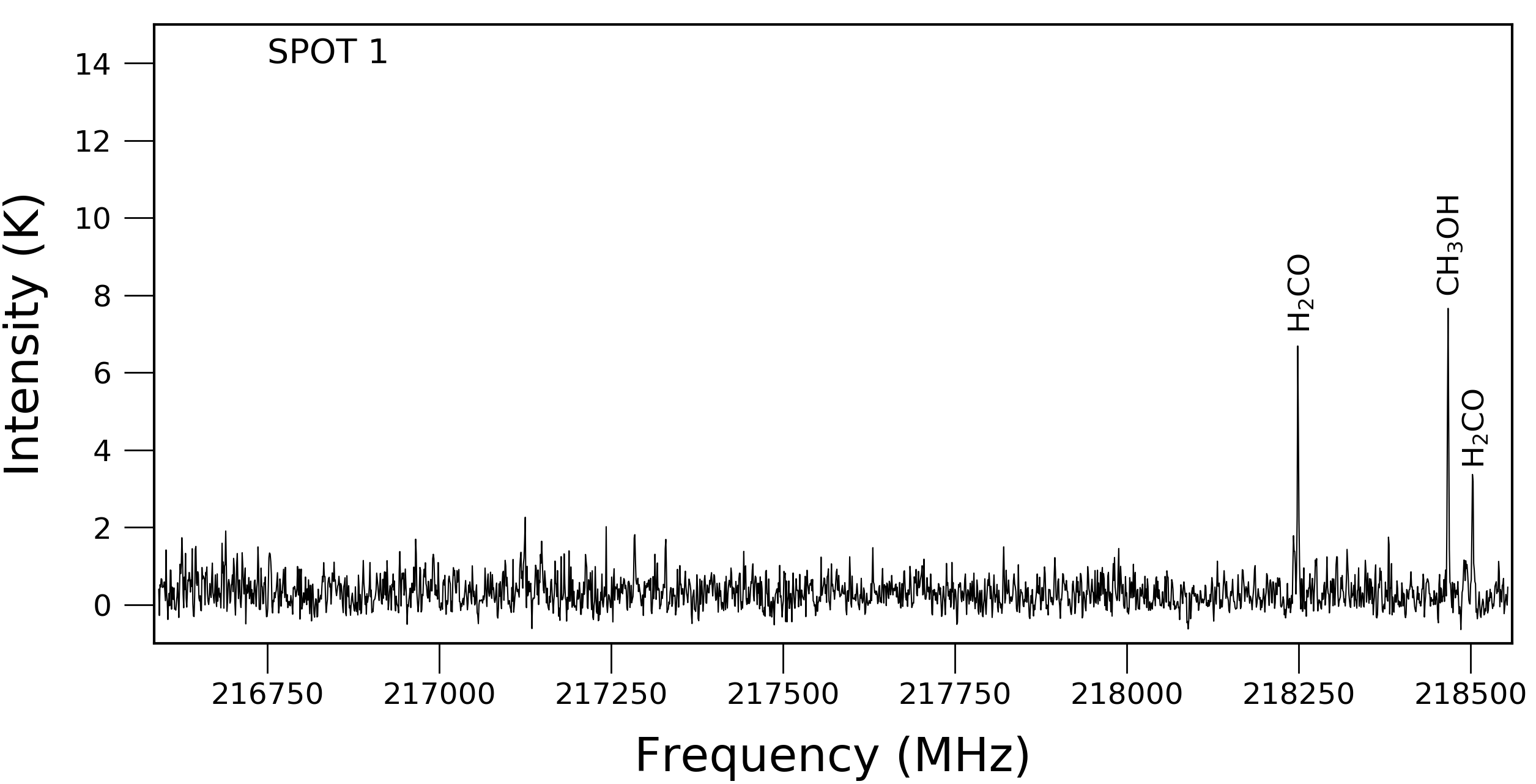}
	&\includegraphics[width=0.5\textwidth]{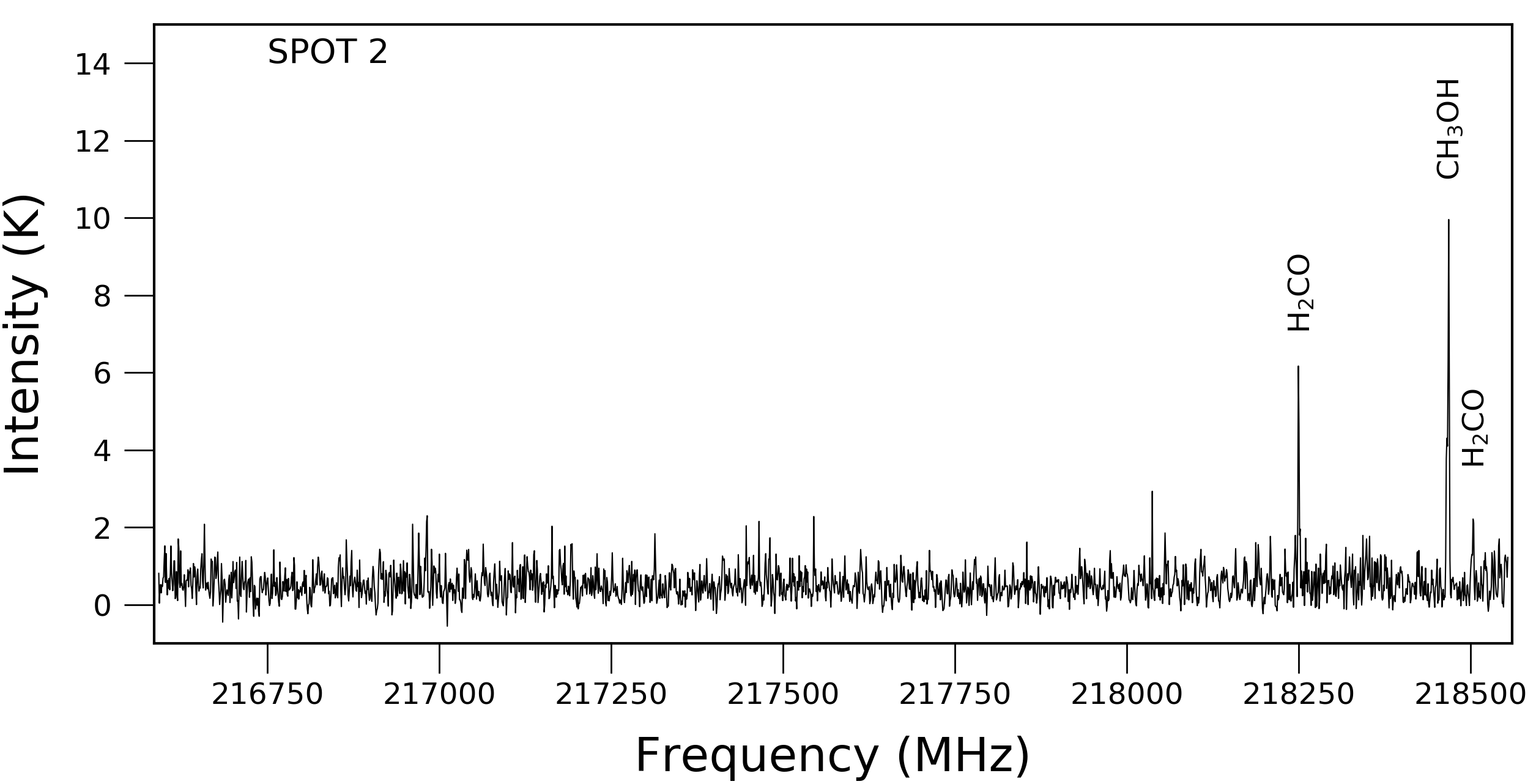}\\
	
	\includegraphics[width=0.5\textwidth]{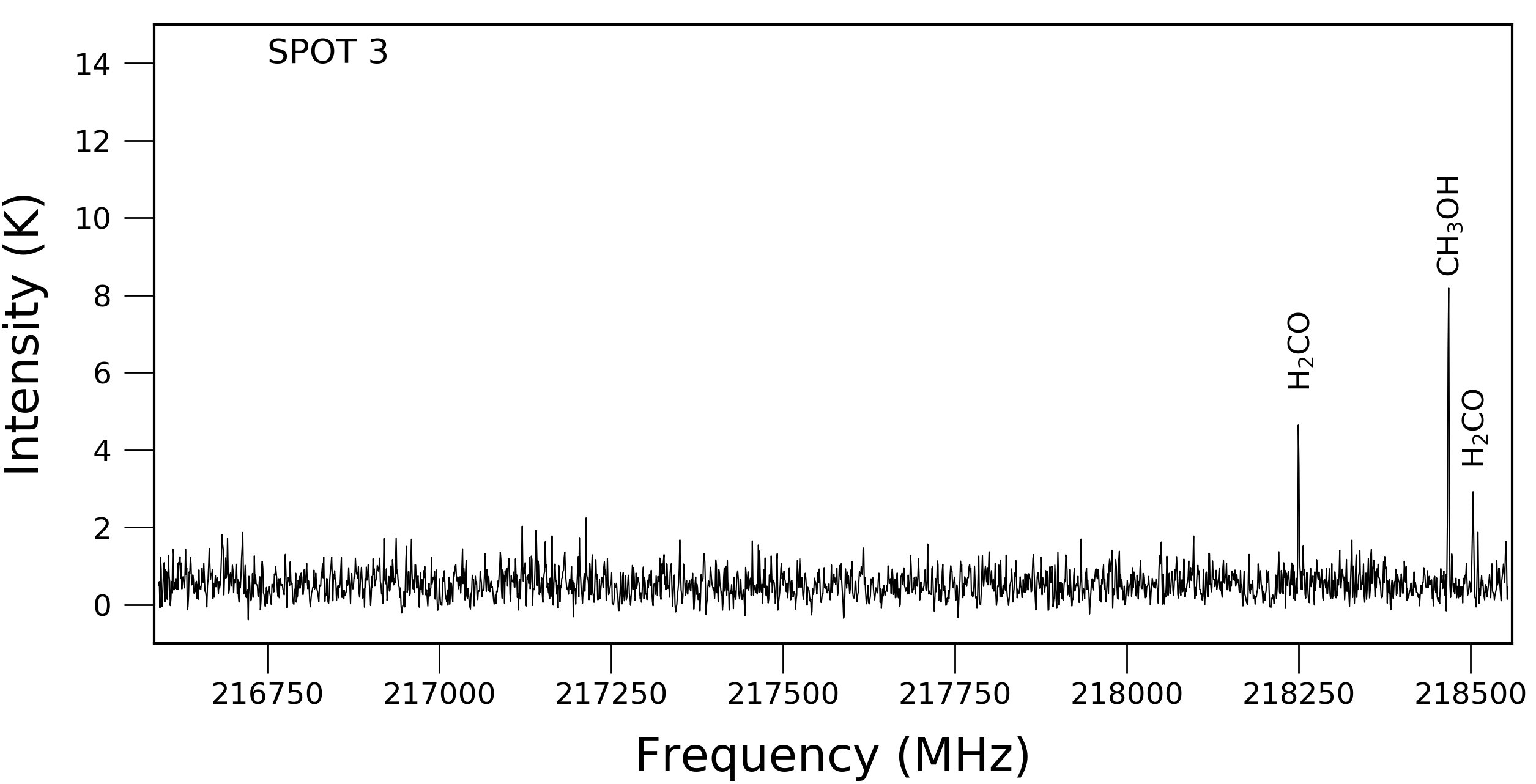}
	&\includegraphics[width=0.5\textwidth]{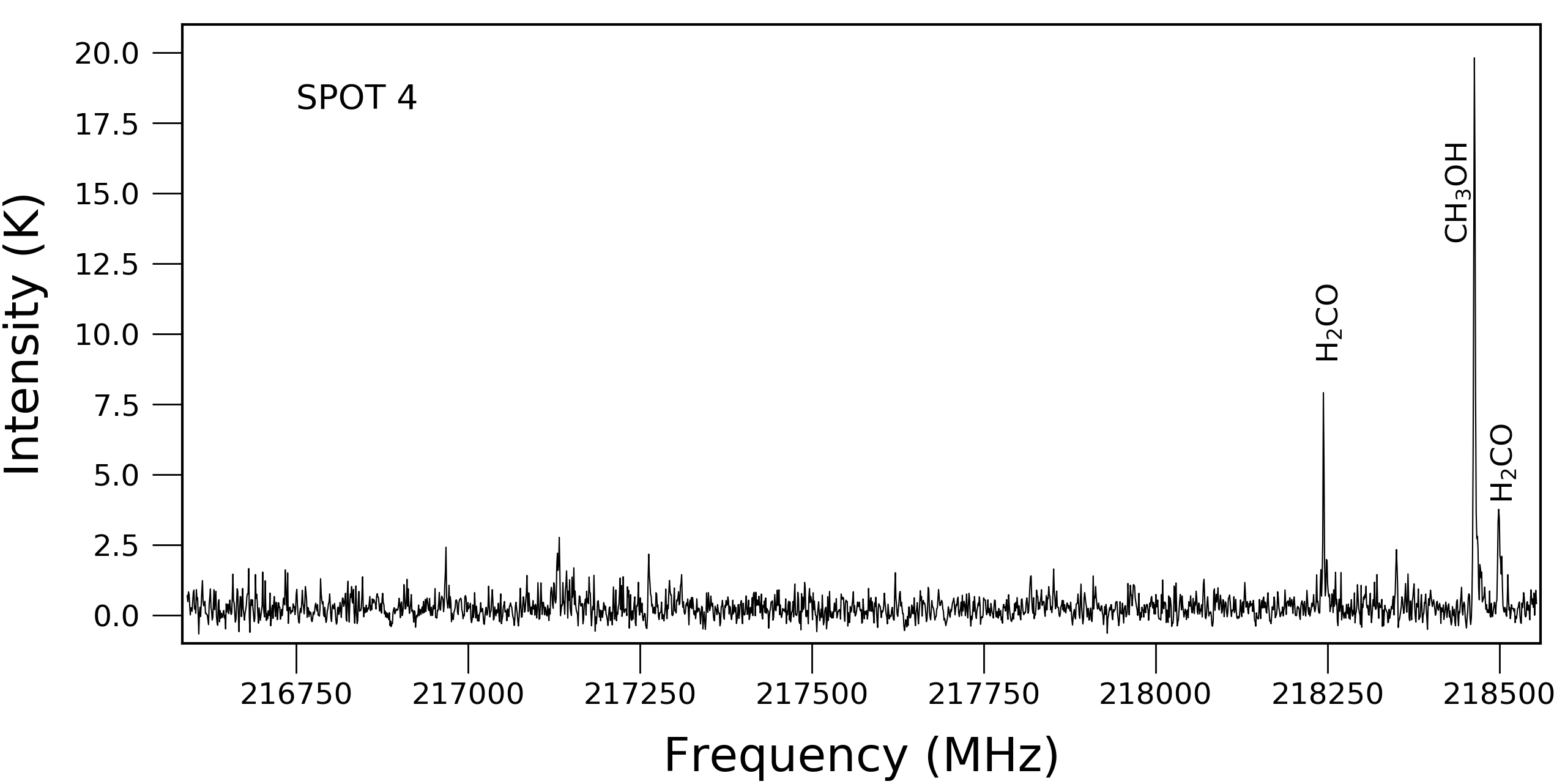}\\
	
	\includegraphics[width=0.5\textwidth]{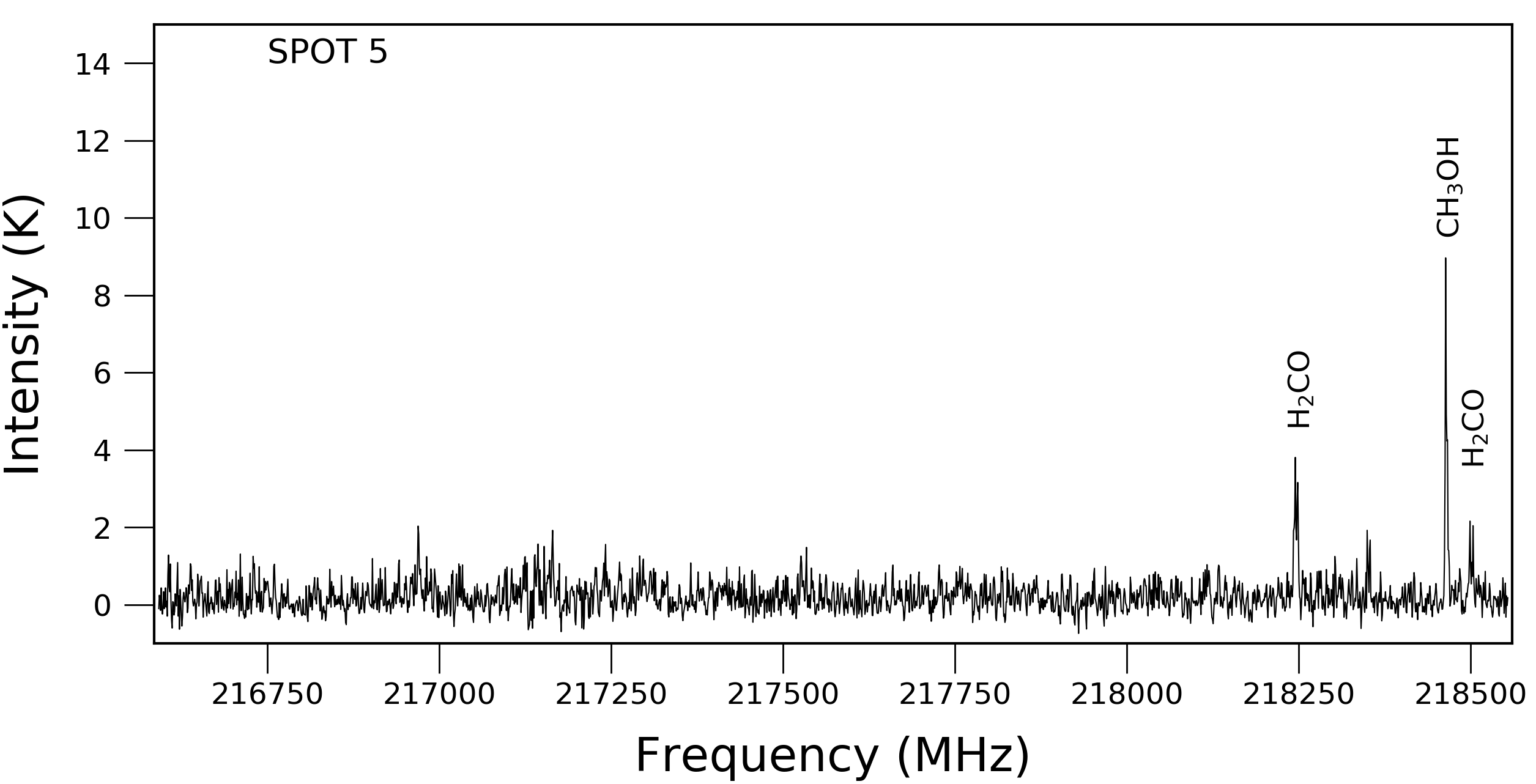}  
		&\includegraphics[width=0.5\textwidth]{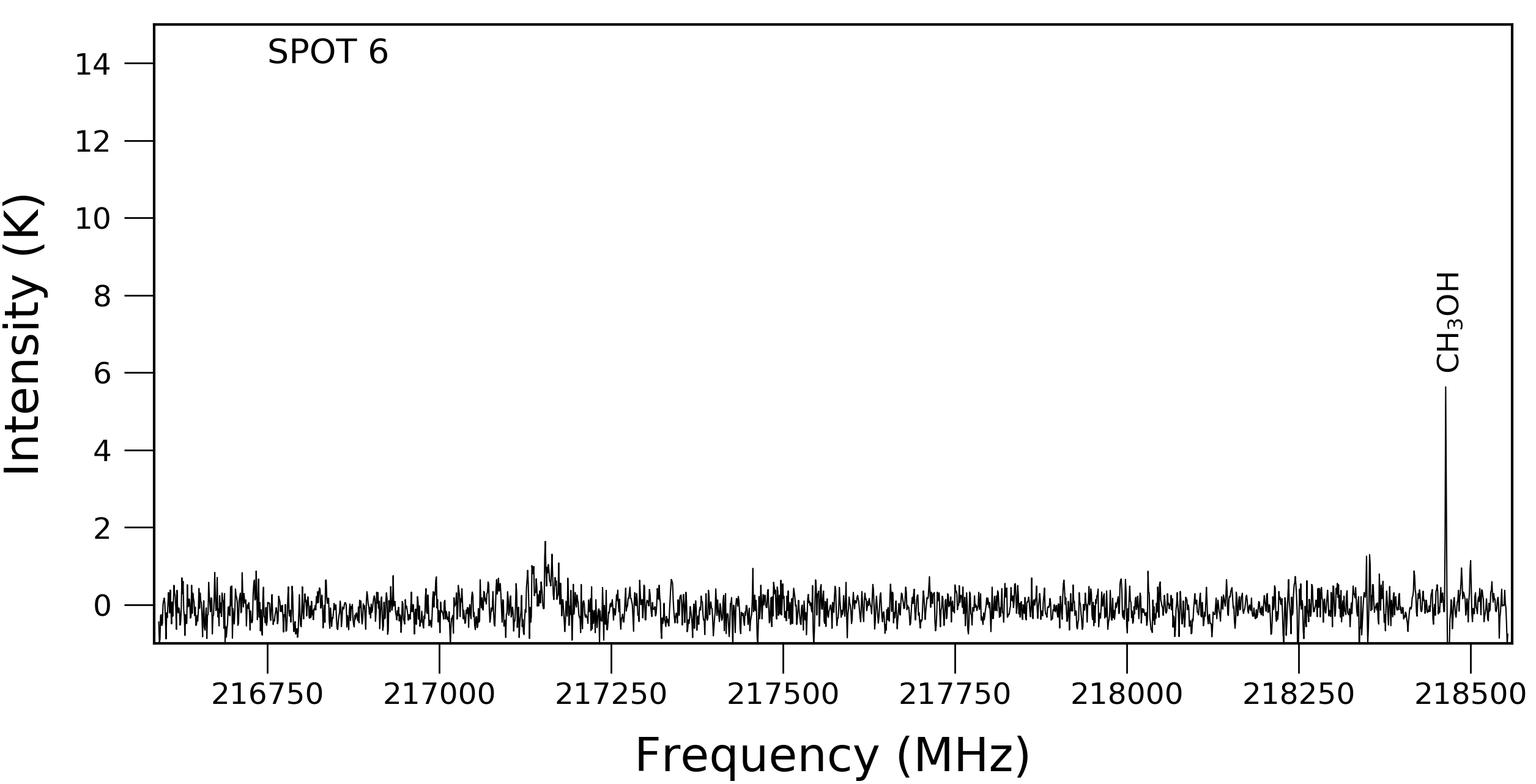}\\
	
	\includegraphics[width=0.5\textwidth]{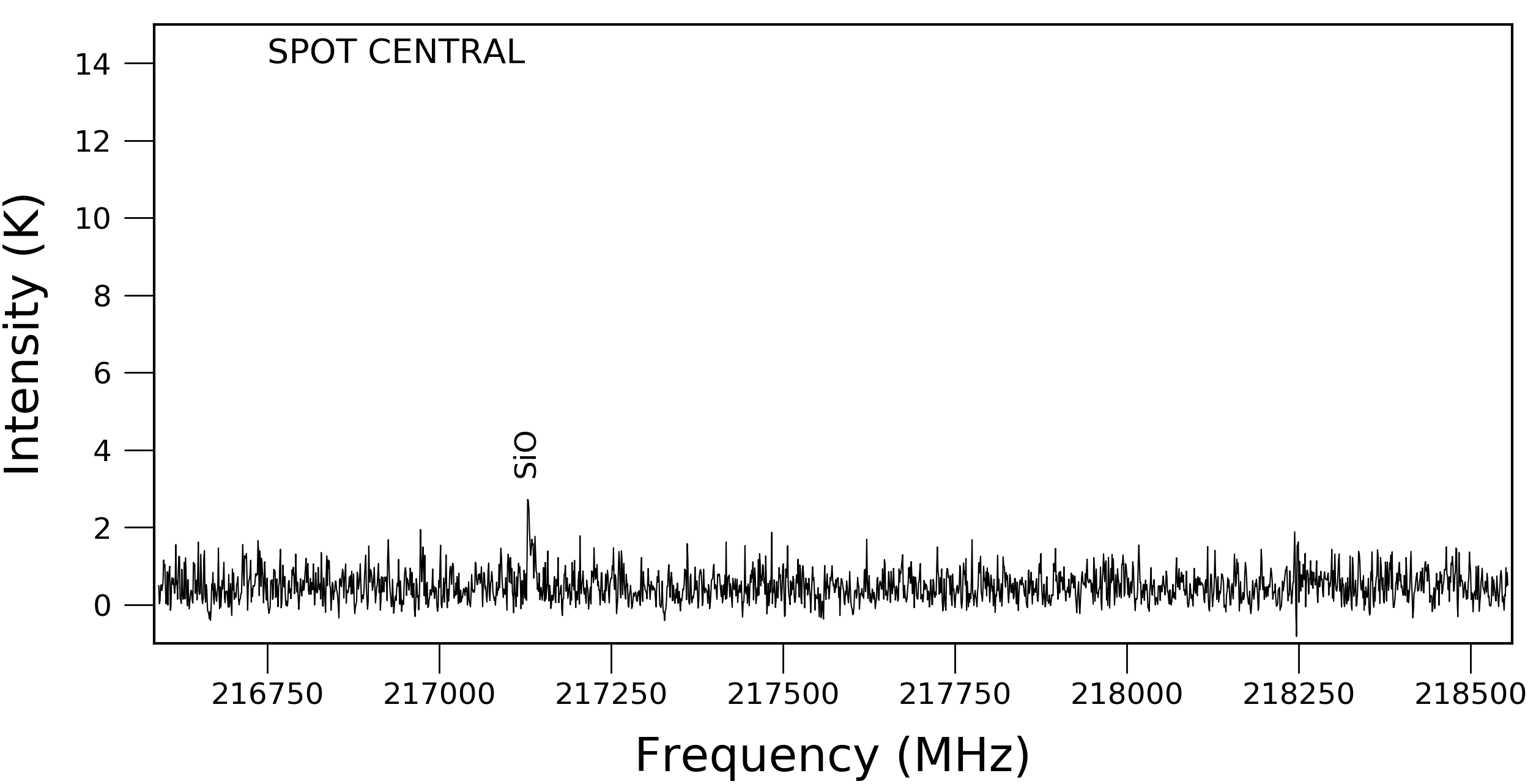}\\
	
	\end{tabular}
  \caption{SMA spectra for all spots in the window of 216-218 GHz.}

  \label{spots217}
\end{figure*}

\begin{figure*}[ht!]
  \centering
	\begin{tabular}{ll}
	\includegraphics[width=0.5\textwidth]{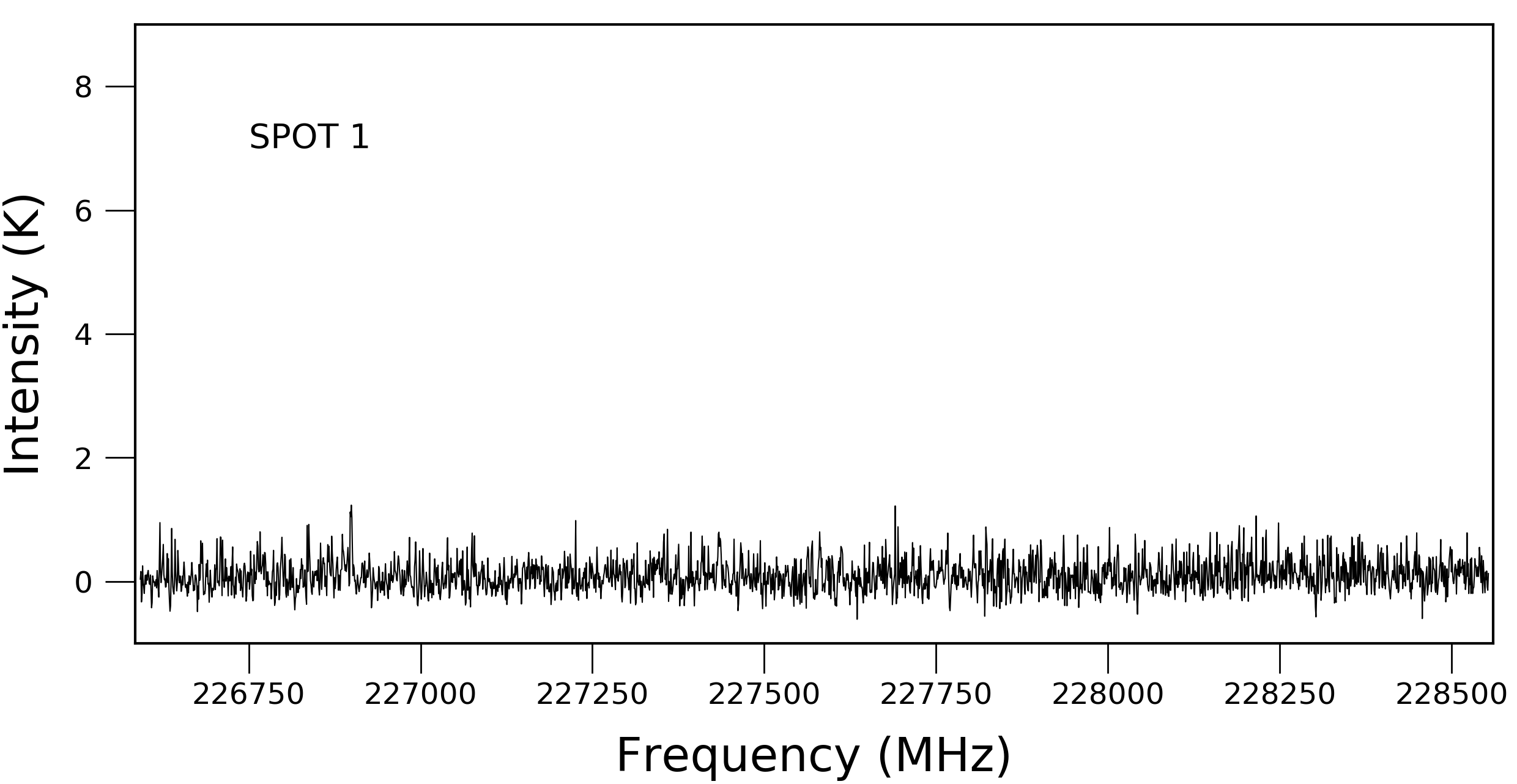}
	&\includegraphics[width=0.5\textwidth]{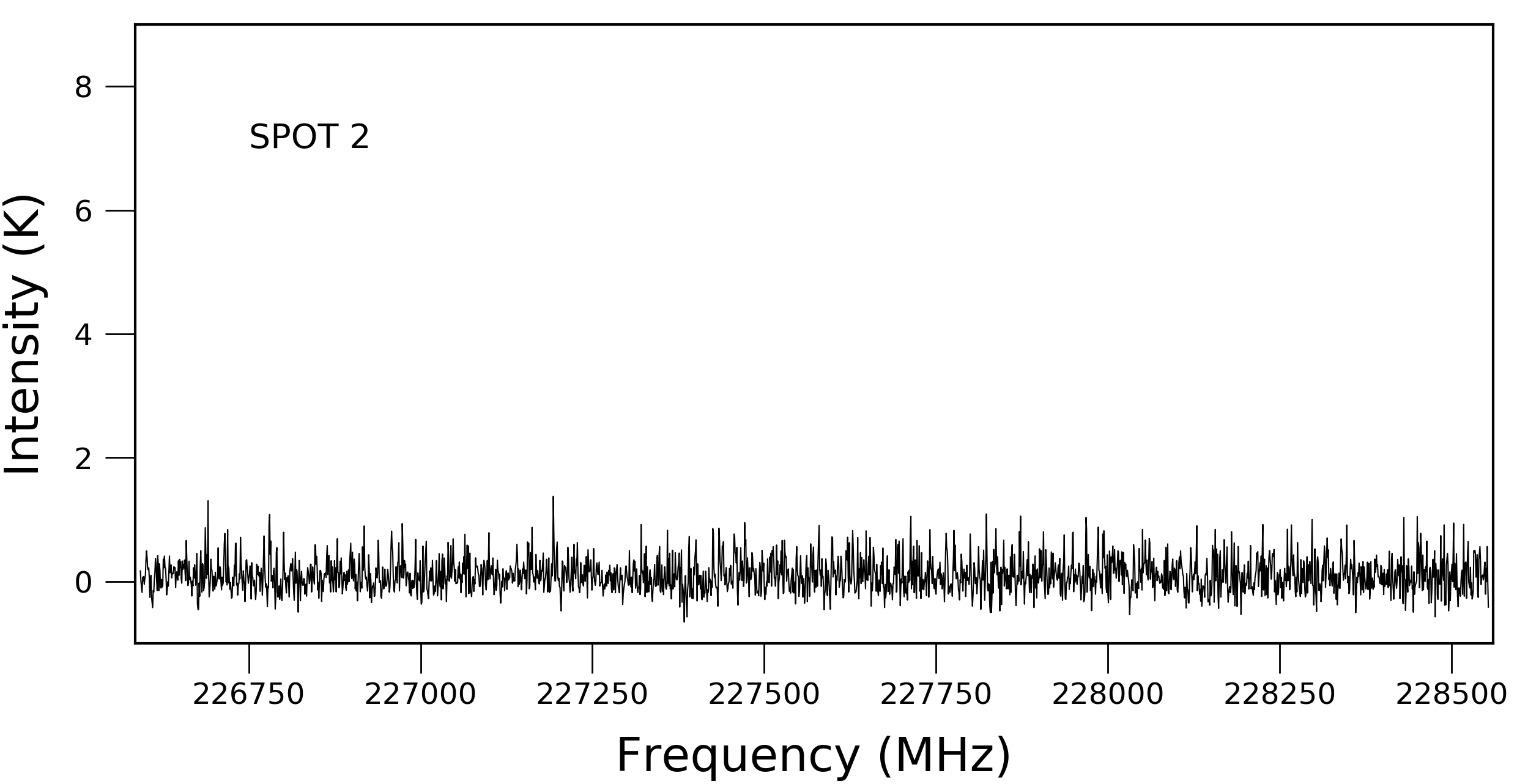}\\
	
	\includegraphics[width=0.5\textwidth]{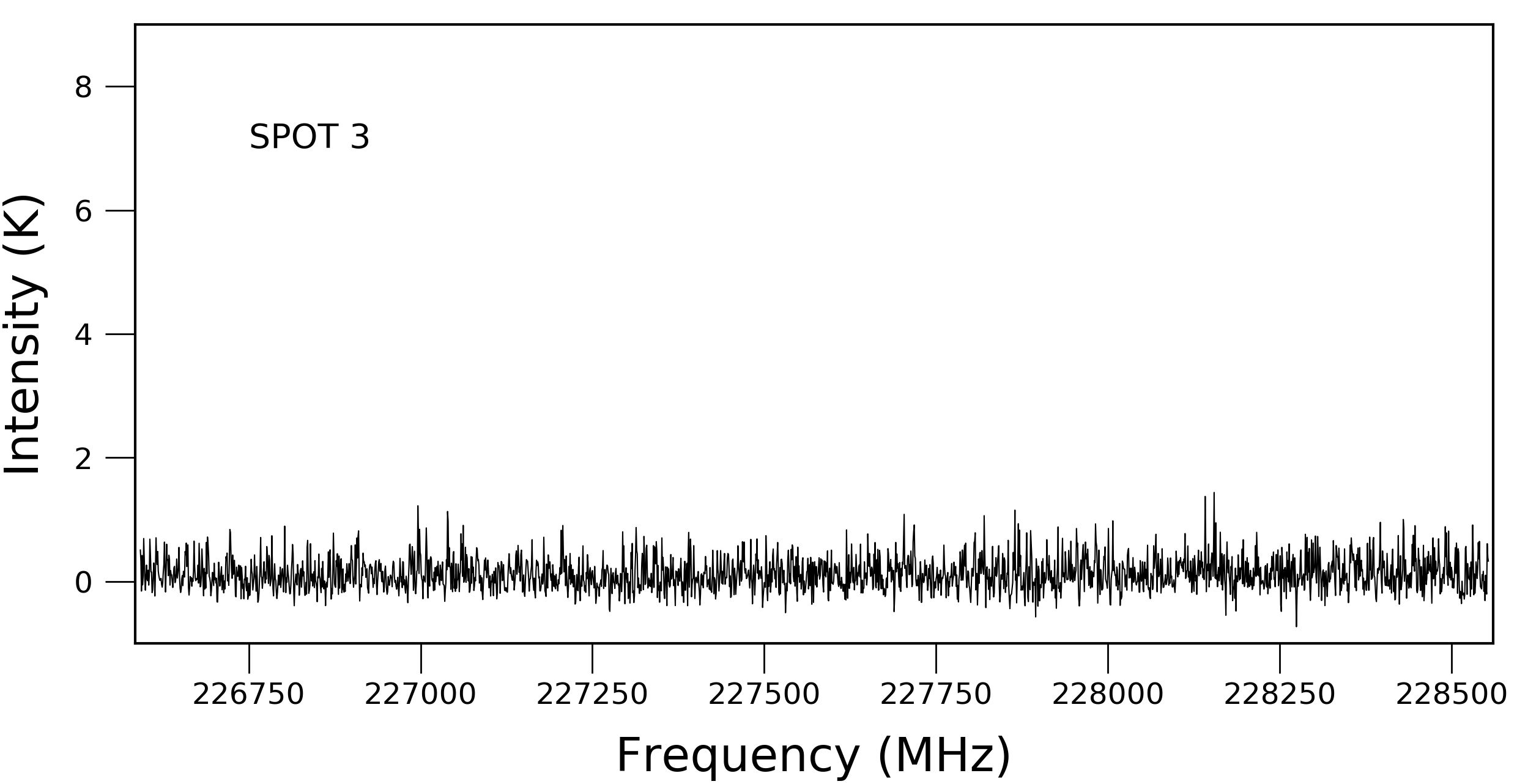}
	&\includegraphics[width=0.5\textwidth]{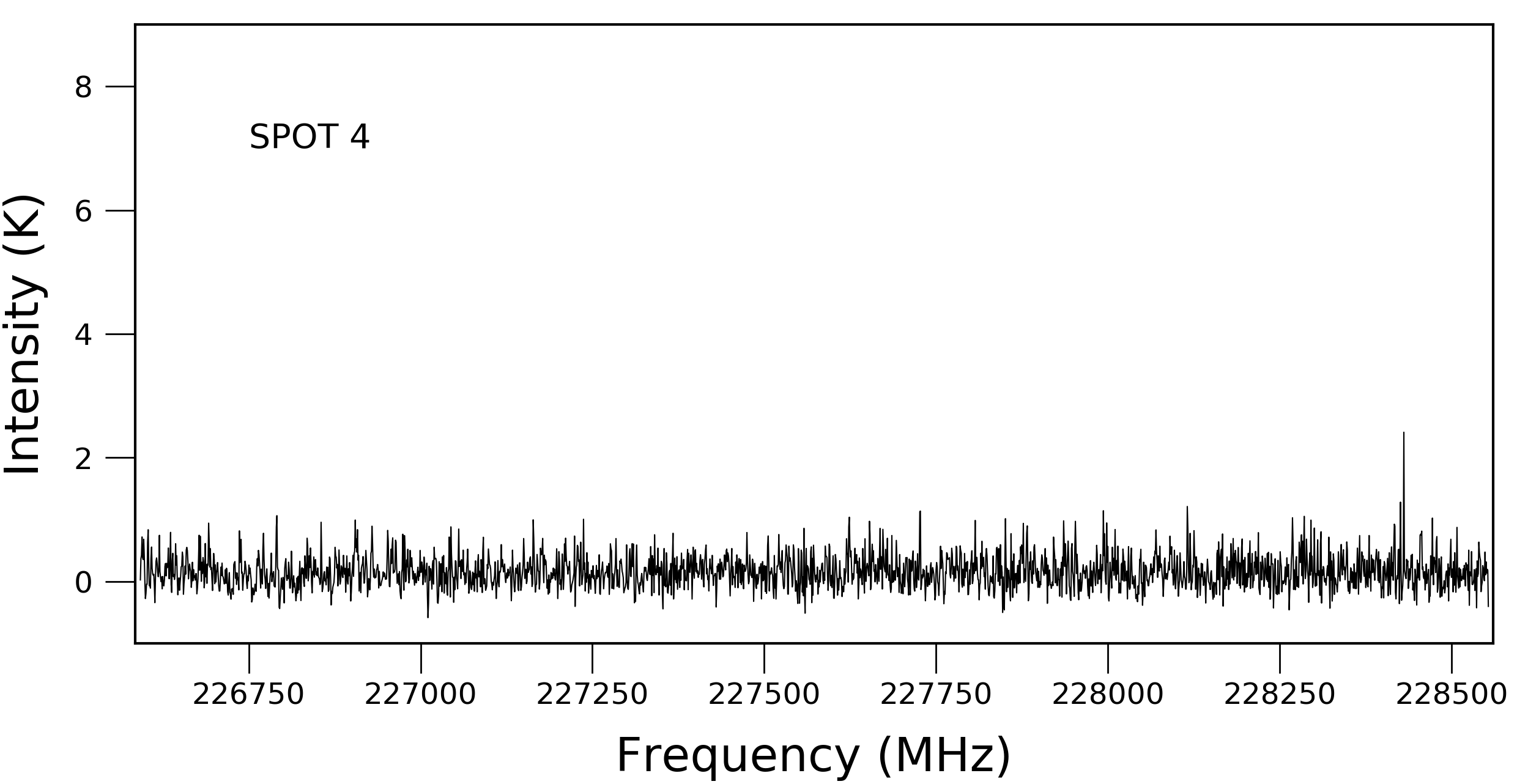}\\
	
	\includegraphics[width=0.5\textwidth]{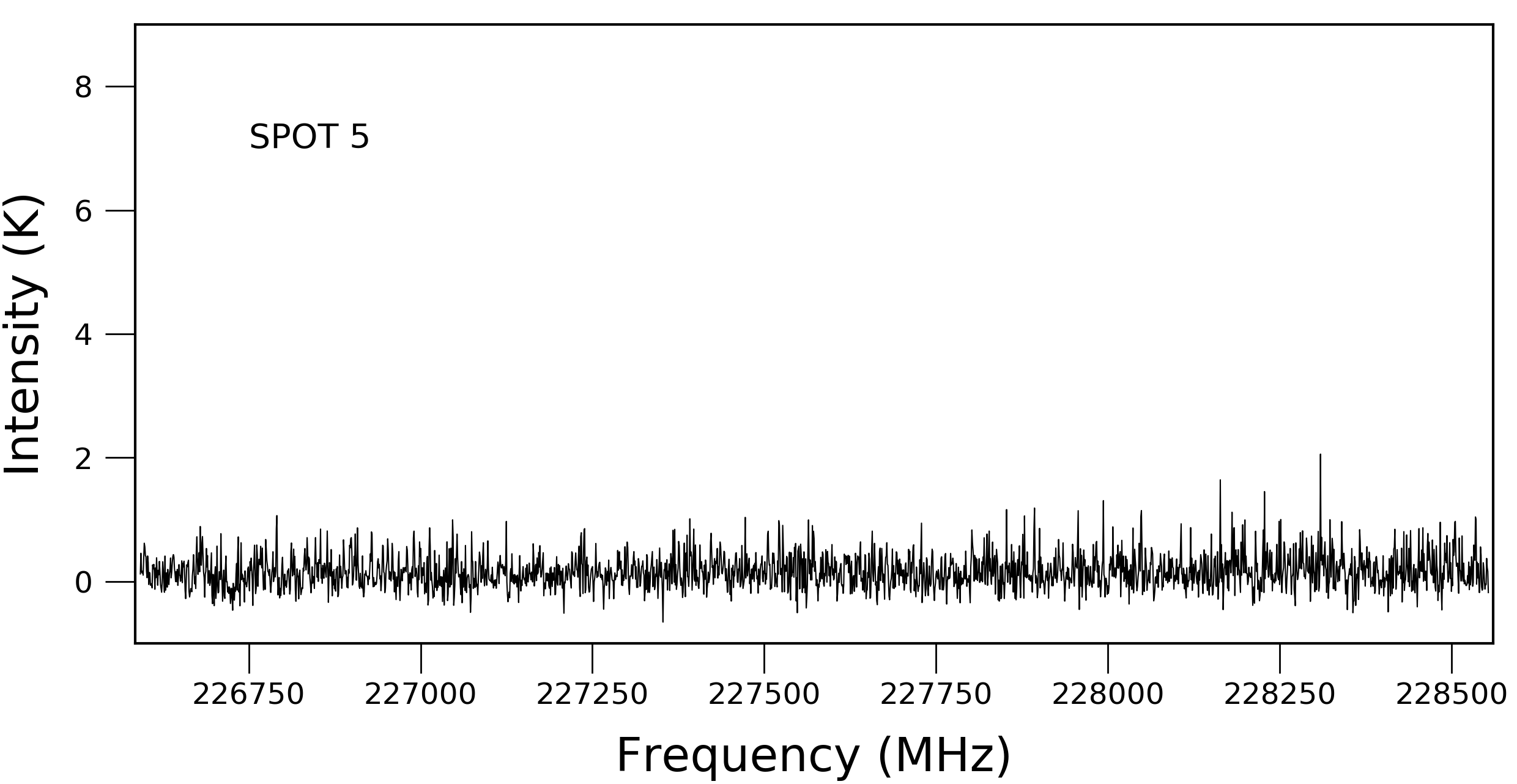}  
		&\includegraphics[width=0.5\textwidth]{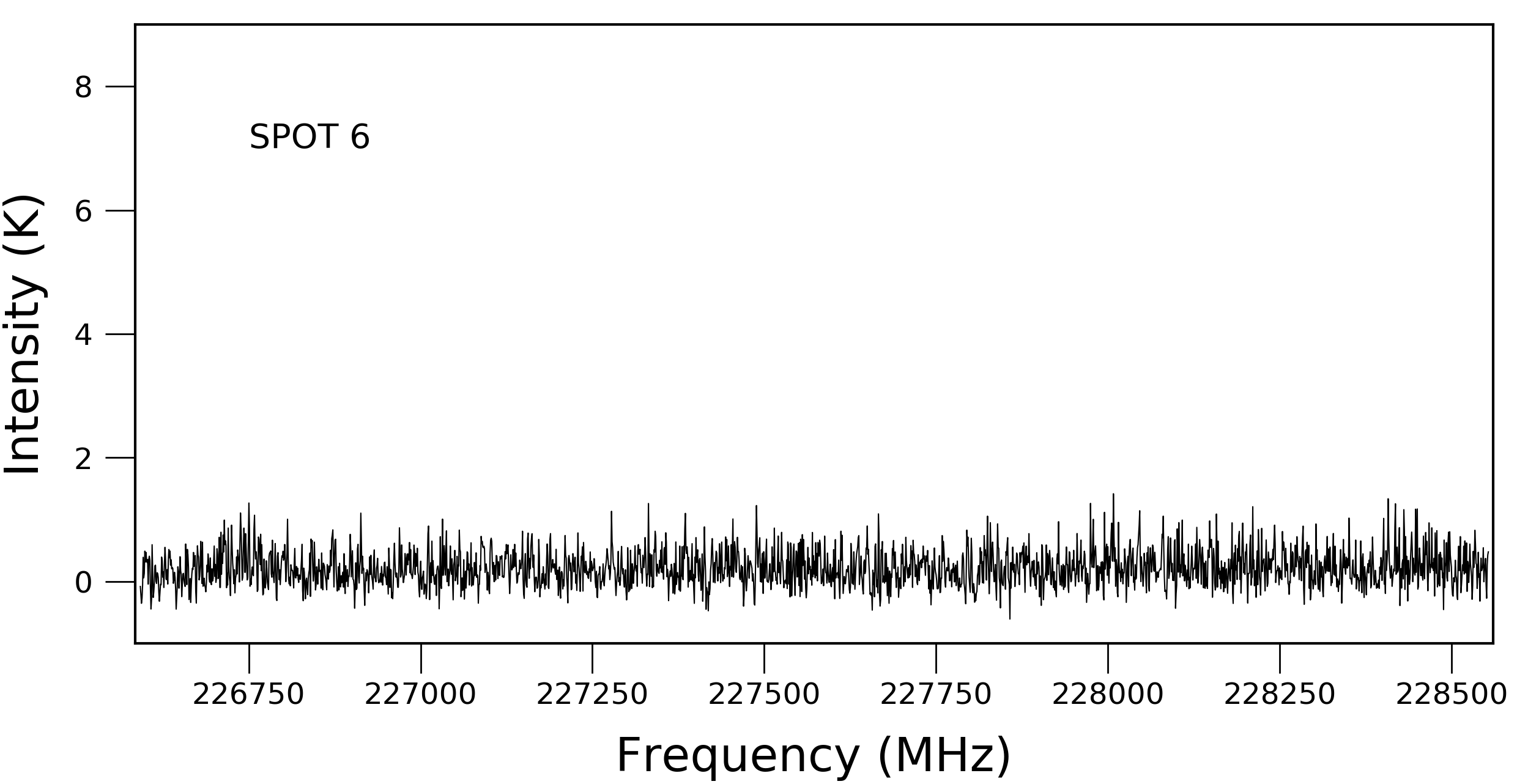}\\
	
	\includegraphics[width=0.5\textwidth]{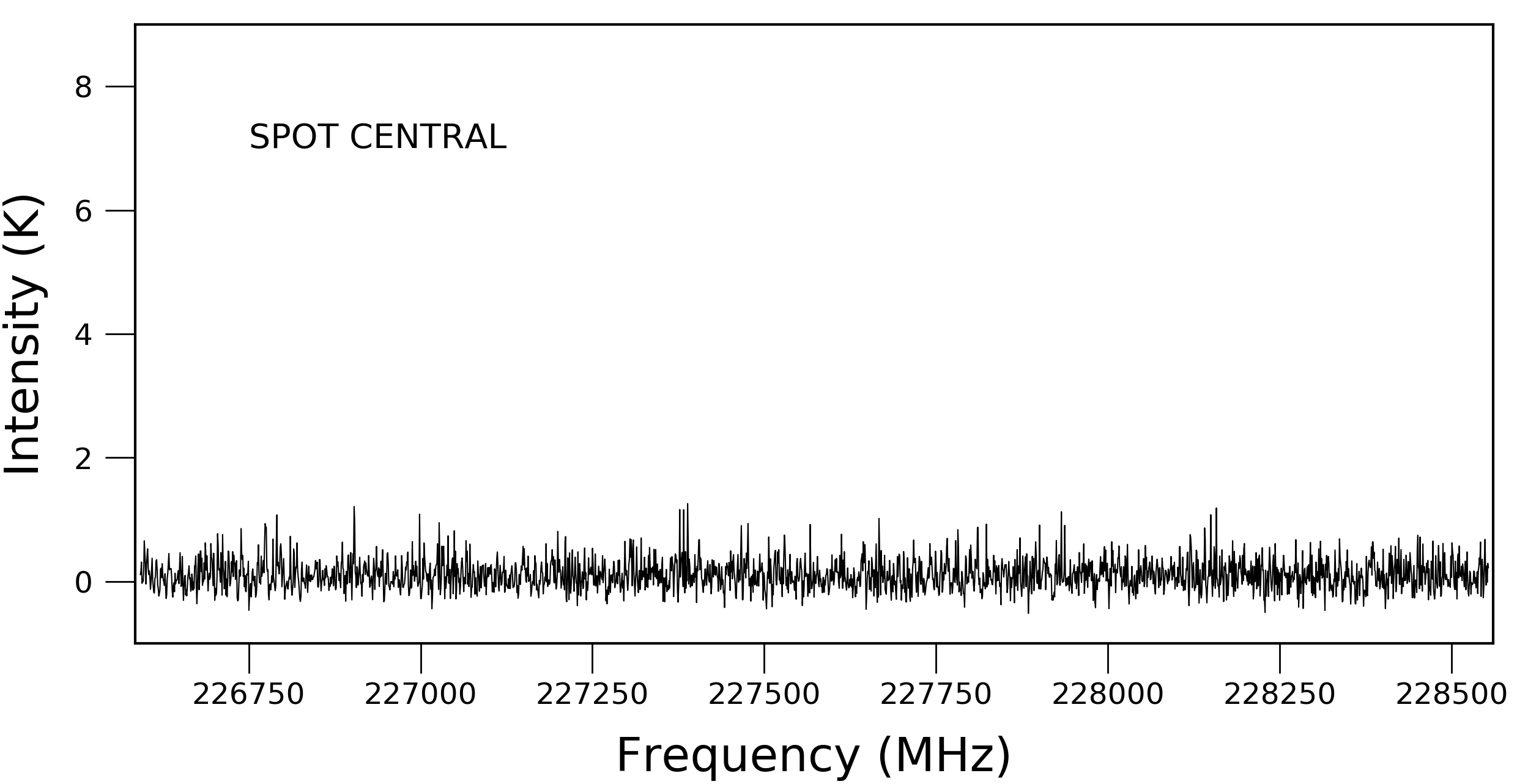}\\
	
	\end{tabular}
  \caption{SMA spectra for all spots in the window of 226-228 GHz.}

  \label{spots227}
\end{figure*}

\begin{figure*}[ht!]
  \centering
	\begin{tabular}{ll}
	\includegraphics[width=0.5\textwidth]{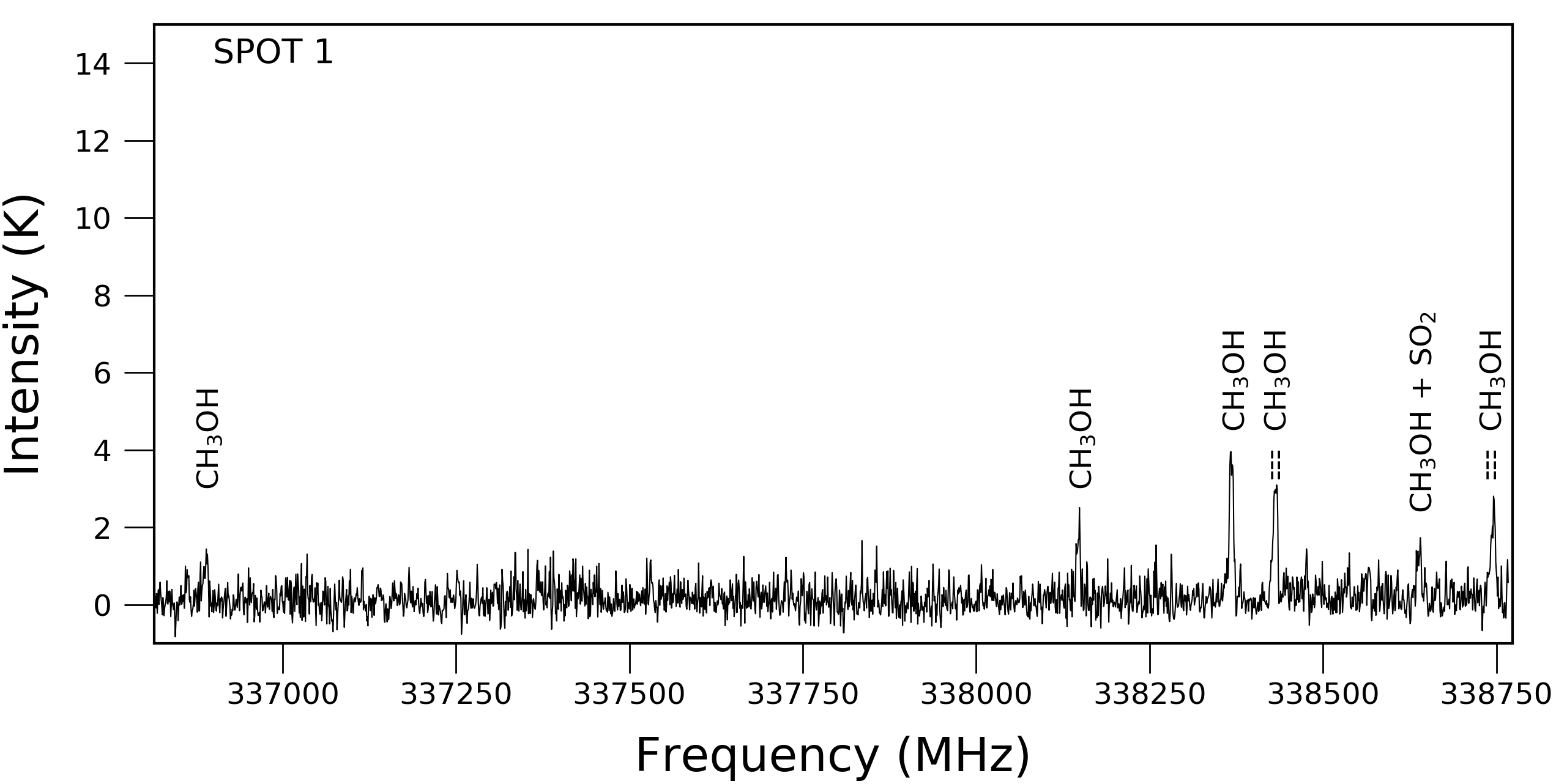}
	&\includegraphics[width=0.5\textwidth]{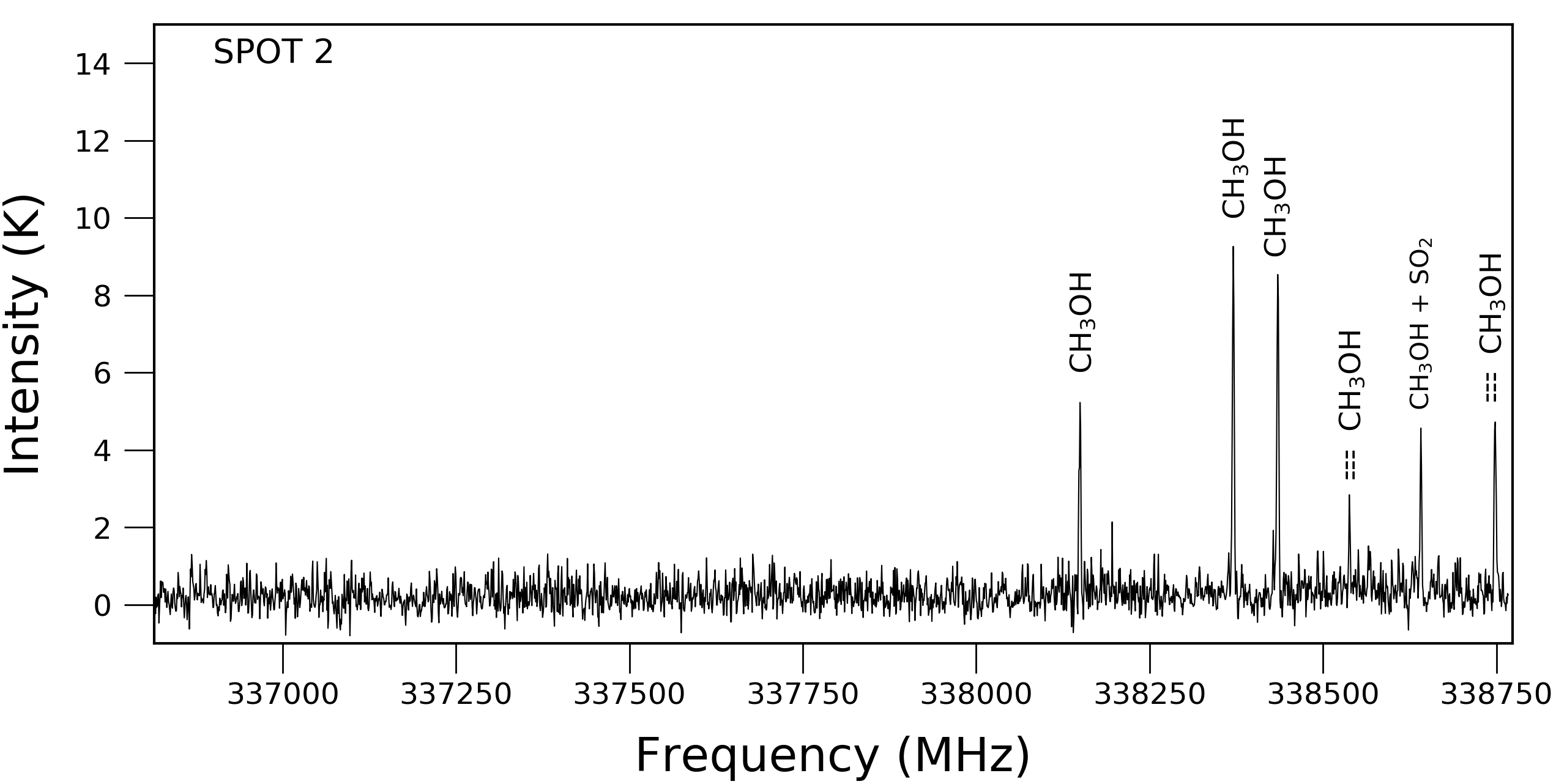}\\
	
	\includegraphics[width=0.5\textwidth]{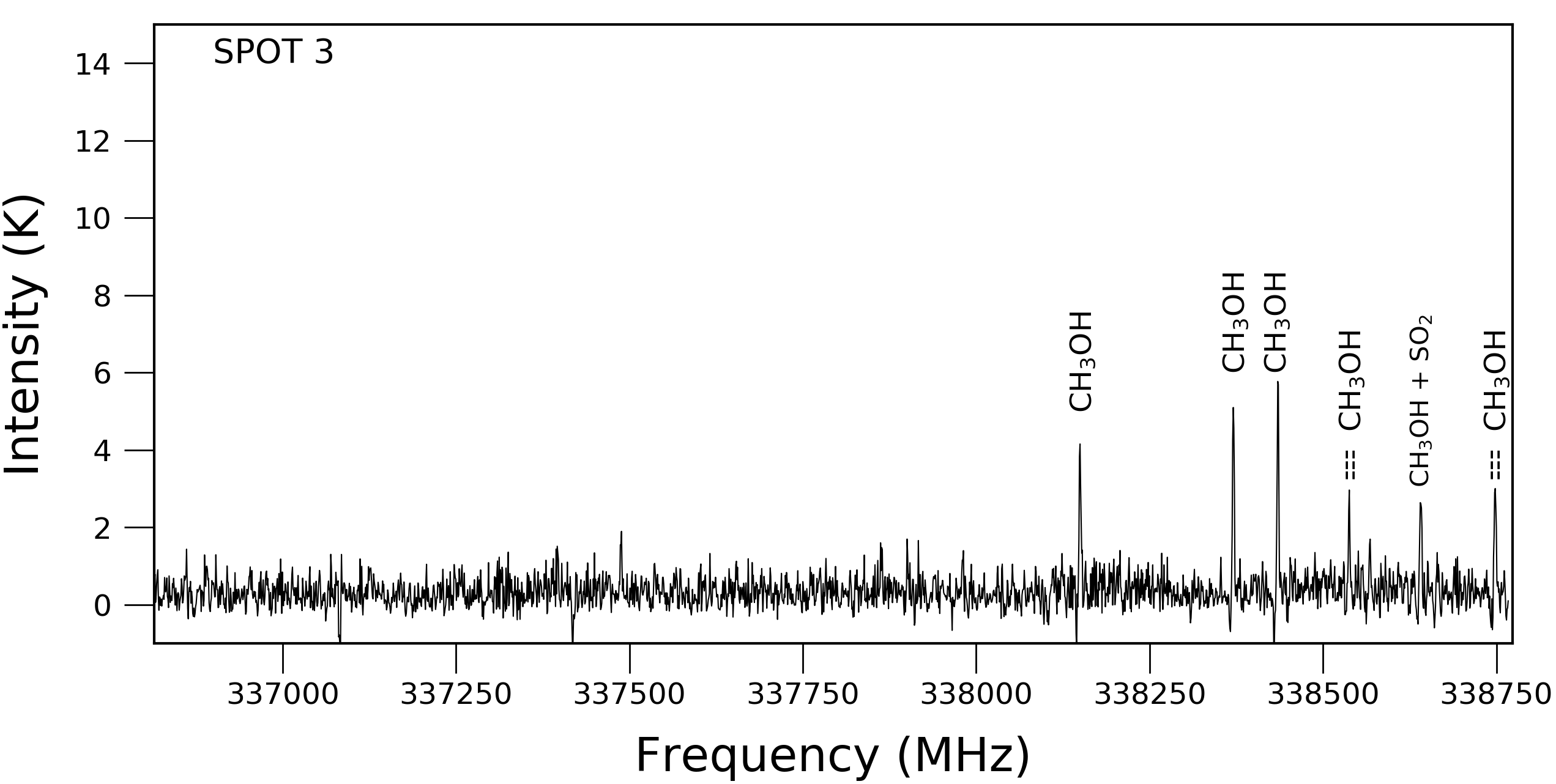}
	&\includegraphics[width=0.5\textwidth]{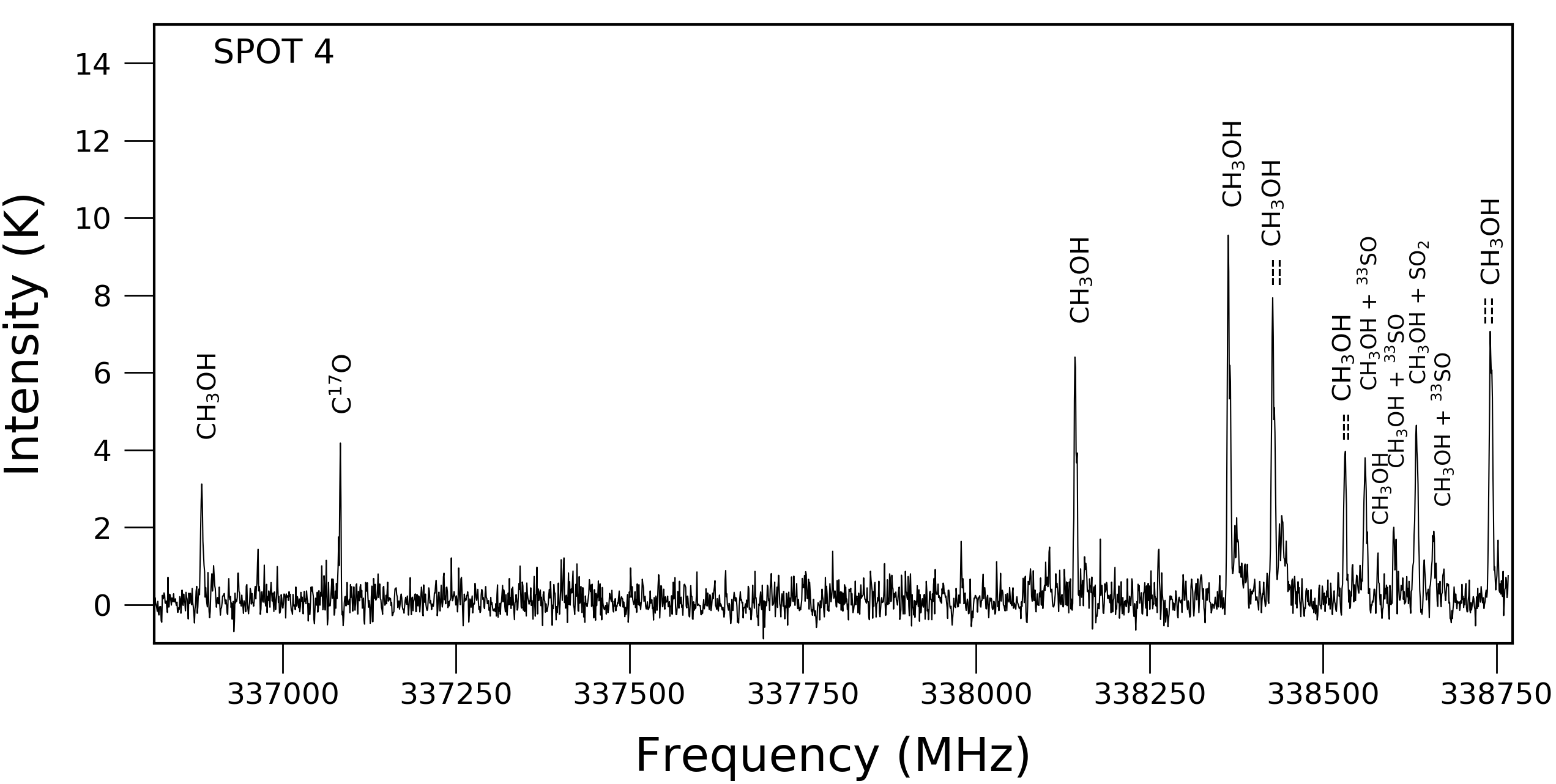}\\
	
	\includegraphics[width=0.5\textwidth]{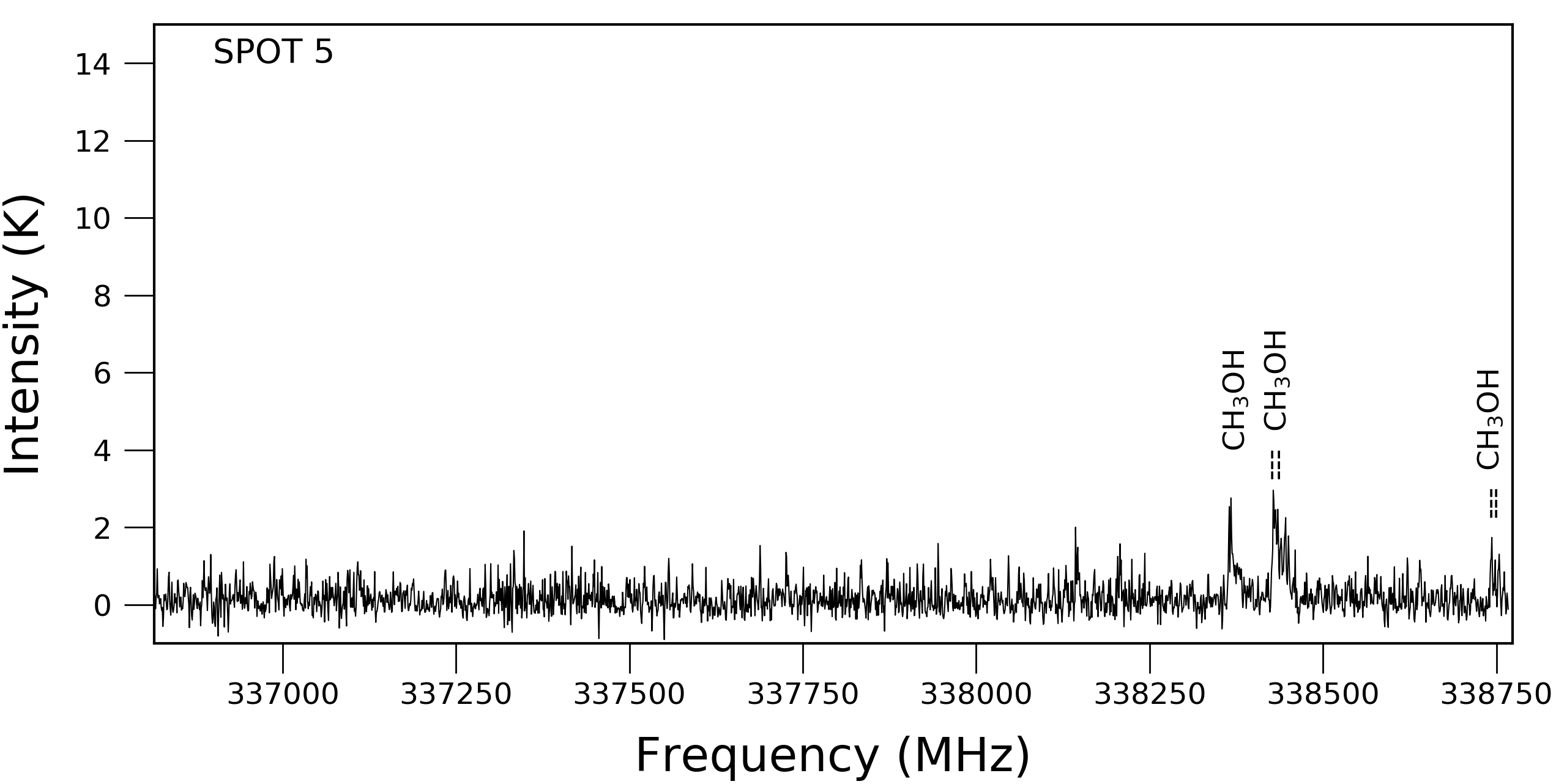}  
		&\includegraphics[width=0.5\textwidth]{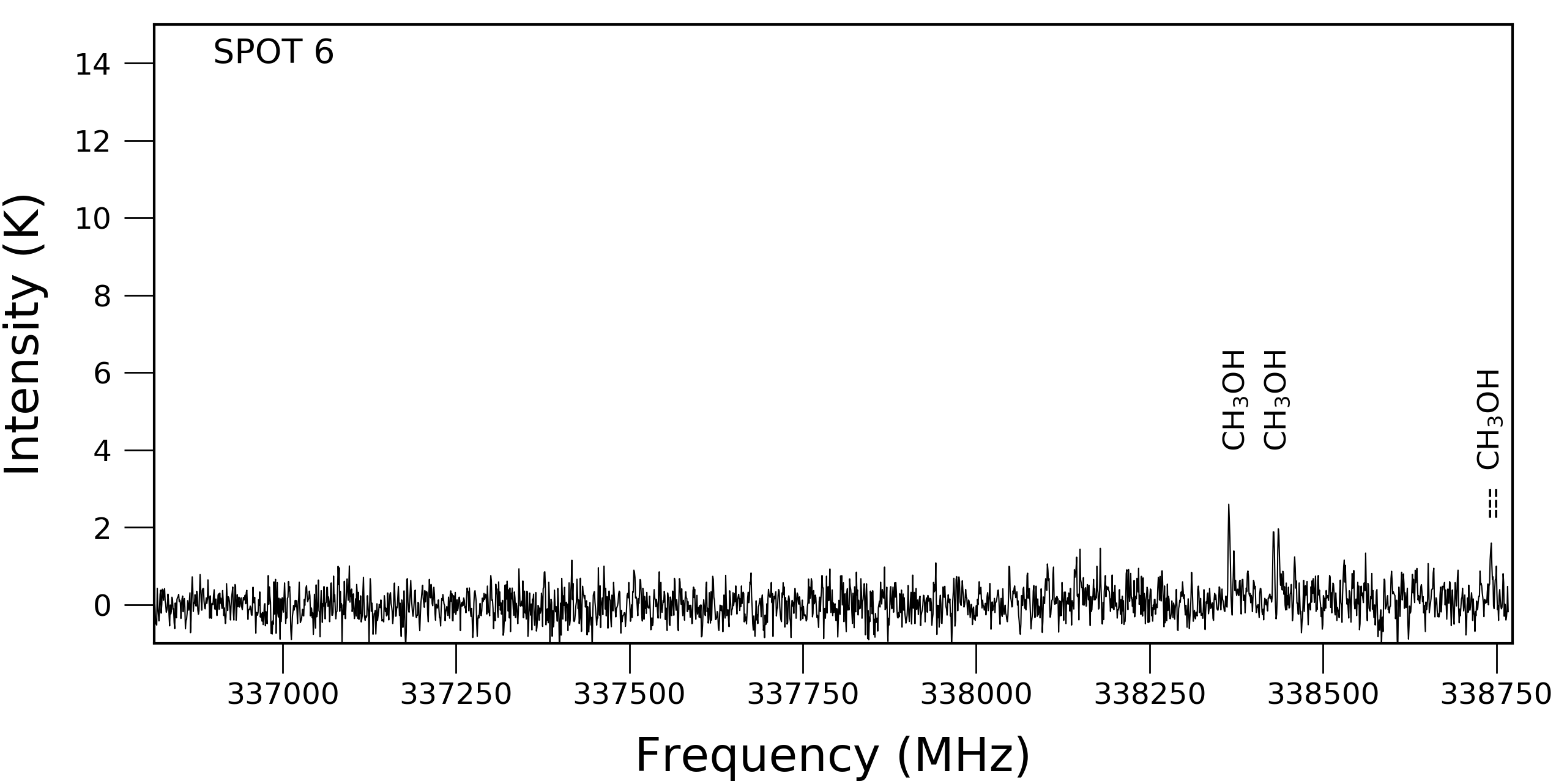}\\
	
	\includegraphics[width=0.5\textwidth]{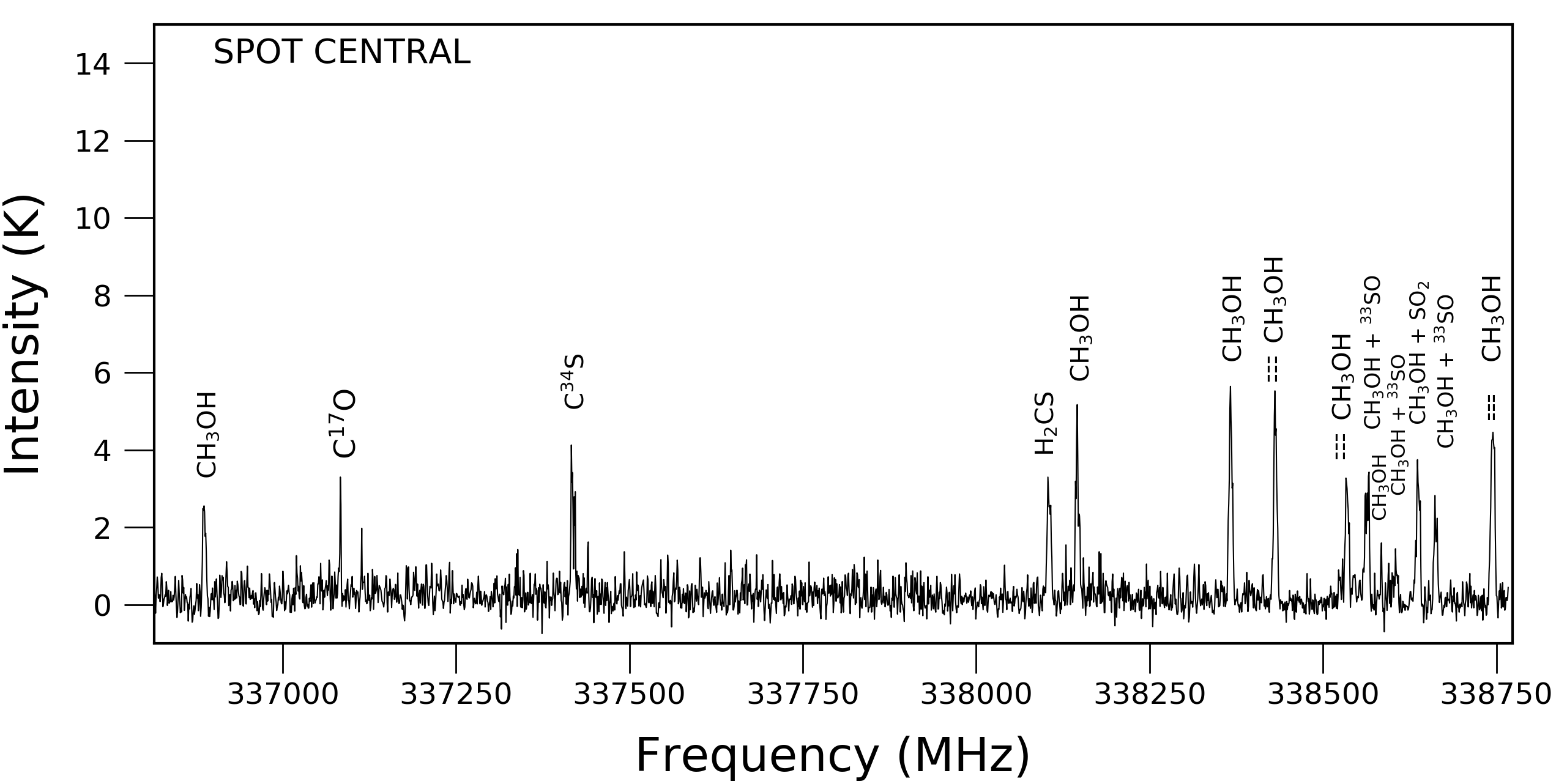}\\
	
	\end{tabular}
  \caption{SMA spectra for all spots in the window of 337-338 GHz.}

  \label{spots337}
\end{figure*}

\begin{figure*}[ht!]
  \centering
	\begin{tabular}{ll}
	\includegraphics[width=0.5\textwidth]{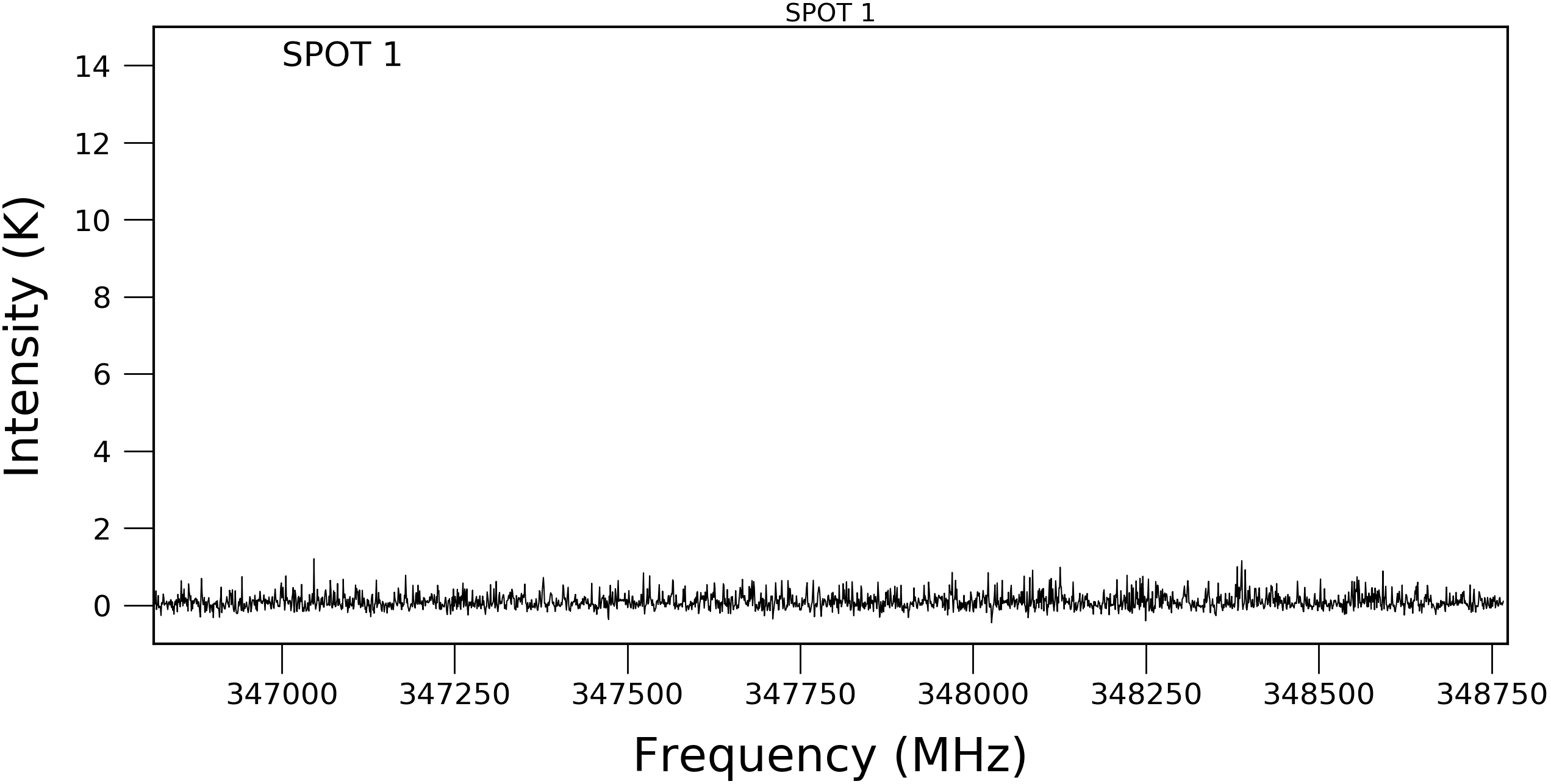}
	&\includegraphics[width=0.5\textwidth]{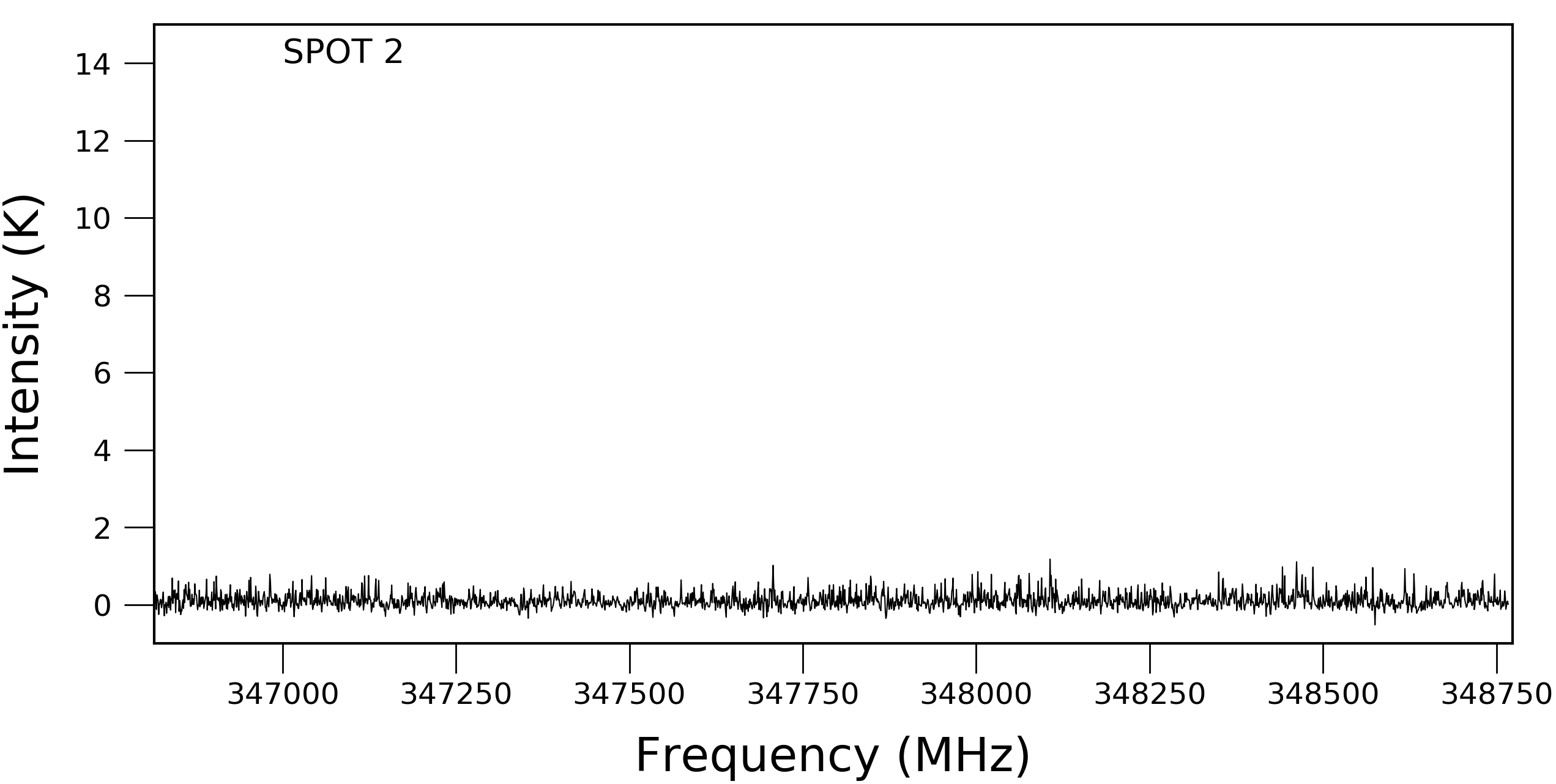}\\
	
	\includegraphics[width=0.5\textwidth]{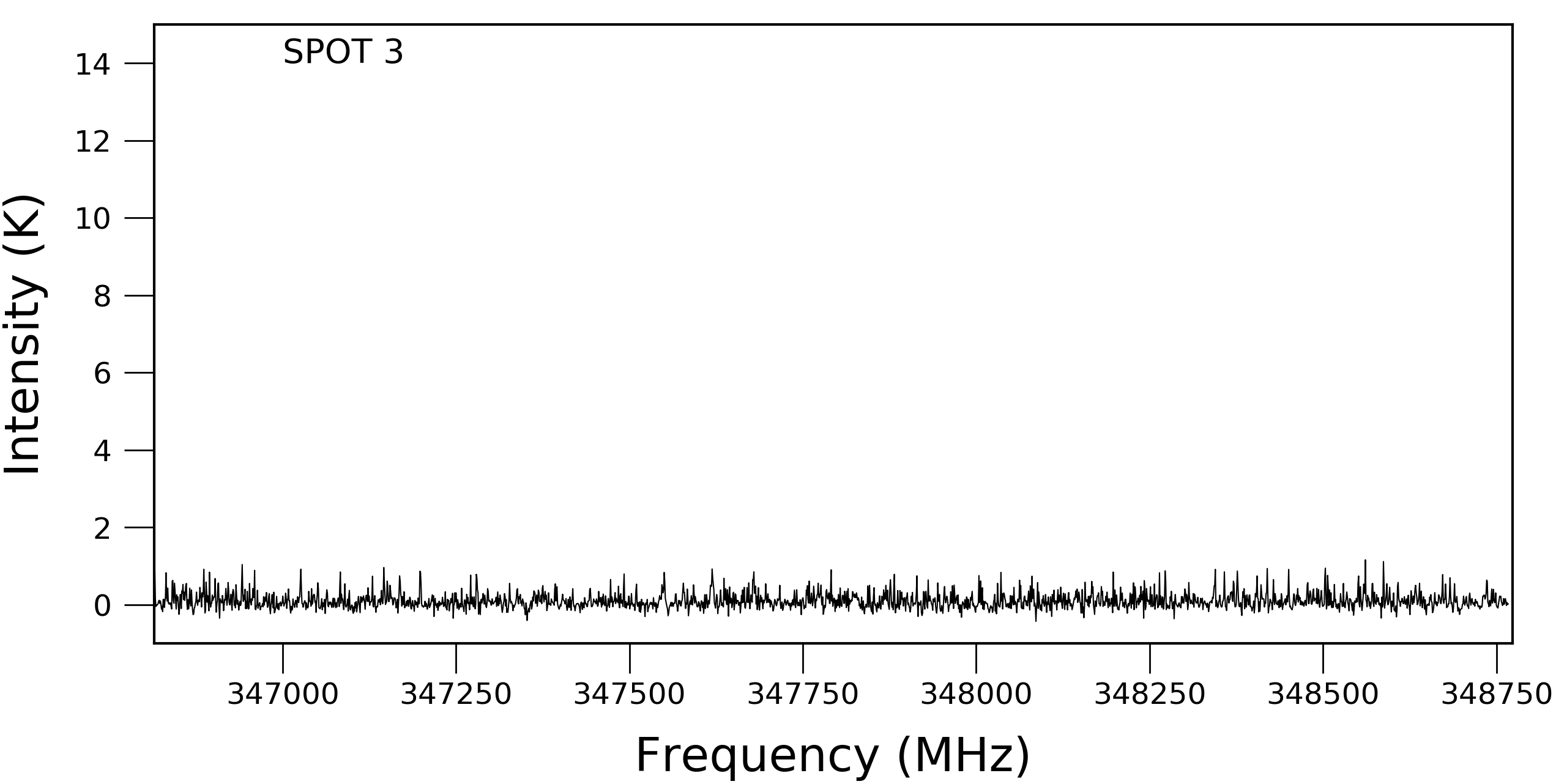}
	&\includegraphics[width=0.5\textwidth]{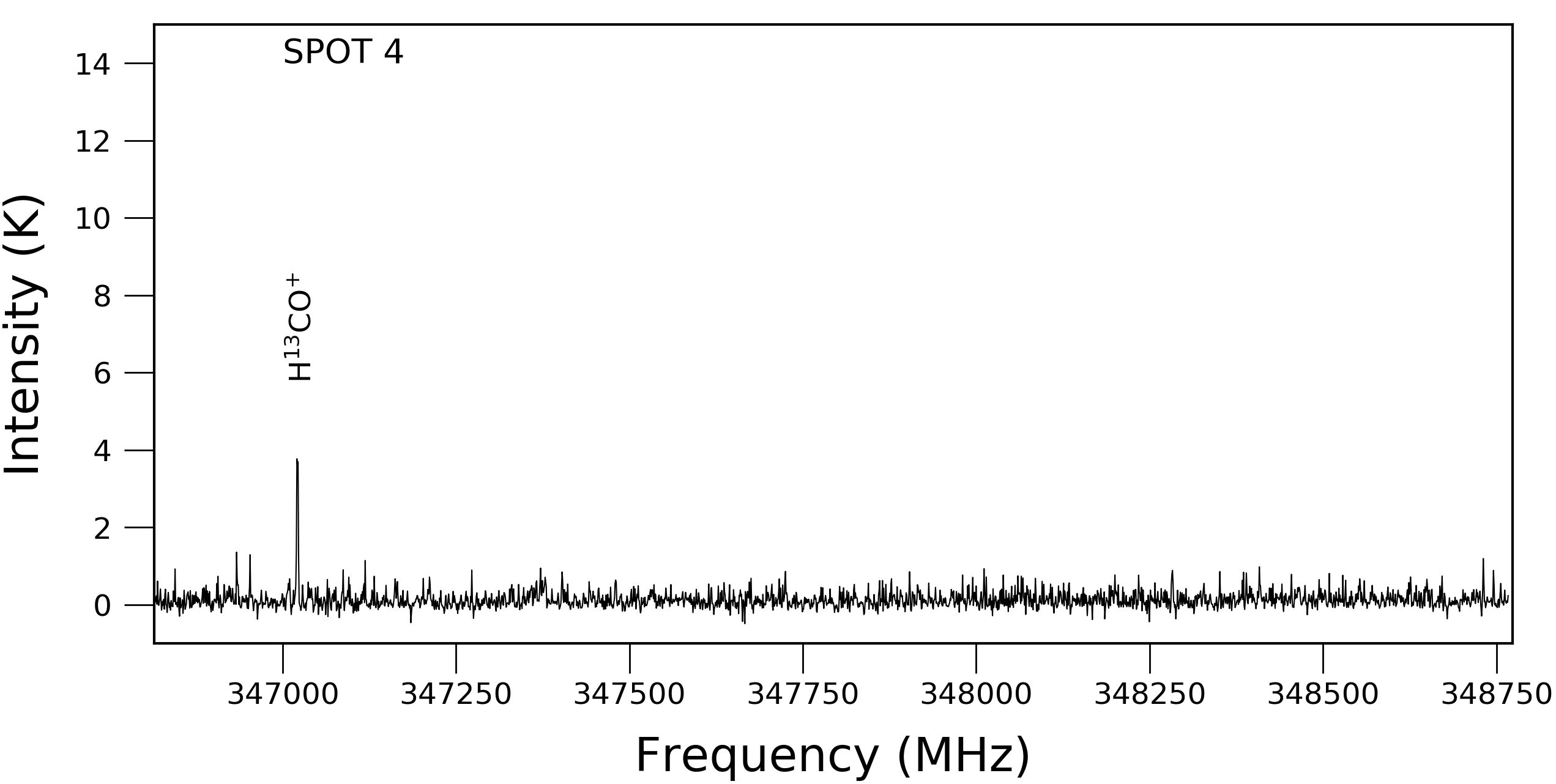}\\
	
	\includegraphics[width=0.5\textwidth]{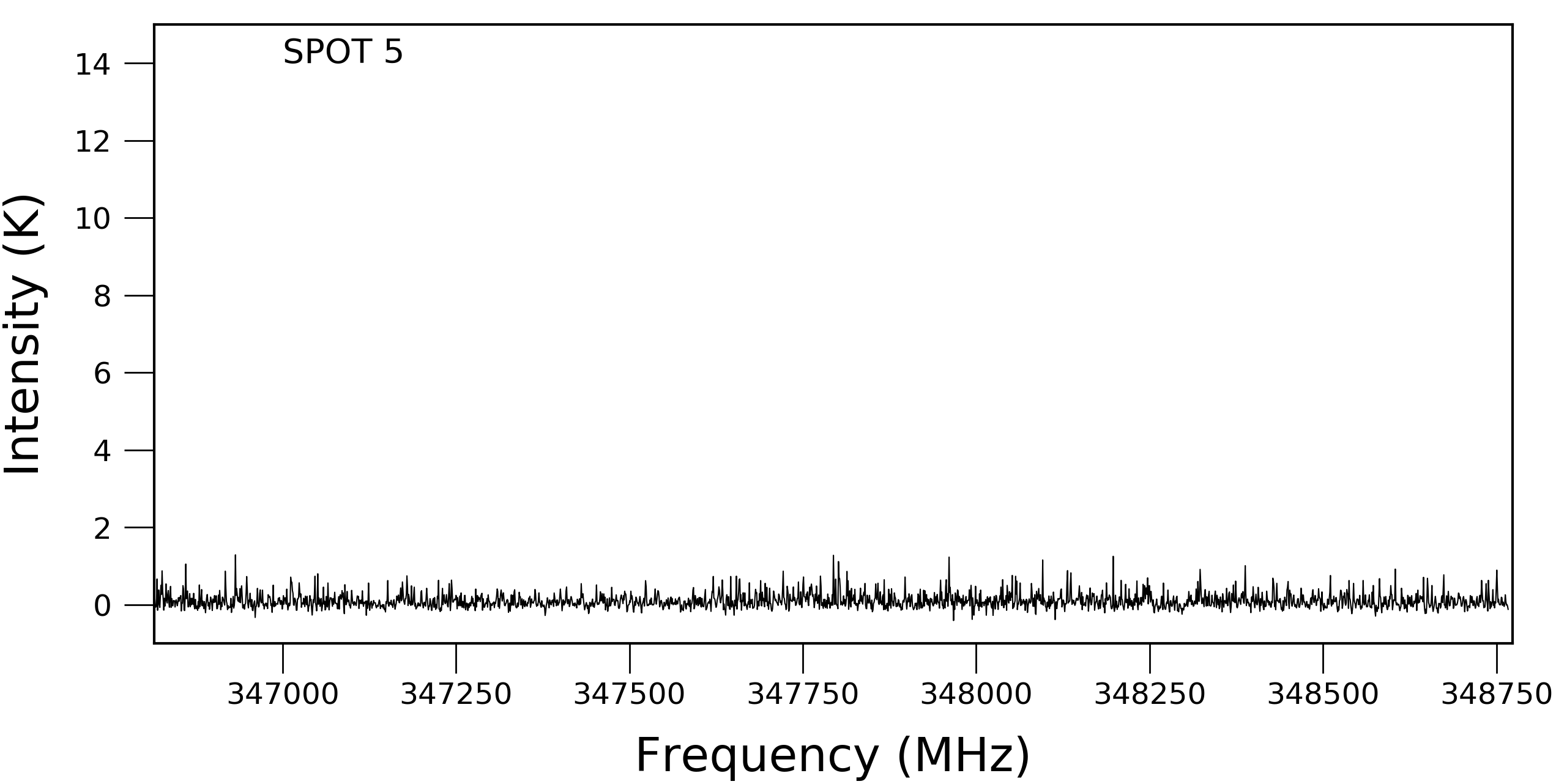}  
		&\includegraphics[width=0.5\textwidth]{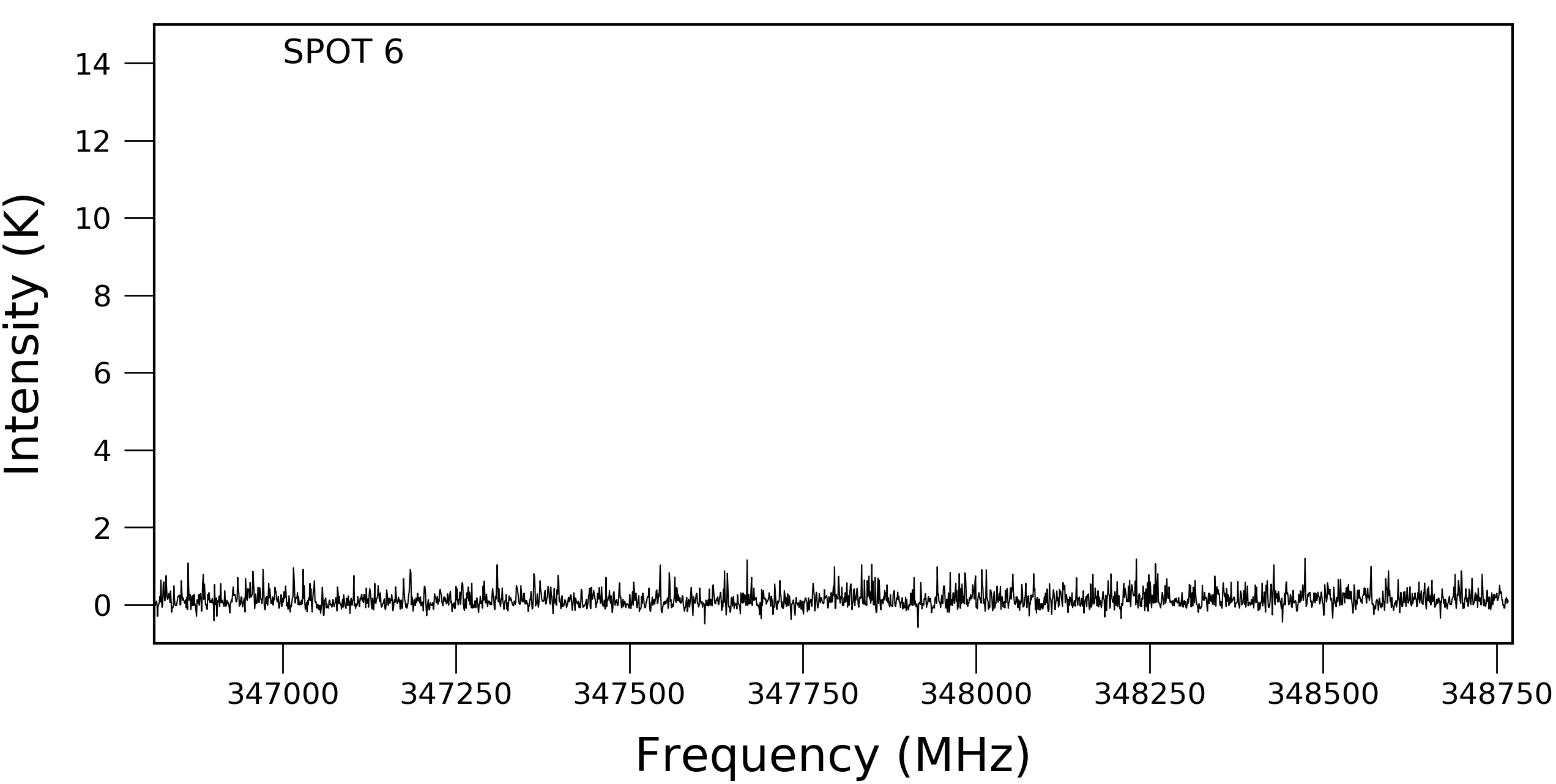}\\
	
	\includegraphics[width=0.5\textwidth]{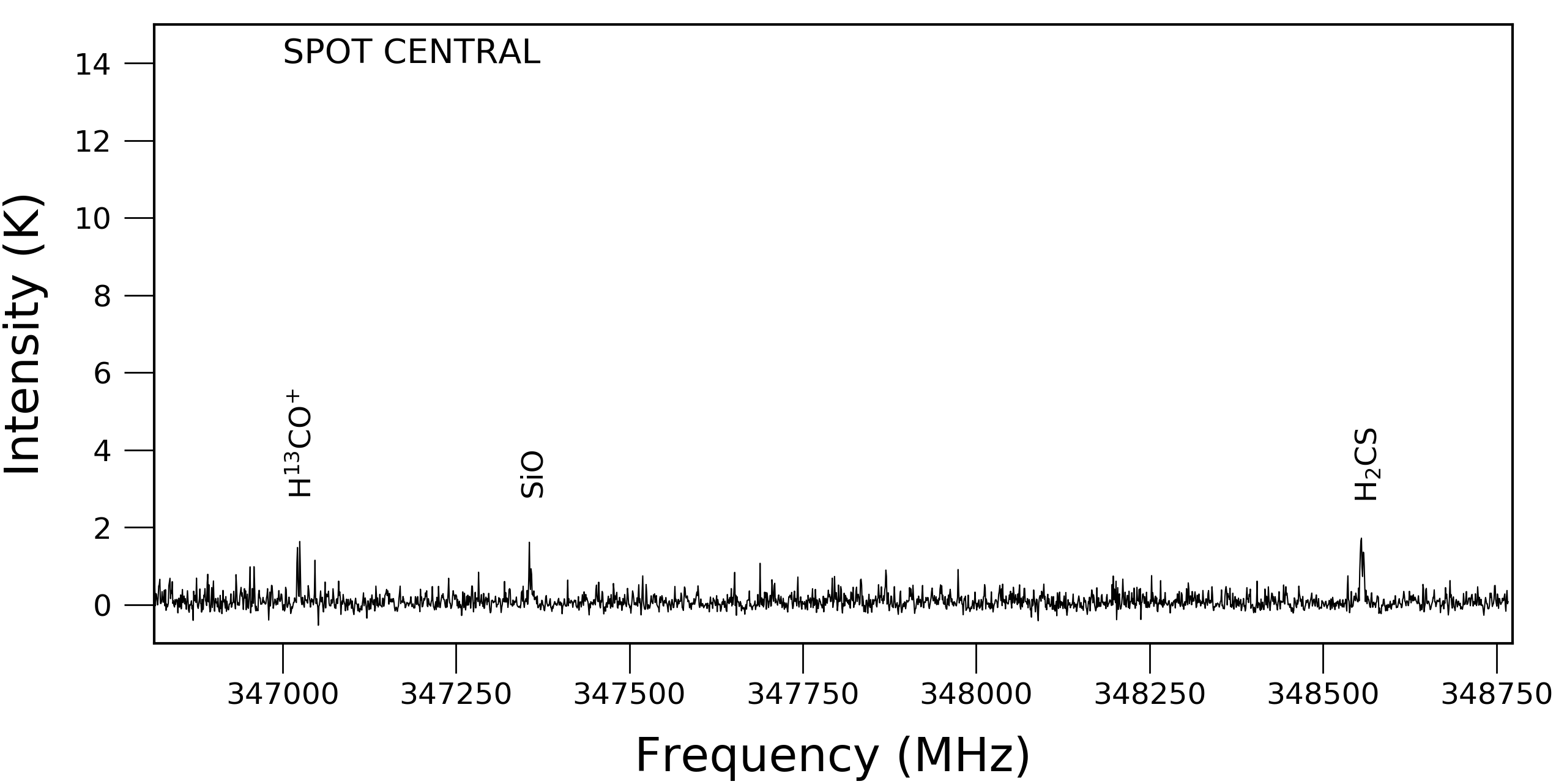}\\
	
	\end{tabular}
  \caption{SMA spectra for all spots in the window of 347-348 GHz.}

  \label{spots347}
\end{figure*}

\clearpage

\subsection{Integrated intensity maps}

In this section we include the moment 0 for all molecules reported in Table \ref{tablaMol}. 
We have used an exponential or logarithmic brightness scale depending on the maximum 
for each map to show the emission distribution.

\subsection{Figures of non-methanol lines.}

\begin{figure*}[ht!]
\includegraphics[width=0.83\textwidth, angle=0, clip=True, trim= 0 0 0 0]{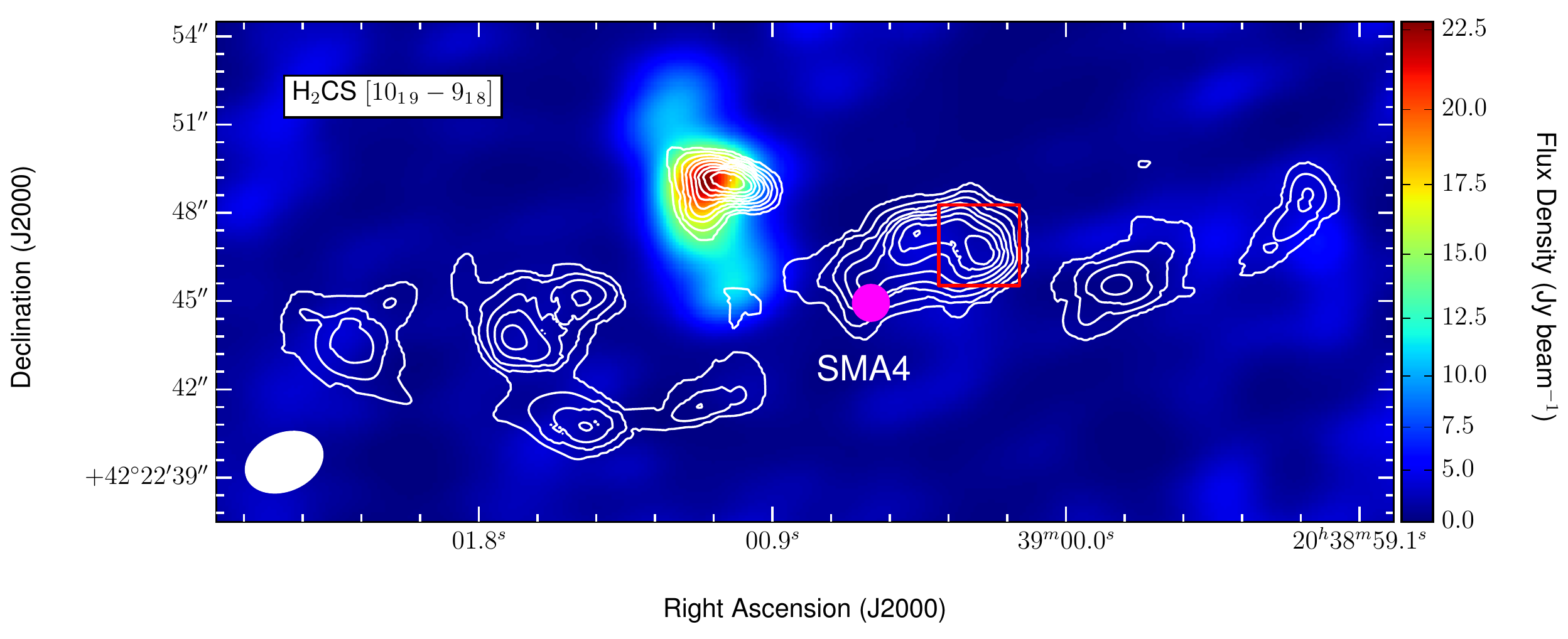}\\
\includegraphics[width=0.83\textwidth, angle=0, clip=True, trim= 0 0 0 0]{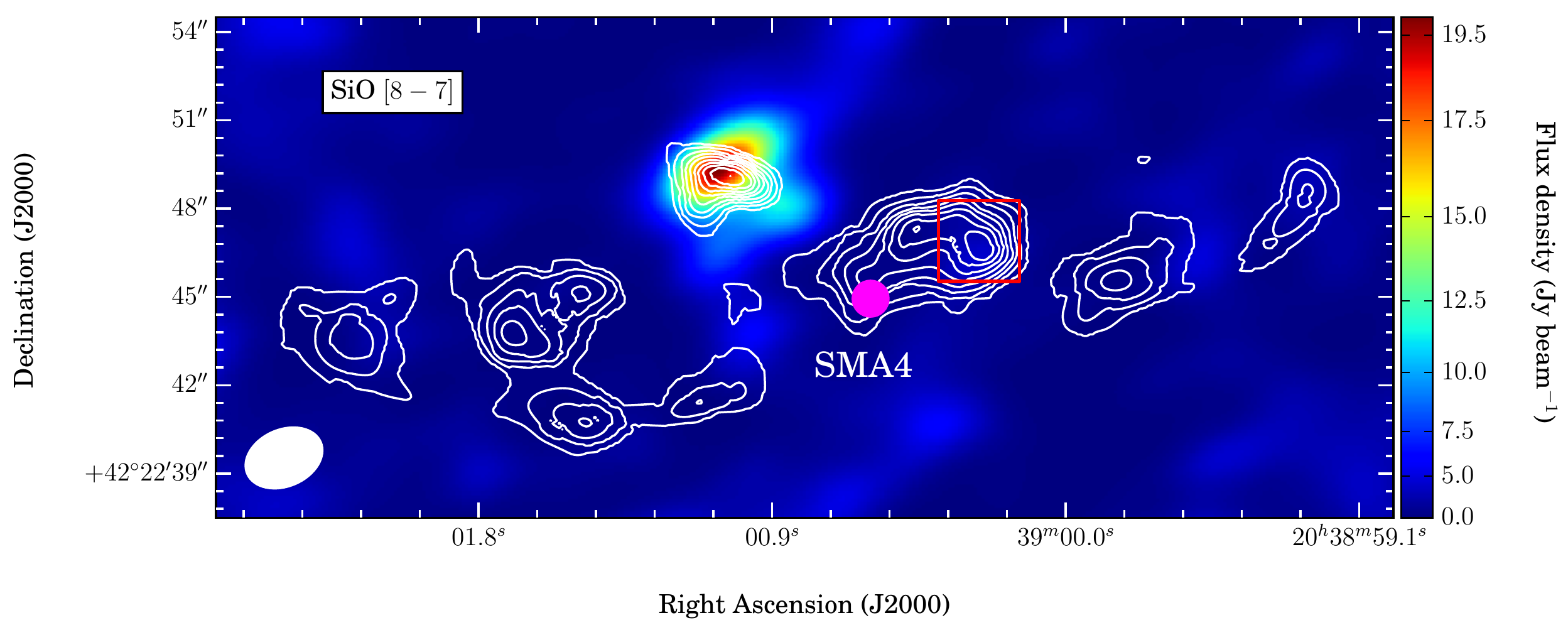}\\
\includegraphics[width=0.83\textwidth, angle=0, clip=True, trim= 0 0 0 0]{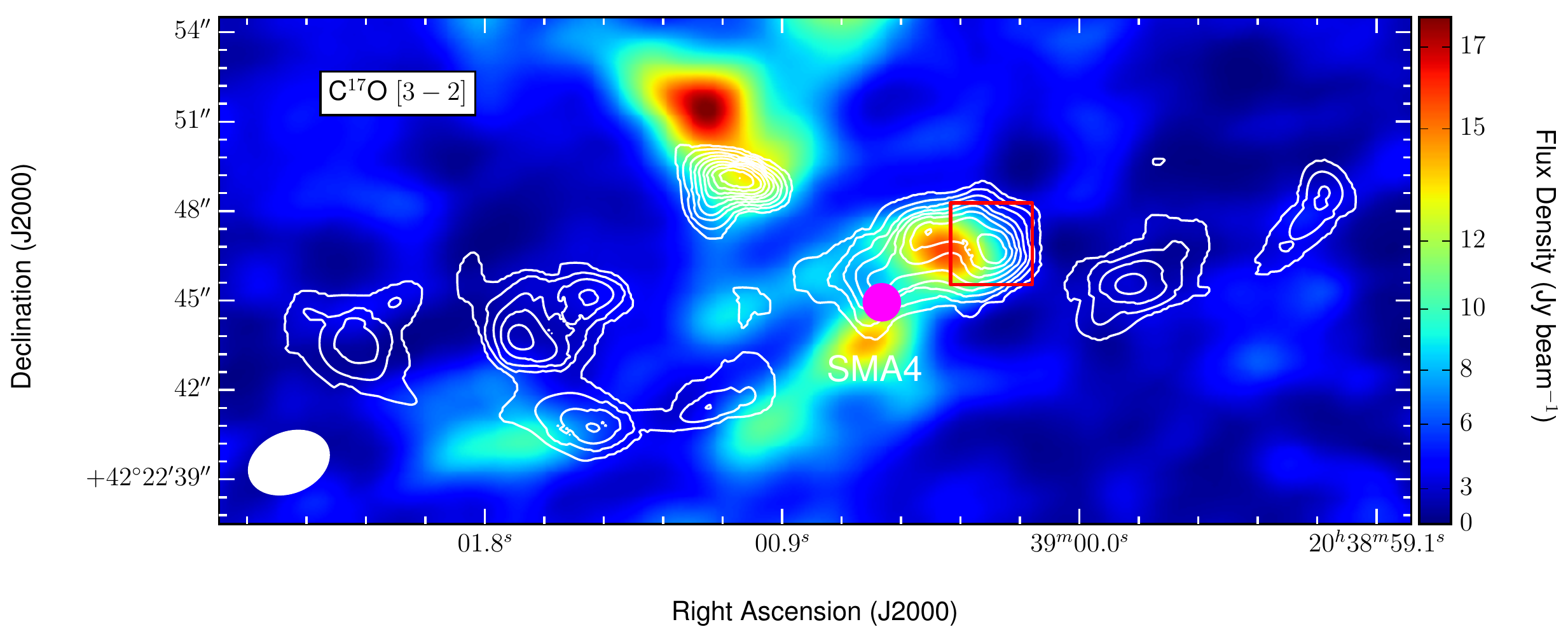}
\caption{Same as Figure 2, but with the SiO, C$^{17}$O and H$_2$CO molecules.}
\label{figMom0}
\end{figure*}

\begin{figure*}[ht!]
\includegraphics[width=0.83\textwidth, angle=0, clip=True, trim= 0 0 0 0]{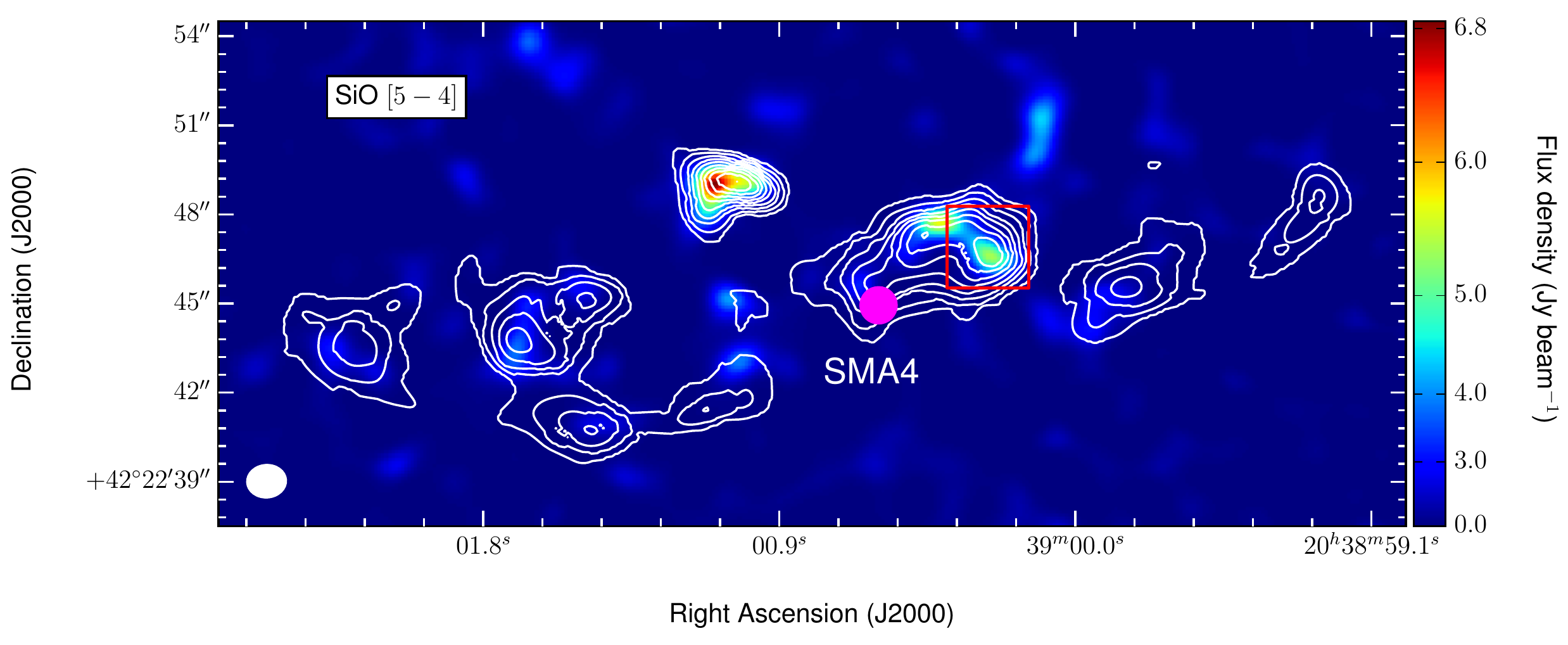}\\
\includegraphics[width=0.83\textwidth, angle=0, clip=True, trim= 0 0 0 0]{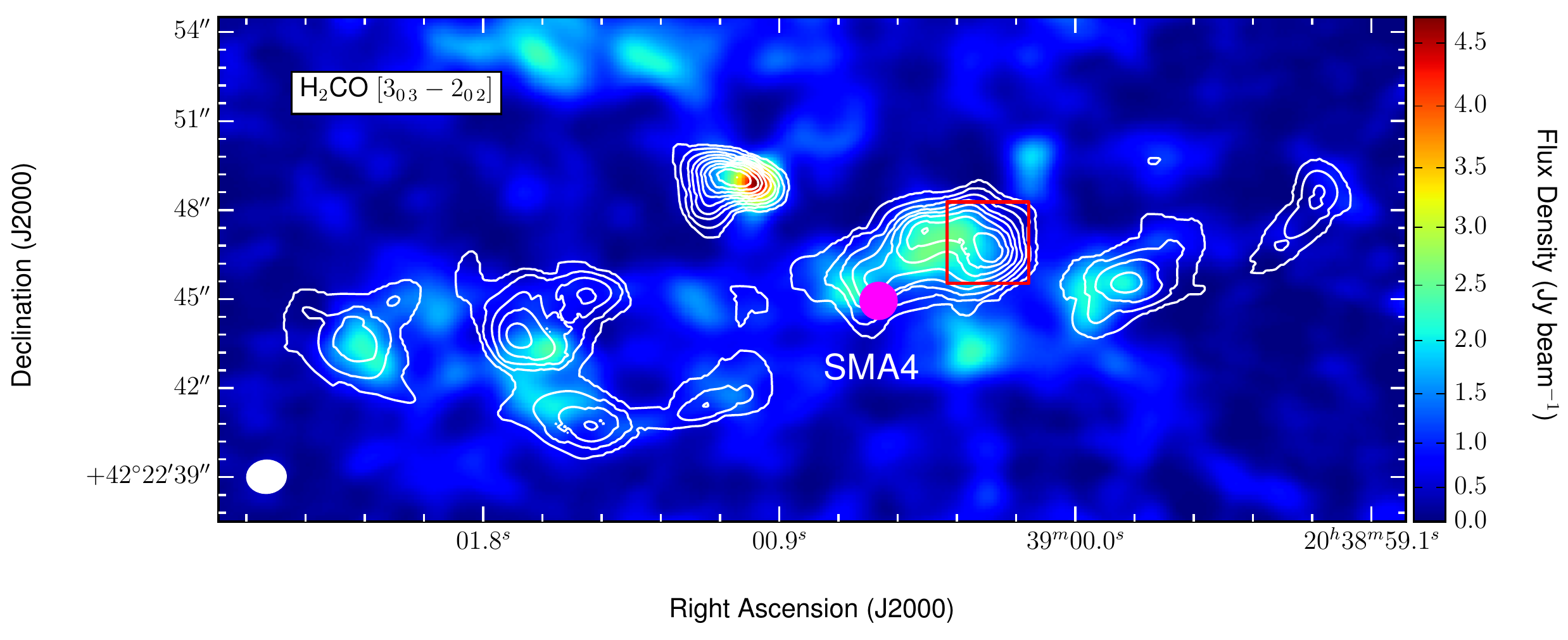}\\
\includegraphics[width=0.83\textwidth, angle=0, clip=True, trim= 0 0 0 0]{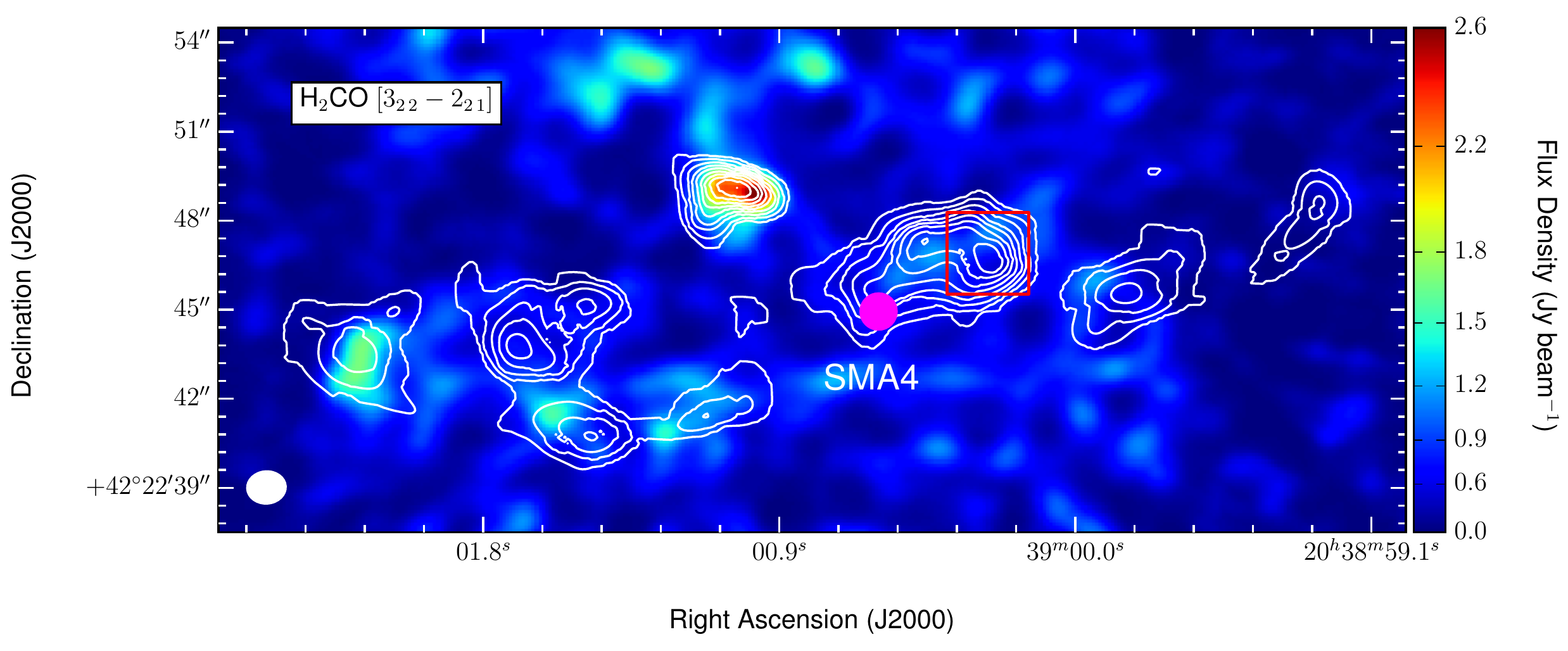}
\caption{Same as Figure 2, but with the SiO and H$_2$CO molecules.}
\label{figMom0}
\end{figure*}

\begin{figure*}[ht!]
\includegraphics[width=0.83\textwidth, angle=0, clip=True, trim= 0 0 0 0]{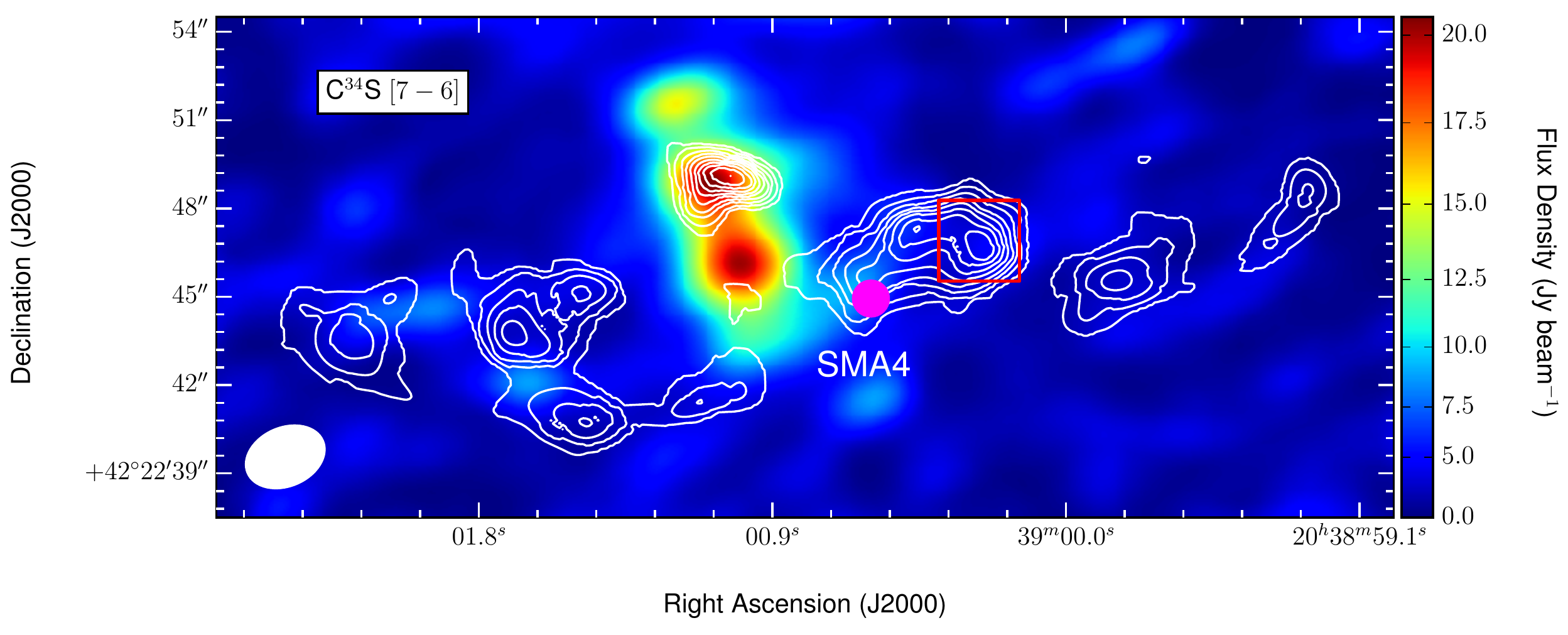}\\
\includegraphics[width=0.83\textwidth, angle=0, clip=True, trim= 0 0 0 0]{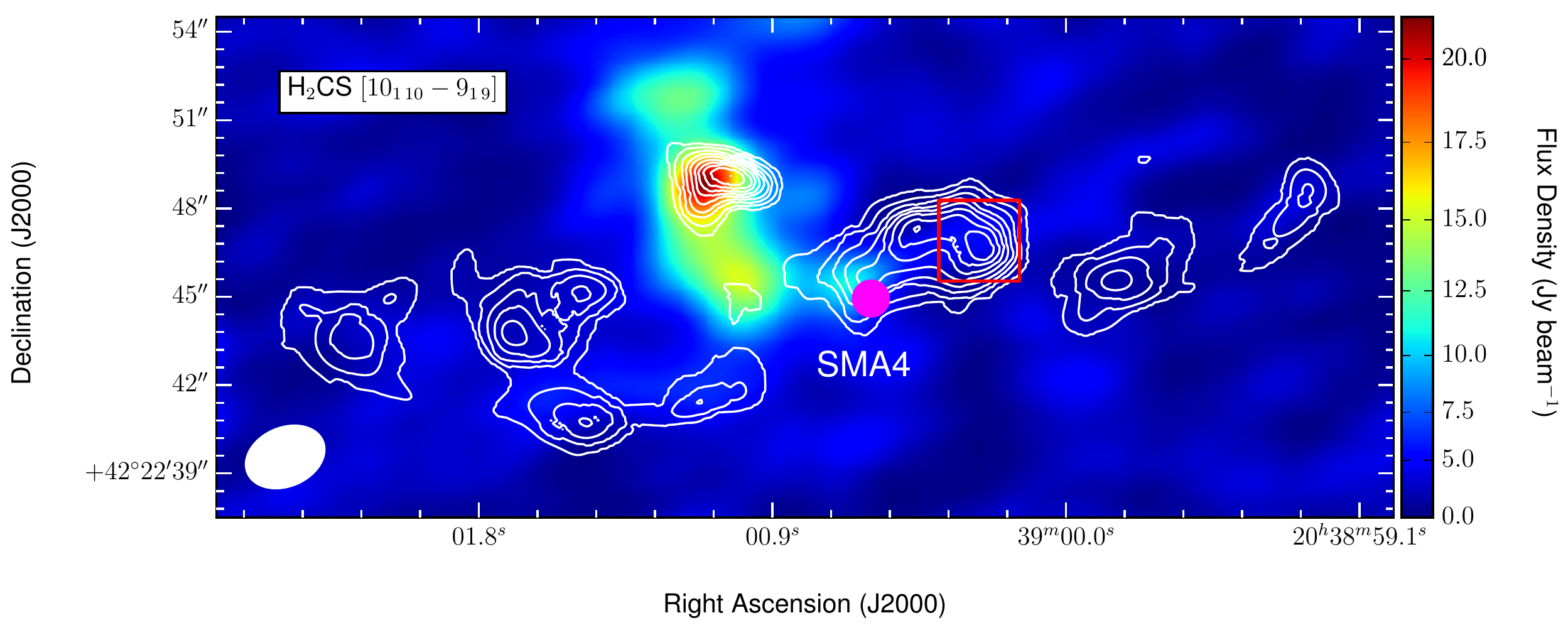}\\
\includegraphics[width=0.83\textwidth, angle=0, clip=True, trim= 0 0 0 0]{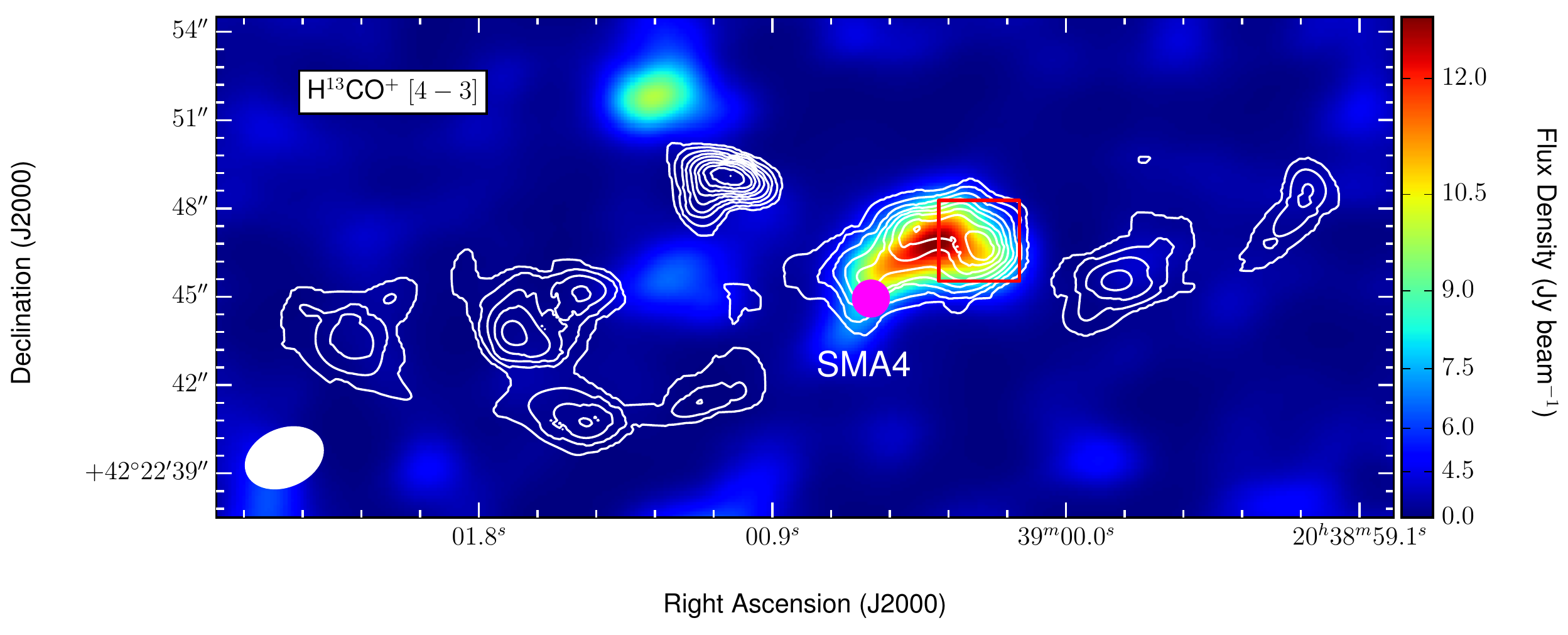}
\caption{Same as Figure 2, but with the C$^{34}$S, H$_2$CS, and H$^{13}$CO$^+$ molecules.}
\label{figMom0}
\end{figure*}

\clearpage

\subsection{Figures of methanol lines.}


\begin{figure*}[ht!]
\includegraphics[width=0.84\textwidth, angle=0, clip=True, trim= 0 0 0 0]{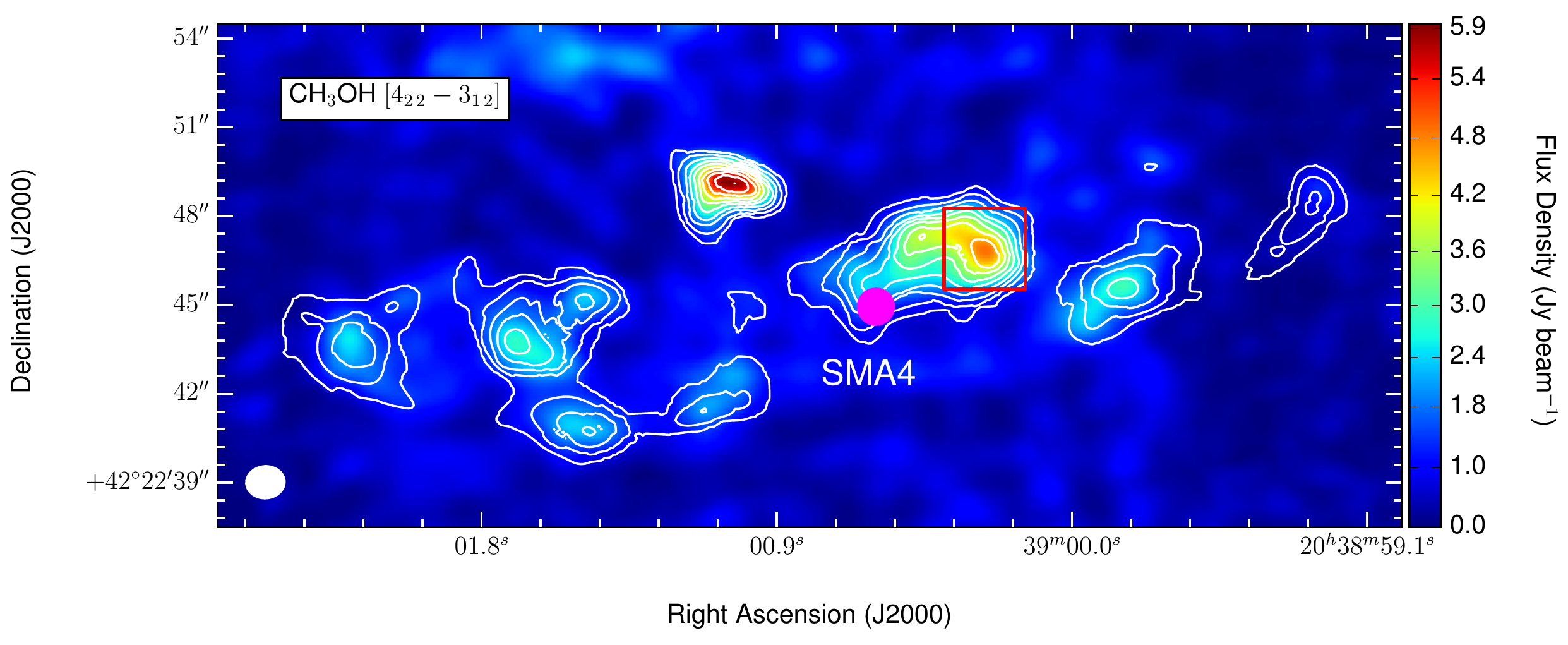}\\
\includegraphics[width=0.83\textwidth, angle=0, clip=True, trim= 0 0 0 0]{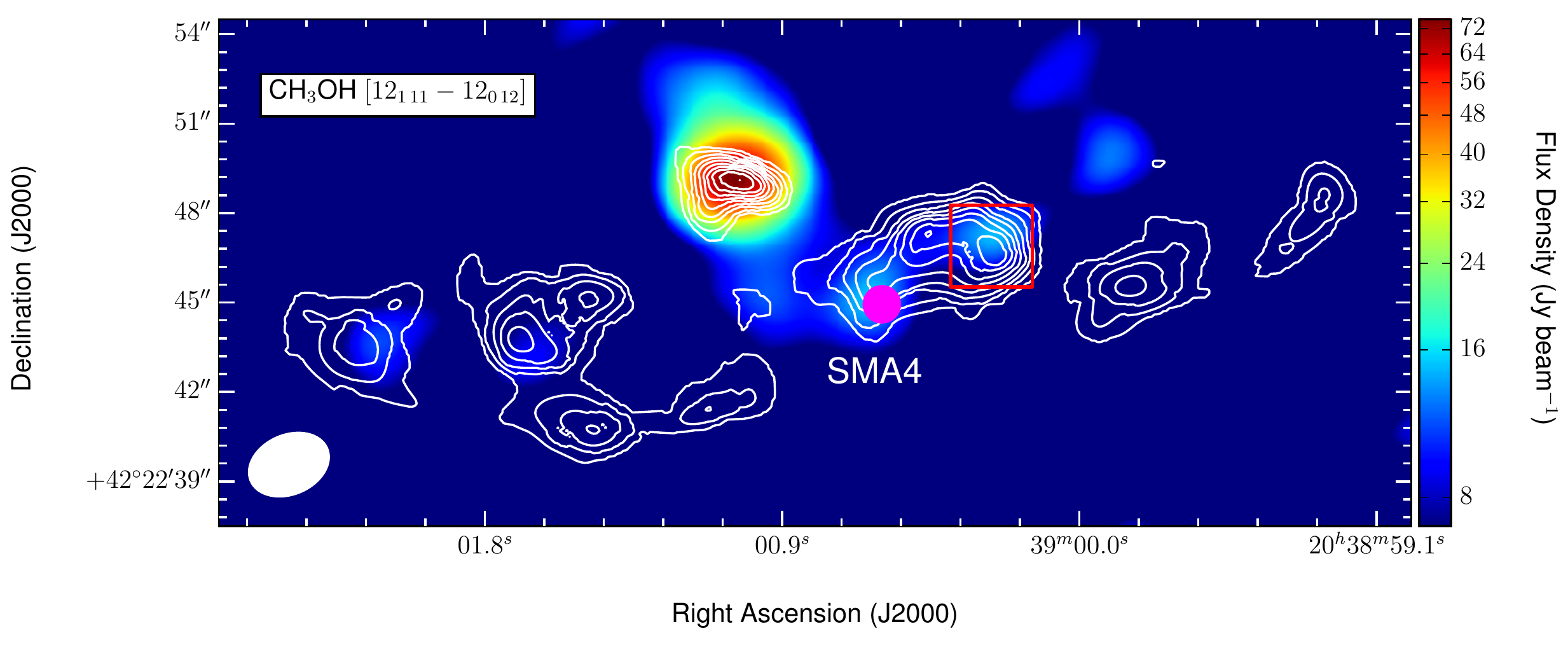}\\
\includegraphics[width=0.83\textwidth, angle=0, clip=True, trim= 0 0 0 0]{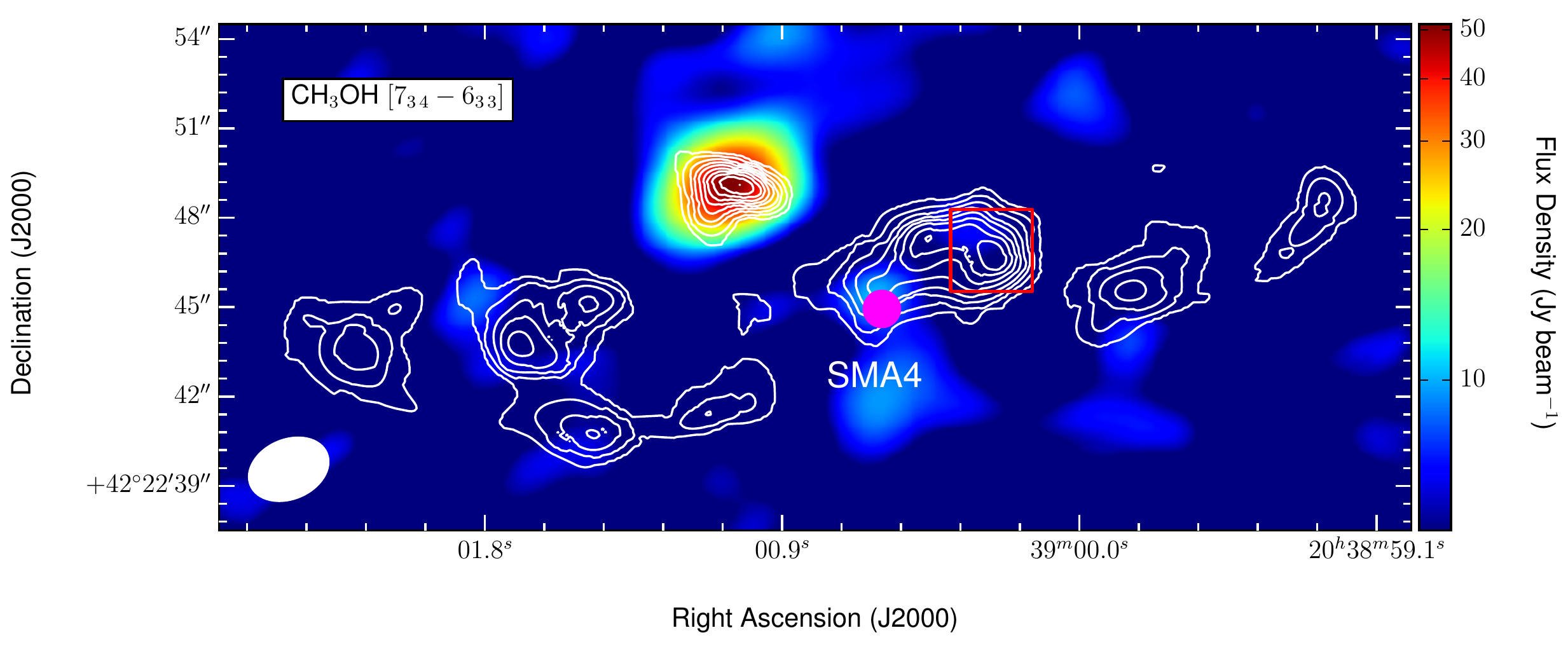}
\caption{Same as Figure 2, but with the CH$_3$OH molecule from different transitions.}

\label{figMom0}
\end{figure*}

\begin{figure*}[ht!]
\includegraphics[width=0.83\textwidth, angle=0, clip=True, trim= 0 0 0 0]{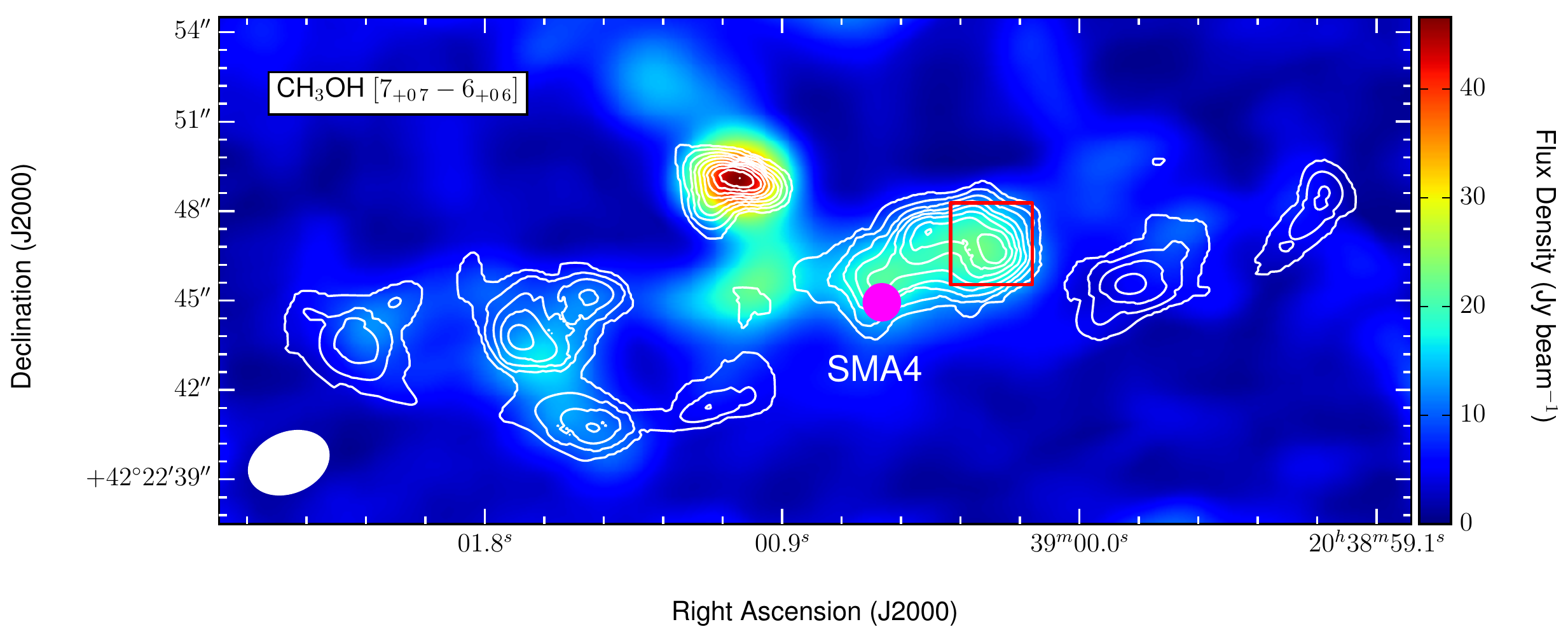}\\
\includegraphics[width=0.83\textwidth, angle=0, clip=True, trim= 0 0 0 0]{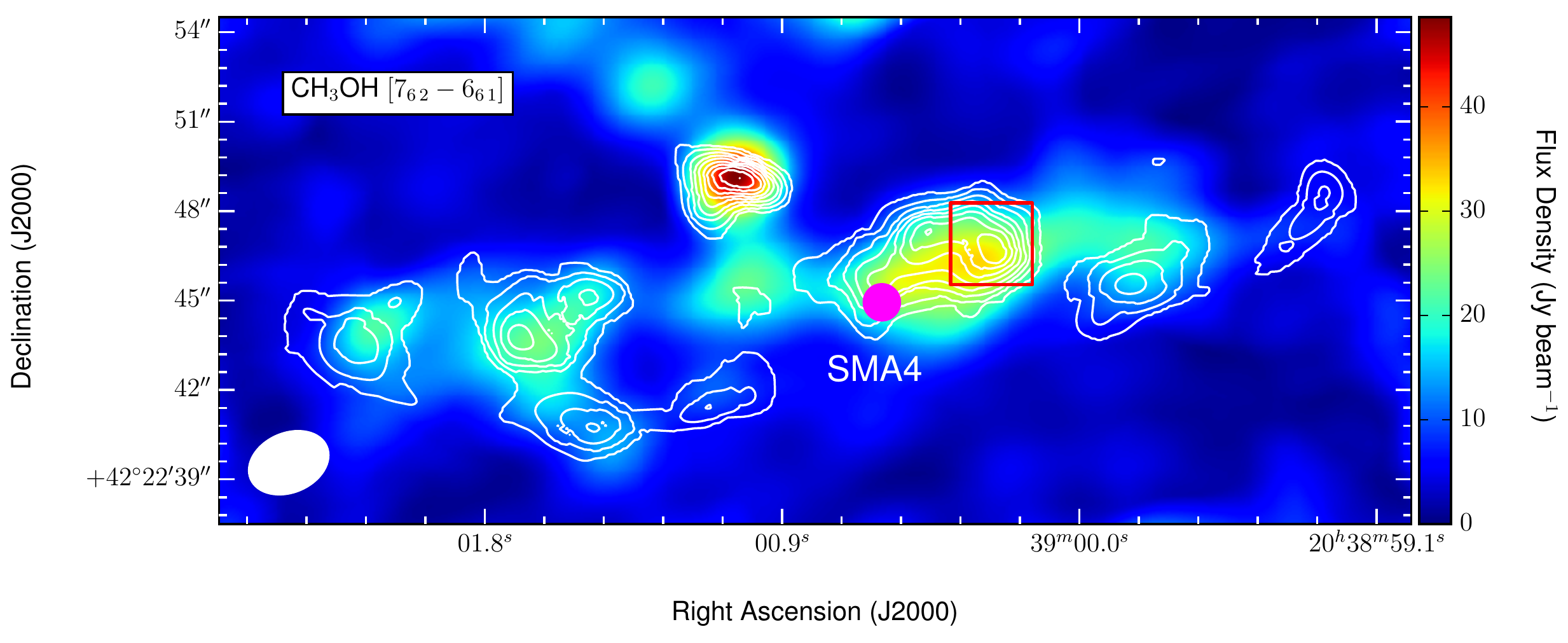}\\
\includegraphics[width=0.84\textwidth, angle=0, clip=True, trim= 0 0 0 0]{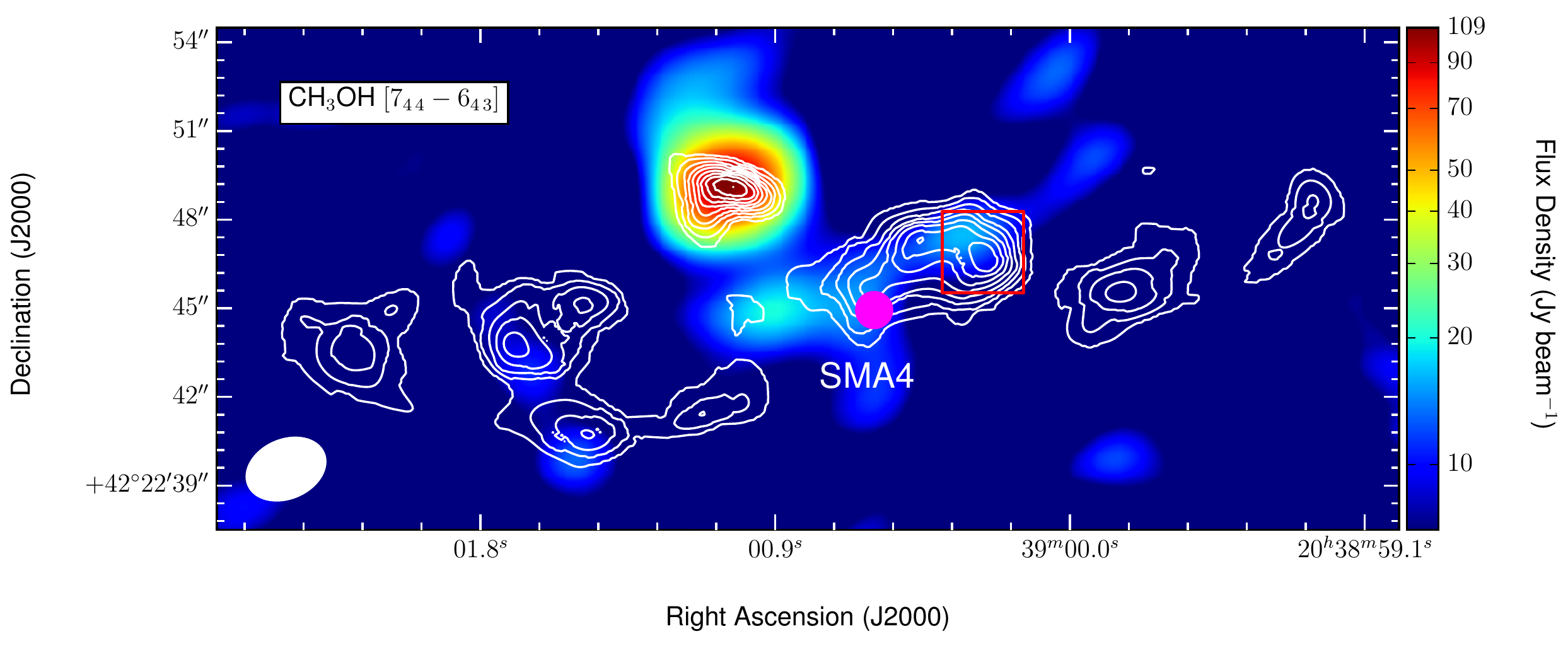}
\caption{Same as Figure 2, but with the CH$_3$OH molecule from different transitions.}
\label{figMom0}
\end{figure*}

\begin{figure*}[ht!]
\includegraphics[width=0.84\textwidth, angle=0, clip=True, trim= 0 0 0 0]{CH3OH_338540.eps}\\
\includegraphics[width=0.83\textwidth, angle=0, clip=True, trim= 0 0 0 0]{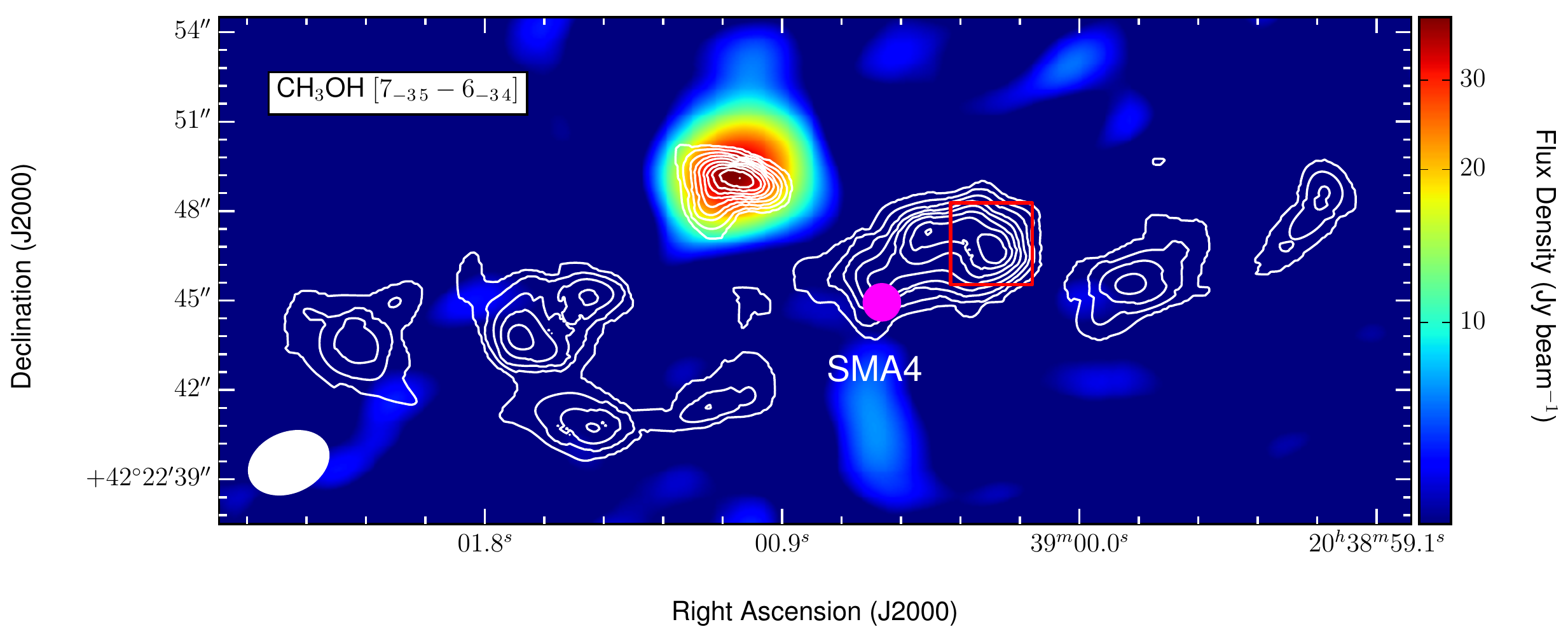}\\
\includegraphics[width=0.83\textwidth, angle=0, clip=True, trim= 0 0 0 0]{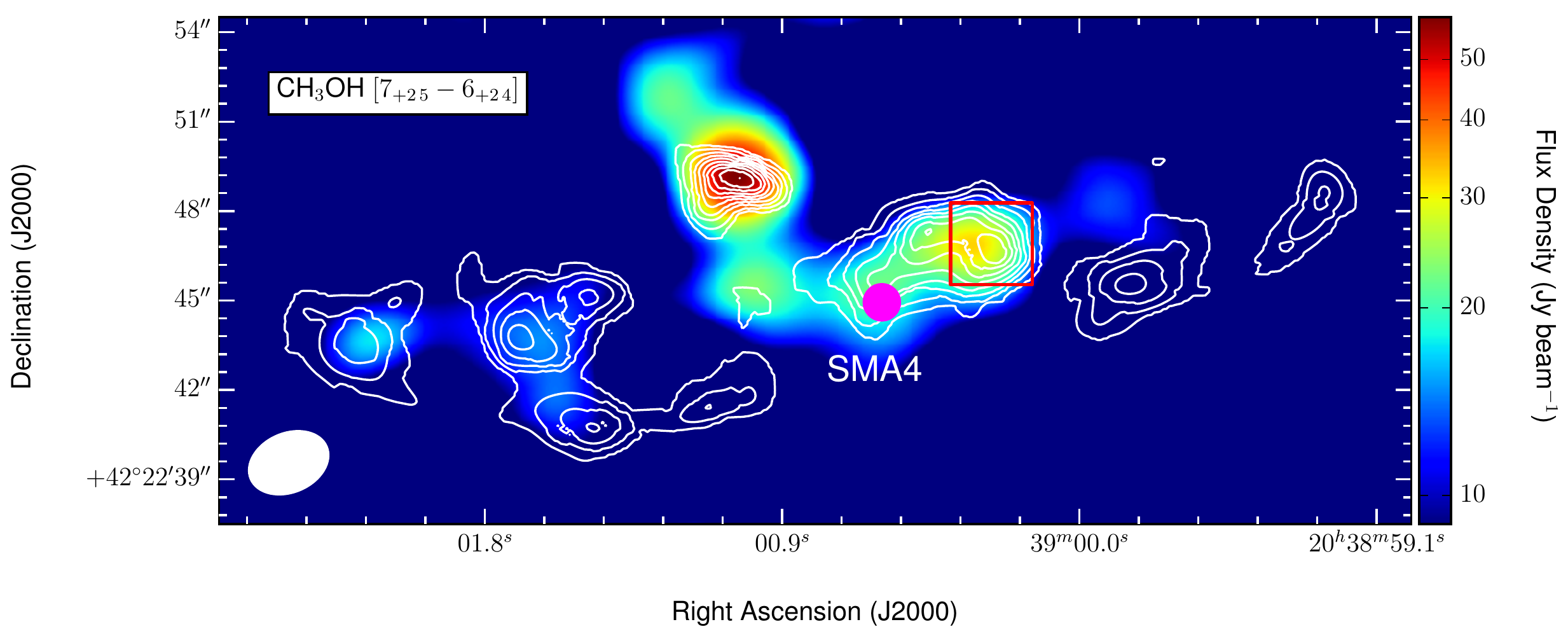}
\caption{Same as Figure 2, but with the CH$_3$OH molecule  from different transitions.}

\label{figMom0}
\end{figure*}

\begin{figure*}[ht!]
\includegraphics[width=0.84\textwidth, angle=0, clip=True, trim= 0 0 0 0]{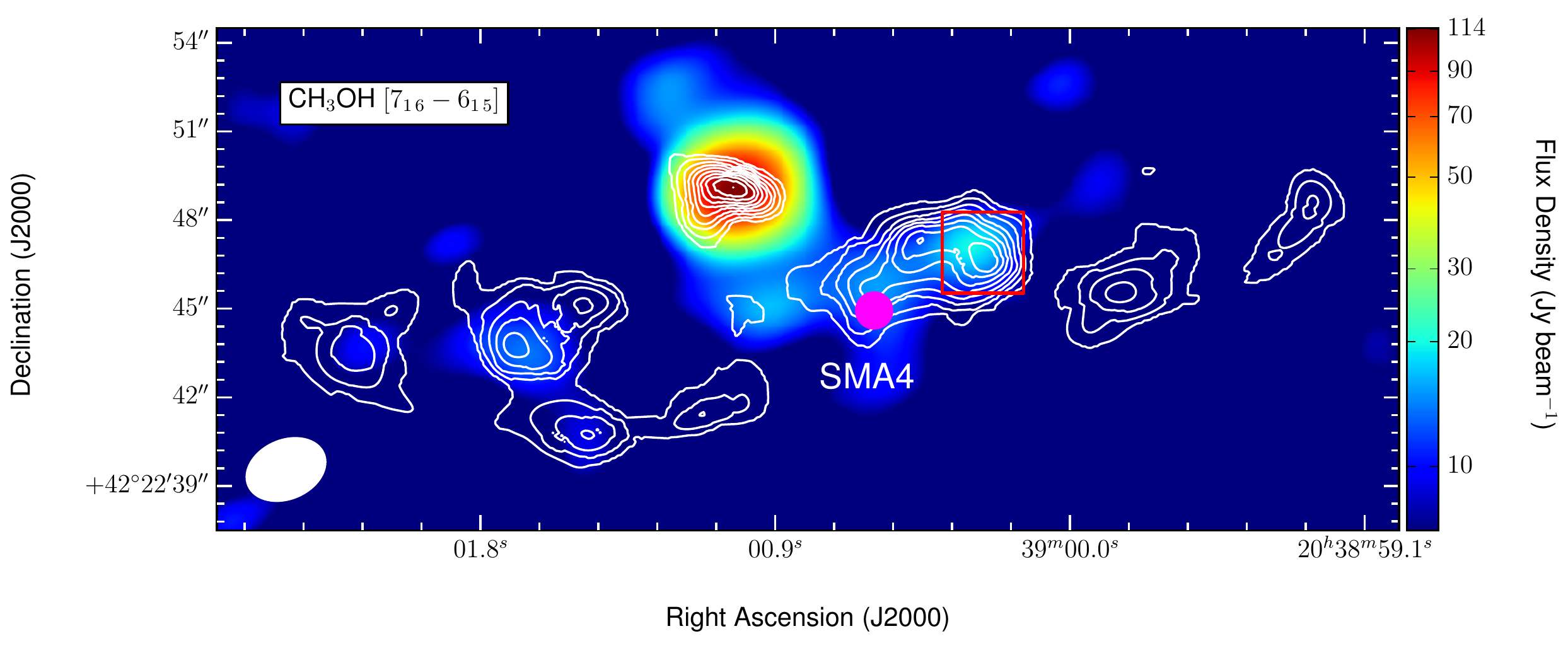}\\
\includegraphics[width=0.83\textwidth, angle=0, clip=True, trim= 0 0 0 0]{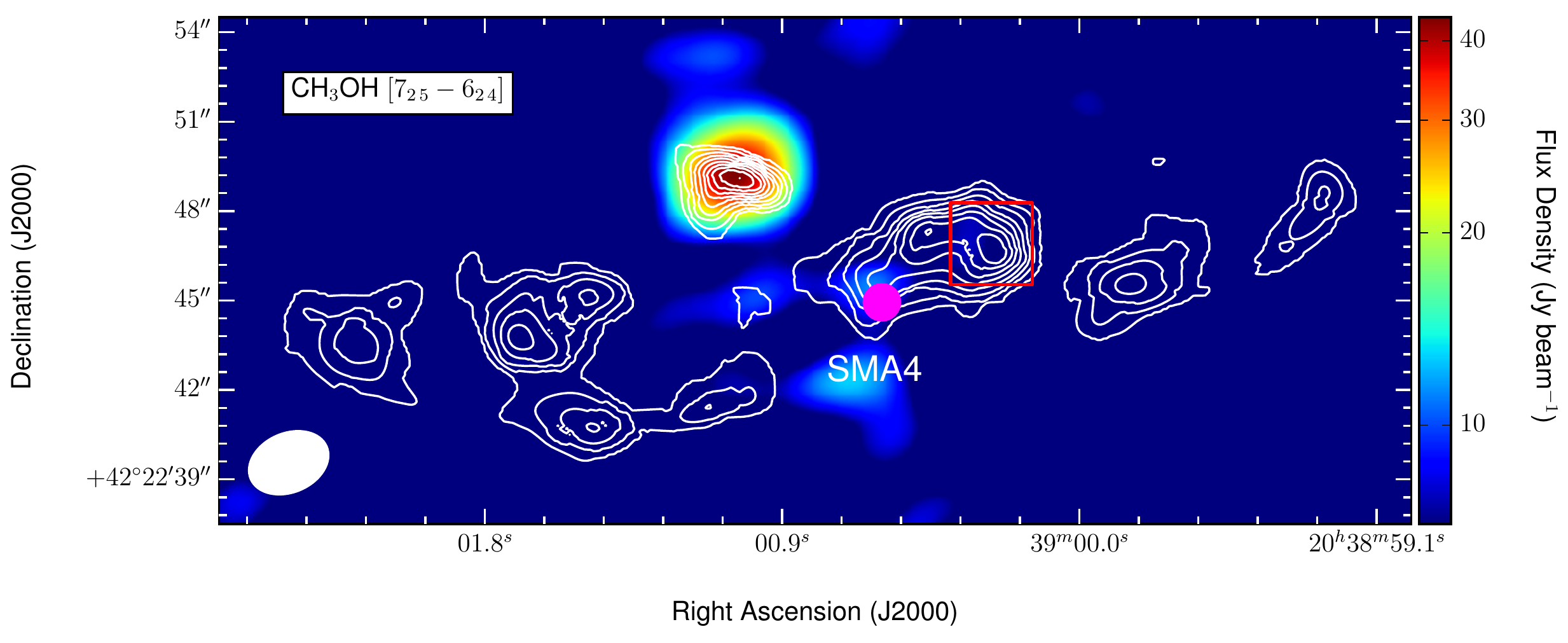}\\

\caption{Same as Figure 2, but with the CH$_3$OH molecule from different transitions.}
\label{figMom0}
\end{figure*}





\begin{thebibliography}{}
\bibitem[Araya et al.(2009)]{araya2009} Araya, E.~D., Kurtz, S., Hofner, P., \& Linz, H.\ 2009, \apj, 698, 1321 
\bibitem[Arce et al.(2008)]{arce2008} Arce, H.~G., Santiago-Garc{\'{\i}}a, J., J{\o}rgensen, J.~K., Tafalla, M., \& Bachiller, R.\ 2008, \apjl, 681, L21 
\bibitem[Bachiller et al.(2001)]{bachiller2001} Bachiller, R., P{\'e}rez Guti{\'e}rrez, M., Kumar, M.~S.~N., \& Tafalla, M.\ 2001, \aap, 372, 899 
\bibitem[Caux et al.(2011)]{caux2011} Caux, E., Bottinelli, S., Vastel, C., \& Glorian, J.~M.\ 2011, The Molecular Universe, 280, 120 
\bibitem[Chandler et al.(1993)]{chan1993} Chandler, C.~J., Gear, W.~K., \& Chini, R.\ 1993, \mnras, 260, 337 
\bibitem[Csengeri et al.(2011)]{csengeri2011} Csengeri, T., Bontemps, S., Schneider, N., et al.\ 2011, \apjl, 740, L5 
\bibitem[Davis et al.(2007)]{davis2007} Davis, C.~J., Kumar, M.~S.~N., Sandell, G., et al.\ 2007, \mnras, 374, 29 
\bibitem[Downes \& Rinehart(1966)]{downes1966} Downes, D., \& Rinehart, R.\ 1966, \apj, 144, 937 
\bibitem[Fish et al.(2011)]{fish2011} Fish, V.~L., Muehlbrad, T.~C., Pratap, P., et al.\ 2011, \apj, 729, 14 
\bibitem[Garay et al.(1998)]{garay1998} Garay, G., K{\"o}hnenkamp, I., Bourke, T.~L., Rodr{\'{\i}}guez, L.~F., \& Lehtinen, K.~K.\ 1998, \apj, 509, 768
\bibitem[Garrod \& Herbst(2006)]{gar2006} Garrod, R.~T., \& Herbst, E.\ 2006, \aap, 457, 927
\bibitem[Girart et al.(2013)]{girart2013} Girart, J.~M., Frau, P., Zhang, Q., et al.\ 2013, \apj, 772, 69 
\bibitem[Gooch(1996)]{goo96} Gooch, R.\ 1996, Astronomical Data Analysis Software and Systems V, 101, 80
\bibitem[Harvey et al.(1986)]{harvey1986} Harvey, P.~M., Joy, M., Lester, D.~F., \& Wilking, B.~A.\ 1986, \apj, 300, 737
\bibitem[Harvey et al.(1977)]{har1977} Harvey, P.~M., Campbell, M.~F., \& Hoffmann, W.~F.\ 1977, \apj, 211, 786  
\bibitem[Hennemann et al.(2012)]{hennemann2012} Hennemann, M., Motte, F., Schneider, N., et al.\ 2012, \aap, 543, L3
\bibitem[Herbst \& van Dishoeck(2009)]{her2009} Herbst, E., \& van Dishoeck, E.~F.\ 2009, \araa, 47, 427
\bibitem[Hern{\'a}ndez-Hern{\'a}ndez et al.(2014)]{hernandez2014} Hern{\'a}ndez-Hern{\'a}ndez, V., Zapata, L., Kurtz, S., \& Garay, G.\ 2014, \apj, 786, 38 
\bibitem[Ho et al.(2004)]{ho2004} Ho, P.~T.~P., Moran, J.~M., \& Lo, K.~Y.\ 2004, \apjl, 616, L1 
\bibitem[Jakob et al.(2007)]{jakob2007} Jakob, H., Kramer, C., Simon, R., et al.\ 2007, \aap, 461, 999
\bibitem[J{\o}rgensen et al.(2007)]{jorgensen2007} J{\o}rgensen, J.~K., Bourke, T.~L., Myers, P.~C., et al.\ 2007, \apj, 659, 479 
\bibitem[Kogan \& Slysh(1998)]{kogan1998} Kogan, L., \& Slysh, V.\ 1998, \apj, 497, 800 
\bibitem[Kumar et al.(2007)]{kum2007} Kumar, M.~S.~N., Davis, C.~J., Grave, J.~M.~C., Ferreira, B., \& Froebrich, D.\ 2007, \mnras, 374, 54 
\bibitem[Kurtz et al.(2004)]{kurtz2004} Kurtz, S., Hofner, P., \& {\'A}lvarez, C.~V.\ 2004, \apjs, 155, 149 
\bibitem[Lai et al.(2003)]{lai2003} Lai, S.-P., Girart, J.~M., \& Crutcher, R.~M.\ 2003, \apj, 598, 392 
\bibitem[Mangum et al.(1991)]{man1991} Mangum, J.~G., Wootten, A., \& Mundy, L.~G.\ 1991, \apj, 378, 576 
\bibitem[Minh et al.(2012)]{minh2012} Minh, Y.~C., Chen, H.-R., Su, Y.-N., \& Liu, S.-Y.\ 2012, Journal of Korean Astronomical Society, 45, 157 
\bibitem[Minh et al.(2011)]{minh2011} Minh, Y.~C., Liu, S.-Y., Chen, H.-R., \& Su, Y.-N.\ 2011, \apjl, 737, L25 
\bibitem[Motte et al.(2007)]{motte2007} Motte, F., Bontemps, S., Schilke, P., et al.\ 2007, \aap, 476, 1243 
\bibitem[M{\"u}ller et al.(2001)]{muller2001} M{\"u}ller, H.~S.~P., Thorwirth, S., Roth, D.~A., \& Winnewisser, G.\ 2001, \aap, 370, L49 
\bibitem[M{\"u}ller et al.(2005)]{muller2005} M{\"u}ller, H.~S.~P., Schl{\"o}der, F., Stutzki, J., \& Winnewisser, G.\ 2005, Journal of Molecular Structure, 742, 215 
\bibitem[M{\"o}ller, et al.(2017)]{muller2017} M{\"o}ller, T., Endres, C. \& Schilke, P.\ 2017, \aap, 598, A7.
\bibitem[{\"O}berg et al.(2011)]{ober2011} {\"O}berg, K.~I., van der Marel, N., Kristensen, L.~E., \& van Dishoeck, E.~F.\ 2011, \apj, 740, 14 
\bibitem[Palau et al.(2017)]{palau2017} Palau, A., Walsh, C., S{\'a}nchez-Monge, {\'A}., et al.\ 2017, \mnras, 467, 2723 
\bibitem[Plambeck \& Menten(1990)]{plambeck1990} Plambeck, R.~L., \& Menten, K.~M.\ 1990, \apj, 364, 555 
\bibitem[Reipurth \& Schneider(2008)]{reipurth2008} Reipurth, B., \& Schneider, N.\ 2008, Handbook of Star Forming Regions, Volume I, 36 
\bibitem[Rygl et al.(2012)]{rygl2012} Rygl, K.~L.~J., Brunthaler, A., Sanna, A., et al.\ 2012, \aap, 539, A79
\bibitem[Sault et al.(1995)]{sau1995} Sault, R.~J., Teuben, P.~J., \& Wright, M.~C.~H.\ 1995, Astronomical Data Analysis Software and Systems IV, 77, 433
\bibitem[Schneider et al.(2010)]{schneider2010} Schneider, N., Csengeri, T., Bontemps, S., et al.\ 2010, \aap, 520, A49 
\bibitem[Tafalla \& Hacar(2013)]{taf2013} Tafalla, M., \& Hacar, A.\ 2013, \aap, 552, L9
\bibitem[Vall{\'e}e \& Fiege(2006)]{vallee2006} Vall{\'e}e, J.~P., \& Fiege, J.~D.\ 2006, \apj, 636, 332 
\bibitem[Zapata et al.(2012)]{zapata2012} Zapata, L.~A., Loinard, L., Su, Y.-N., et al.\ 2012, \apj, 744, 86
\bibitem[Zapata et al.(2013)]{zapata2013} Zapata, L.~A., Schmid-Burgk, J., P{\'e}rez-Goytia, N., et al.\ 2013, \apjl, 765, L29 
\bibitem[Zapata et al.(2017)]{zapata2017} Zapata, L.~A., Schmid-Burgk, J., Rodr{\'{\i}}guez, L.~F., Palau, A., \& Loinard, L.\ 2017, \apj, 836, 133 
\end{thebibliography}
\end{document}